\documentclass[12pt,letter,american,intoc,bibliography=totoc,index=totoc,BCOR10mm,captions=tableheading,nottitlepage,fleqn]{article}

\usepackage{setspace, geometry}
\usepackage{etex}
\usepackage{babel,newcent,textcomp}
\usepackage{scrextend}
\usepackage{booktabs, threeparttable}
\usepackage{array}
\usepackage[multiple,bottom]{footmisc}
\usepackage{microtype}

\usepackage[T1]{fontenc}
\usepackage[latin9]{inputenc}
\usepackage[autostyle]{csquotes}
\usepackage[style=bwl-FU, citestyle=authoryear, backend=biber, maxbibnames=99, uniquename=false, uniquelist=false]{biblatex}
\usepackage{lmodern}

\usepackage{sgame}
\usepackage{fancyhdr}
\usepackage{amsmath,amssymb,bbm,bm}
\usepackage[amsthm]{ntheorem}
\usepackage{dsfont}
\usepackage{tablefootnote}
\usepackage[labelfont=bf]{caption}
\usepackage{enumitem}
\usepackage{geometry}
\usepackage{lscape}
\usepackage{amsmath}
\usepackage{amsfonts}
\newcommand{\e}{\mathbb{E}}
\newcommand{\logit}{\text{logit}}
\newcommand{\s}{\mathbb{S}}

\usepackage[unicode=true,
 bookmarks=true,bookmarksnumbered=true,bookmarksopen=true,bookmarksopenlevel=1,
 breaklinks=false,pdfborder={0 0 0},backref=false,colorlinks=true,linkcolor=black,citecolor=black,anchorcolor=black,urlcolor=blue]
 {hyperref}
\hypersetup{pdftitle={Overinference from Weak Signals and Underinference from Strong Signals},
 pdfauthor={Ned Augenblick, Eben Lazarus, and Michael Thaler},
 pdfpagelayout=OneColumn, pdfnewwindow=true, pdfstartview=XYZ, plainpages=false}

\def\seline{\addlinespace[-3pt]}

\usepackage{tikz}

\usetikzlibrary{shapes.geometric, arrows, positioning}
\tikzset{
    >=stealth',
    box/.style={
           rectangle,
           rounded corners,
           draw=black, very thick,
           text width=6.5em,
           minimum height=1.5em,
           text centered},
    boxsmall/.style={
           rectangle,
           rounded corners,
           draw=black, very thick,
           text width=6em,
           minimum height=1.5em,
           text centered},
     simple/.style={
           text centered},
    arrow/.style={
           ->,
           thick,
           shorten <=2pt,
           shorten >=2pt,}
}
\usepackage{graphicx}

\usepackage{verbatim}

\usepackage{cleveref}
\usepackage[figure]{hypcap}
\usepackage{calc}
\let\mySection\section\renewcommand{\section}{\suppressfloats[t]\mySection}
\newcommand{\p}{\mathbb{P}}

\theoremstyle{break}
\newtheorem{theorem}{Theorem}

\newtheorem{proposition}[theorem]{Proposition}
\theoremstyle{definition} 
\newtheorem{assumption2}{Assumption}

\newcommand*{\printdate}{%
   \ifcase \month\or January\or February\or March\or April\or May\or June\or July\or
    August\or September\or October\or November\or December\fi \space \number\year}

\theoremstyle{empty}

\newenvironment{definition}[1][Definition]{\begin{trivlist}
\item[\hskip \labelsep {\bfseries #1}]}{\end{trivlist}}

\makeatother

\begin{document}

\lhead[]{}

\rhead[]{}

\lfoot[\thepage]{}

\cfoot{}

\rfoot[]{\thepage}

\newpage

\begin{titlepage}
\title{\begin{changemargin}{5mm}{5mm} 
\begin{center}
Overinference from Weak Signals and \\ Underinference from Strong Signals
\end{center}
\end{changemargin}}
\author{Ned Augenblick\thanks{\scriptsize{Haas School of Business, University of California, Berkeley. Email: \href{mailto:ned@haas.berkeley.edu}{ned@haas.berkeley.edu}.}},$\,$ Eben Lazarus\thanks{\scriptsize{Haas School of Business, University of California, Berkeley. Email: \href{mailto:lazarus@berkeley.edu}{lazarus@berkeley.edu}.}},$\,$ and Michael Thaler\thanks{\scriptsize{University College London. Email: \href{mailto:michael.thaler@ucl.ac.uk}{michael.thaler@ucl.ac.uk}. \\[.1em] \hspace*{15.5pt} An early version of this paper was circulated as \textcite{T-WP}. We would especially like to thank Matthew Rabin for his advice throughout this project. We are also grateful to the editor (Andrei Shleifer) and four anonymous referees for very helpful feedback, as well as Nick Barberis, Francesca Bastianello, Roland B\'{e}nabou, Pol Campos-Mercade, Stefano DellaVigna, Ben Enke, Christine Exley, Xavier Gabaix, Nicola Gennaioli, Thomas Graeber, Benjamin H\'{e}bert, Spencer Kwon, Martin Lettau, Alessandro Lizzeri, Peter Maxted, Terrance Odean, Pietro Ortoleva, Cameron Peng, Josh Schwartzstein, David Sraer, David Thesmar, Mike Woodford, Leeat Yariv, and seminar participants at BEAM, NBER Behavioral Finance, Princeton, Berkeley Haas, HBS, Lund, Stockholm University, and UBC. The experiments were approved by IRBs at Princeton University (13114-03), UCL (SHSEco-2223-003-1), and UC Berkeley (2023-07-16581). The pre-analysis plans are available at \url{https://aspredicted.org/ax4wg.pdf} (Study~1a), \url{https://aspredicted.org/8Q4_6Y9} (Study~1b), and \url{https://aspredicted.org/SYW_QWF} (Study~2).}}}
\date{First version: September 2021 \\
This version: \printdate} 
\vspace{-3mm}

\maketitle
\abstract{
When people receive new information, sometimes they revise their beliefs too much, and sometimes too little. In this paper, we show that a key driver of whether people overinfer or underinfer is the strength of the information. Based on a model in which people know which direction to update in, but not exactly how much to update, we hypothesize that people will overinfer from weak signals and underinfer from strong signals. We then test this hypothesis across four different environments: abstract experiments, a naturalistic experiment, sports betting markets, and financial markets. In each environment, our consistent and robust finding is overinference from weak signals and underinference from strong signals. Our framework and findings can help harmonize apparently contradictory results from the experimental and empirical literatures.

\vspace{5mm}
}

%\vspace{1mm}
%\noindent \textbf{Keywords:} 

\bigskip

\setcounter{page}{0}
\thispagestyle{empty}
\end{titlepage}
\pagebreak \newpage

\section{Introduction}
\label{introduction}

How do people update their beliefs given new information? This important question has spawned a vast experimental and empirical literature, with seemingly contradictory results. A common finding in the experimental literature is that people often underreact to information in standard updating tasks. But this is seemingly at odds with observational evidence from real-world settings, such as excess volatility in asset prices, which often appears more consistent with overreaction. Updating behavior is clearly context-dependent, but what specific mediating factors help explain how people will respond to a given piece of information?

This paper hypothesizes that people commonly \emph{overinfer} from \emph{weak} information and \emph{underinfer} from \emph{strong} information. We start with a theoretical framework in which we formalize these concepts and provide simple but general conditions under which the effect will arise. We then use a classic experimental paradigm to show that while people do underinfer when provided with strong signals (as commonly studied in the lab), they overinfer from sufficiently weak signals (which have been previously understudied). After replicating and extending this result in a follow-up study, we then demonstrate that this effect is not an artifact of the abstract environment by showing the same results in a novel experiment with more naturalistic information. Finally, we use two empirical settings to show that betting markets and asset prices exhibit excess volatility when information is weak, but this effect reverses with sufficiently strong information.

To understand the intuition for our hypothesis, consider the constant stream of information faced by people every day. People might read a new poll about an election, have a conversation with their boss at work, or see news about daily stock-market movements. In many cases, people understand the \textit{directional} impact this news should have on their beliefs, but are less certain about the \textit{strength} of the information. That is, they know that better polling raises a candidate's election chances, managerial praise raises promotion chances, and positive stock returns raise early retirement chances, but they don't know exactly how much their beliefs should move. How will a person update in this situation? Consider the extreme case in which the person knows that a signal is positive but is completely unsure about the signal's strength. The person only knows that beliefs should rise, and therefore updates as if the news has ``intermediate'' strength. But if a person is always updating an intermediate amount, then they will be overreacting to weak news and underreacting to strong news.\footnote{We largely use the terms ``overinfer'' and ``overreact'' interchangeably. However, we see a subtle difference: a person ``overinfers'' if they perceive a signal as more informative than a Bayesian would, while ``overreaction'' is the resultant behavior of reacting too strongly. We will generally use ``overinfer'' when we are clearly discussing overestimating signal strength (such as in our theory), while we will generally use ``overreact'' when discussing observable behavior. We will only highlight the difference when there is a contaminating force (such as base-rate neglect) that might cause beliefs to react too strongly for a reason other than overinference.}
In other words, when people know the signal's direction, \textit{insensitivity} to signal strength leads to a pattern of over- and underreaction relative to a full Bayesian.

In more realistic scenarios, people will have a rough guess of the strength of signals they receive. This \emph{estimate} might be based on a simplified model, unconscious approximation, or constrained information processing given attention to certain aspects of a signal. Given that the estimate is imperfect, the person should still shrink their estimate toward an intermediate strength. While different people can have different estimates given the same information, the shrinkage will \emph{on average} lead to overreaction to weak signals and underreaction to strong signals. In \Cref{sec:theory}, we model this intuition formally, and show that it holds across a general set of information structures, estimation strategies, and possibly non-Bayesian updating rules. We then use distributional assumptions to obtain a set of simple, parametric updating rules that can be taken to the data.

Our theory relies throughout on four important, high-level assumptions. First, it assumes that people pay attention to the information they receive (instead of, for example, ignoring weak signals altogether). Second, it assumes that the directional meaning of information is unambiguous. Third, it assumes that people form reasonable estimates of the uncertain signal strength. Fourth, our theory requires that people be at least partially aware that their signal-strength estimate is imperfect. These assumptions of course do not apply to all settings.\footnote{First, our theory suggests that people facing weak news will overreact conditional on paying attention, but there may be some (unmodeled) information that is weak enough that it goes unnoticed. Second, the directional impact of some news is ambiguous. In these settings, we broadly predict that people shade their posterior belief toward their prior belief and thus generally underreact. Third, if people have a clear bias in their estimates, that bias can overwhelm our effect. Finally, if people believe that their estimated signal strength is perfectly accurate, then they will not shade it toward an intermediate value.} That said, we think they are present in many important empirical applications, including the settings we consider in our empirical analyses.

To test our theoretical predictions, we study how people's reaction to new information varies when signals are weak versus strong. To do so, we create three controlled lab experiments (with preregistered hypotheses) and study two empirical environments in which signal strengths vary systematically and updating behavior can be measured consistently. While the environments and methods differ, we find consistent results across each of the settings.

The first two experiments (Studies 1a and 1b) employ the classic ``bookbag-and-poker-chips'' paradigm (\cite{GHR65}). This is the most commonly used experimental setup to study belief updating; for example, \citeauthor{B19}'s (\citeyear{B19}) survey of the literature includes 500 experimental treatment blocks across 21 papers that study inference from symmetric binary signals about a binary state, which is our main focus. Belief updating in these settings often features underreaction relative to Bayes' rule, with \citeauthor{B19}'s ``Stylized Fact 1'' stating that ``Underinference is by far the dominant direction of bias.'' The vast majority of the evidence, though, is on strong signals: in all of these papers where people receive one symmetric binary signal, the diagnosticity --- the likelihood of seeing a ``high'' signal conditional on the ``high'' state --- is never lower than $3/5$. Our hypothesis is that people will overinfer given lower signal strengths. There is a hint of the importance of signal strength for underreaction: \citeauthor{B19} notes that ``Underinference...is more severe the larger is the diagnosticity,'' suggesting that the pattern may flip. We hypothesize that this is indeed the case.

To test our hypothesis, in Study 1a we run this standard experiment with 500 participants using our much wider range of signal strengths. In the main treatment, participants are presented with two decks of cards: a green deck containing more spades than diamonds, and a purple deck with more diamonds than spades. Participants see a single card drawn from one of the two decks, and they must then estimate probabilities for which deck was chosen based on the suit of the drawn card. We vary signal strength by changing the number of spades and diamonds in each deck. This design broadly aligns with our theoretical setup: the direction to update is fairly clear (e.g., a spade is evidence for the green deck), but the correct magnitude is less obvious (requiring clear understanding of the data-generating process, correct use of Bayes' rule, and exact calculation of the proportion of suits in each deck).

We find that almost all participants update their beliefs in the right direction, but there is substantial heterogeneity in how much they revise their beliefs. We interpret this as showing that participants know to update in a particular direction, but differ in how they perceive the strength of the signal.
Notably, participants' answers are not random: the average perceived signal strength rises monotonically with the true strength. Our main result, however, is that this relationship is muted, leading to overreaction to weak signals and underreaction to strong signals in a manner consistent with our theory. 
Reassuringly, our estimates of the magnitudes of underinference for the high-strength signals are in line with the previous literature. It is only in the previously understudied low-strength signals that we find overinference: for very weak signals, participants act as if signals are twice as strong as they truly are.\footnote{In particular, following nearly all existing evidence, we find robust evidence for underinference for strong signals with diagnosticity $p \geq 2/3$. For $p \in [3/5, 2/3]$, we do not find consistent evidence for over- or underinference, as is common in the literature (\cite{B19}). Finally, for $p \in (1/2, 3/5)$, we find clear and robust evidence for overinference.}

Study 1a focuses on the case in which both decks are equally likely to be drawn ex ante, so we conduct a follow-up in Study 1b in which we systematically vary the prior (considering values of 1/4, 1/3, and 1/2) in addition to the signal strength. All of our main findings continue to hold. Participants again overreact to weak signals and underreact to strong signals. While we estimate that people exhibit modest base-rate neglect, our core findings about inference are not substantially affected. In other words, although people's biases in using base rates can impact how they react to new information, disentangling these biases from our effects does not impact our conclusion that people overinfer from weak signals and underinfer from strong signals.

Exploring heterogeneity in updating, both experiments provide further evidence in line with the theoretical framework. Intuitively, the theory suggests that our effect will be stronger for people with less precise estimates of the signal strength. Consistent with this prediction, we find that our effect is stronger for people who exhibit more variance in their level of under- and overinference in 1a and 1b, have less task experience in 1a and 1b, have lower scores on a cognitive reflection test (adapted from \cite{F05}) in 1a, and state that they are more uncertain about their answers in 1b (adapted from \cite{EG-WP}). 

Studies 1a and 1b provide clean evidence for our effect, as the bookbags-and-poker-chips setting allows us to manipulate the DGP and compare people's behavior to an objective benchmark. But this control comes with some costs: the setting is quite abstract, and signals are difficult to understand largely as a result of numerical and calculation-related complexity. As with many experiments, one may be concerned that people treat this math-exam-like situation in a different way than in real-life scenarios.

Given this concern, our Study 2 analyzes belief updating in a more naturalistic setting, where participants are not provided with precise numbers representing likelihoods or signal strengths. Since naturalistic DGPs are often highly complicated, it is challenging to find an appropriate environment. Such an environment must (1) be reasonably understood by participants, (2) allow for clean variation in signal strength, and (3) allow some way to estimate the correct answer in order to calculate under- and overreaction. To address these challenges, we design a new experiment in which we ask basketball fans to predict the outcome of an NBA basketball game given sequences of game scenarios. For example, we elicit the probability that a team wins when they are ahead by 1 point with 2 minutes left in the game, and then we elicit it again given a scenario in which they have just made a shot to go ahead by 3 points a few seconds later. Although the DGP itself is complex, (1) the scenario is simple enough for basketball fans to immediately understand it, (2) the strength of the same news (like a scored basket) changes over the course of the game, and (3) we can use a data-driven, third-party benchmark estimate of signal strength. Note again that as in our theory, the direction of the news is clear (a made shot increases the probability of winning), but the exact change in probability in different scenarios is less clear (requiring some estimation process given personal experience and understanding of basketball games). While there are costs in moving away from a fully controlled DGP, this environment provides a much more naturalistic source of uncertainty about signal strengths.

To implement Study 2, we recruited 500 basketball fans, providing them with sequences of events over the course of four quarters of a hypothetical NBA game. Here, the variation in information strength is largely driven by timing: making a basket to take a lead in the fourth quarter is a much stronger signal than making a basket in the first quarter. As in the abstract experiments, we find that the vast majority of participants update in the right direction, but there is dispersion in the perceived strength of each signal; not surprisingly, people agree that a basket raises winning probability, but differ on their view of how much. Crucially, people again are not answering randomly: on average, a basket is seen as a stronger signal in the fourth quarter than in the first. But just as before, the relationship is muted, such that people on average overreact to weak signals (in the first quarter) and underreact to strong signals (in the fourth quarter), switching from over- to underreaction on average in the third quarter. Overall, these findings replicate the core findings from Studies 1a and 1b in a more realistic setting.

While Study 2 is more naturalistic than Studies 1a and 1b, it still places participants in a new experimental paradigm with fictional scenarios and relatively low stakes. In light of these concerns, we turn to evidence from more realistic high-stakes settings by studying the movement of market-implied probability distributions in both (1) sports betting markets and (2) financial markets.  For (1), we use over 5~million transactions from a large sports prediction market for five major sports, corresponding to about 260,000 sporting events. The market-implied beliefs for these sporting events --- particularly the subsample of NBA games~--- provide an empirical analogue to our Study 2. For (2), we study S\&P~500 index option markets, using option-implied beliefs regarding the future value of the S\&P from daily option prices observed over a roughly 20-year span.

These settings allow us to examine external validity but come with their own challenges. 
Perhaps the most important one is that we can no longer create credible estimates of the Bayesian probability for a given situation, as we see neither the 
full information set of participants nor the structure of the DGP.\footnote{In our finance data, it is clear that creating a ``correct'' forecast of the distribution of future outcomes is infeasible. In the sports data, one could create a reasonable forecast given observables (like score and game time), but this would not reflect the observer's full information set (injury or foul issues, game importance, whether Drake is courtside, etc.), and therefore stating that the observer's beliefs are wrong is dubious. This is not an issue in the experiment because participants' information sets are limited and controlled by the experimenter. We also face challenges related to the use of prices (which reflect the marginal trader's beliefs and risk preferences) instead of individual beliefs. We discuss how we deal with these in Section~\ref{sec:markets}.} 
To overcome this challenge, we develop a new empirical method based on  theoretical results from \textcite{AR21} and \textcite{AL-WP}. The core intuition of these papers is that, when a Bayesian is changing their beliefs over time about some event, they must be learning something and thus on average must reduce their uncertainty correspondingly. This intuition can be formalized by defining \textit{movement} as the sum of the squared deviations of changes in beliefs over time, and \textit{uncertainty reduction} as the drop in perceived variance in the outcome. 
While movement and uncertainty reduction may differ for a given signal realization, they must be equal in expectation across signal realizations, \emph{regardless of the DGP}. This insight allows for a DGP-agnostic test of Bayesian updating in observational data. And crucially, these statistics are intuitively and theoretically related to over- and underinference: overinference will lead to positive excess movement relative to uncertainty reduction on average, while underinference will lead to too little movement relative to the reduction in uncertainty.

While this allows for an intuitive test of over- vs.\ underinference with an unknown DGP, to test our theory, we also need to distinguish situations in which signals are weak versus strong. Given that the signal strength is also unobservable, we turn to the same separating variable from Study 2: time to resolution. As in the experiment, our insight is that when a person is predicting the value of the S\&P 500 in three months, information today should generally not lead to much belief movement; meanwhile, information today is highly informative for the value of the S\&P tomorrow, and we should accordingly observe more movement of short-horizon beliefs in response to information.\footnote{The relationship between the time horizon and signal strength of course depends on the exact DGP. We show that the predicted  relationship holds strongly in simulations of game-like DGPs; it also holds in standard option-pricing models. More importantly, it clearly holds in our empirical settings.} Our theory then intuitively suggests that there should be too much movement (evidence for overinference and overreaction) at long forecast horizons, and too little movement (vice versa) at short horizons.

Turning to the data, we find strong and consistent evidence for the hypothesized effect in both sports betting and financial markets. 
Both uncertainty reduction and movement  increase over time as resolution approaches, but movement is generally higher than uncertainty reduction early on (i.e., far from resolution), and lower toward the end of the event. 
For example, in the options data, there is very little daily uncertainty reduction until a few weeks before the contract expires, but beliefs consistently move back and forth, generating positive movement. In other words, news today appears to hold relatively little information about the value of the S\&P in multiple months, but the market acts as if it has more diagnosticity. However, within two weeks of a contract's resolution, the relationship reverses: movement is either less than or equal to uncertainty reduction. That is, as signals become stronger, the market begins to underreact. On net, total movement averaged over an entire option contract is too high, matching the finding of excess movement in \textcite{AL-WP}. But this overall average masks meaningful heterogeneity as one varies the signal strength, in the manner predicted by our theory. The same broad pattern holds in the sports-betting data we consider. In both cases, the results are clear both visually and in formal statistical tests on  movement and uncertainty reduction.

Given that we cannot observe true signal strength, we must rely on our indirect measure (time to resolution) to test the relationship between signal strength and over- vs.\ underreaction in our two real-world high-stakes settings. The strength of our experimental settings, meanwhile, is that these variables are observable or plausibly constructable, but the experiments are lower-stakes and less realistic. The multiple settings thus provide complementary evidence for our theory, whose predictions align well with both sets of data.

Our experimental results relate to a large literature on the topic of belief updating, including many papers providing evidence for other forms of over- and underinference; we provide a brief and incomplete review here. Classically, our paper is most closely related to \textcite{GT92}, who show that experimental participants underreact to the discriminability (which we call signal strength) and weight (the sample size) of a signal, and \textcite{PE66}, which is the first paper we know of to consider the effect of signal strength on underinference (in an unincentivized task with many sequences of signals). More recently, \textcite{GLW-WP} find underreaction to strong signals but even further underreaction to the retraction of those signals, while \textcite{KMW-WP} find overreaction to disconfirming signals. \textcite{BCGKS-WP} also find evidence for insensitivity to signal strength, along with a range of other results (including multimodality and instability in updating) across tasks; we discuss how our modeling approach complements and contrasts with theirs at the end of \Cref{sec:theory}.\footnote{Other recent papers (\cite{BGMS20}, \cite{AKLMT-WP}, \cite{FLP-WP}) provide evidence that forecasting problems may generally induce overreaction, while we focus on inference problems. That said, there is evidence that more-persistent series generate less overreaction, suggesting a connection to our results that we discuss further in our conclusion.} 
\textcite{BBI-WP} run an experiment that confirms many of the patterns that we originally documented in our Study~1a, and argue with additional studies that the patterns they observe are consistent with a two-stage model of channeled attention, followed by cognitively imprecise updating. We see these results as complementary.  

Two important recent influences on our paper are \textcite{KLW21} and \textcite{EG-WP}. \textcite{KLW21} present a model of cognitive noise that connects mental errors in perceiving and encoding information 
with insensitivity to information. We build off the structure of this model, but under the premise that people use some cognitive process to generate an imperfect estimate of the meaning of a signal in complicated problems. \textcite{EG-WP} present a related model of cognitive uncertainty in which people's perception of new information is noisy, such that people shade their answer toward their 
prior. 
This leads people to be insensitive to new information overall, such that they underinfer on average. Our argument follows similar logic with one important distinction: we focus on environments in which people have no issue determining the direction of the signal, but perceive the strength of the signal imperfectly. Consequently, people do not shade toward their prior belief, but rather shade toward the belief given a signal with an ``average'' strength. That is, the perception of signal strength determines whether people underinfer or overinfer, and we predict overinference when signals are weak.\footnote{\citeauthor{EG-WP} run a variety of experiments, including one mirroring our abstract experiment. As in their paper, we also find that cognitive uncertainty correlates with insensitivity to signal strength, but this now leads to greater overinference from weak signals (which they did not include in their experiment).}

Our results are further related to a large literature using asset prices for evidence on beliefs, as surveyed in \textcite{B18}.\footnote{A smaller, growing literature uses sports-betting data to similar ends. As a relevant recent example, \textcite{M21} shows that betting returns from the open of betting to the start of a game predict reversals from there until the end of the game. We focus instead on variation within a game.} For the overall market, a long literature (building from \cite{S81}, with more recent work including \cite{BGJS15} and \cite{GK18}) argues for a link between apparent excess volatility and overreaction. For individual firms, earnings news seems to provide strong information about near-term firm fundamentals (\cite{KL87}, \cite{BKLT19}), and multiple papers (e.g., \cite{BT89}, \cite{DP09}) provide evidence that post-earnings announcement drift arises from the market underreacting to such news. A host of other factors, including uninformative news content (\cite{T14}) and a string of good fundamental news (\cite{BGLS-WP}), are predictive of apparent overreaction and return reversals.\footnote{While we do not provide direct evidence, our theory suggests an interpretation that earnings surprises are strong news about short-term fundamentals (generating underreaction), while even a string of news gives relatively {weak} information about the long-run or aggregate regime (leading to overreaction), loosely in the spirit of Barberis et al.\ (\citeyear{BSV98}). Separately, \textcite{GS14} document underreaction to the passage of time. We view this as underattentiveness to certain relevant aspects of information, as modeled in \Cref{subsec:multiple}.} \textcite{KT-WP} reconcile some of these findings by considering the distribution of past outcomes for the given category of news; they argue that categories with more extreme outliers tend to generate greater overreaction. Our focus on the informativeness of a given signal is conceptually somewhat different.\footnote{That said, we provide only a high-level theory of what default ``intermediate'' signal strength people shrink toward. The results of \textcite{KT-WP} suggest that salience of outliers in past data may be important for determining this default strength for a given type of signal.} 
While signal strength is clearly not the only relevant factor for belief behavior, we contribute by isolating it as a simple, powerful determinant in a range of settings, with complementary evidence from both a new set of experiments and market-price data.

We proceed as follows. \Cref{sec:theory} provides our theoretical framework; \Cref{sec:experiment} presents the three experiments; \Cref{sec:markets} analyzes the sports betting and finance data; and \Cref{sec:discussion} discusses and concludes. The Appendix contains model proofs and additional empirical details and results, and screenshots of the pages in the experiments are provided in the Supplementary Appendix.

\section{Theory}
\label{sec:theory}

\subsubsection*{Overview}

We consider a setting in which people can easily understand the \emph{direction} they should update their beliefs after seeing a signal, but where it may be challenging to understand the \emph{strength} of the signal, even if the signal is perfectly observed. There are a variety of reasons why a person may find it difficult to fully comprehend the signal strength. In contexts where the signal strength and correct posterior can in theory be calculated directly (e.g., in controlled experiments), the person might have issues undertaking a set of potentially complex mental calculations, but nonetheless have the ability to generate a rough estimate. In real-world settings, the person may not fully understand the exact data-generating process, but nonetheless have a simplified model of the process. Similarly, the person may only be able to appreciate parts of a complicated signal and thus generate an incomplete estimate of its strength. In each case, the person is using a cognitive process --- whether conscious and deliberative or unconscious and automatic --- to form an estimate of the signal strength. Our goal is to provide a framework that is broad enough to capture these different situations.

After setting up the model in \ref{sec:theory-setup}, we show how overinference from weak signals and underinference from strong signals arises from a set of simple and intuitive (potentially non-Bayesian) updating rules. In \ref{subsection:FunctionalForm}, we study a parameterized model to derive a more concrete relationship between strength and reaction, which we then use in our experimental analysis. In \ref{sec:relax}, we consider how incorrect priors, base-rate neglect, or uncertainty about direction may affect the analysis. In \ref{subsec:multiple}, we broaden the analysis to consider multiple people with possibly correlated estimates, providing a specific example arising from limited attention.

\subsection{Setup and Main Results} \label{sec:theory-setup}

\subsubsection*{Setup}

We consider a person who receives a signal $s$ about a binary state $\theta\in\{0,1\}$, with $s\in\mathcal{S}$ generated according to the likelihood function $p(s|\theta)$. As a benchmark for comparison, we denote the correct prior that $\theta=1$ by $\pi_0$ and the Bayesian posterior given $s$ as $\pi_1(s)$, or $\pi_1$.

To formalize the idea that some aspects of a signal are easier to understand than others, we break the signal into two \emph{components}, $s=(s_d,s_m).$ The first component, $s_d$, determines the \emph{direction} of updating and accordingly can only take two values, ``positive'' or ``negative.'' Given a positive (negative) directional signal, the Bayesian posterior is always above (below) the prior.\footnote{Formally, $s_d$ is such that $\pi_1(s_d=\text{positive}, s_m) \geq \pi_0 \geq \pi_1(s_d=\text{negative}, s'_m)$ for any $s_m$ and $s'_m$. Note also that all $p(\cdot)$ can be understood either as mass functions or densities, while $P(\cdot)$ refers to a probability.} Given the direction, the second component $s_m\in \mathbb{R}$ determines the \emph{magnitude} or \emph{strength} of the signal. We define signal strength $\mathbb{S}$ formally as 
 \begin{equation} 
\mathbb{S}(s) \equiv \left\lvert \log\!\left(\frac{p(s|\theta=1)}{p(s|\theta=0)}\right)\right\rvert, \label{eq:strength}
\end{equation}
which is the magnitude of the log odds ratio of the signal. Defining $\text{logit}(x)\equiv\log\!\left(\frac{x}{1-x}\right)$, a Bayesian updates such that
\begin{align}\label{bayesianUpdating}
    \underbrace{\text{logit}(\pi_1(s))}_{\substack{ \text{Logit of} \\ \text{Posterior}}} = \underbrace{\text{logit}(\pi_0)}_{\substack{ \text{Logit of} \\ \text{Prior}}} \hspace{4mm} \underbrace{\vphantom{\text{logit}(\pi_0)} \pm}_{\substack{ \text{Signal} \\ \text{Direction} \\ \text{(from $s_d$)}}} \hspace{4mm}  \underbrace{\mathbb{S}(s)}_{\substack{ \text{Signal} \\ \text{Strength} \\ \text{(from $s_m|s_d$)}}}.
\end{align}
Consequently, fixing $\pi_0$, a signal $s$ with a greater signal strength $\mathbb{S}(s)$ will lead to a larger absolute change in beliefs $|\pi_1(s)-\pi_0|$.

\subsubsection*{Estimates of Signal Strength}

Our main behavioral assumption is that a person fully understands the direction of the signal, but does not fully understand the magnitude.\footnote{We view our empirical settings as ones in which the direction of updating is clear, but we discuss below how our theoretical predictions are altered when the direction is also uncertain.} Instead, we assume that people use some internal process to form a guess about $\mathbb{S}$, which we call an \emph{estimate} $e\in\mathbb{R}$. While the Bayesian uses the information in the signal $s=(s_d,s_m)$, the person we consider uses the information in $\hat{s}\equiv (s_d,e)$. Estimates may be formed using a simplified model, unconscious approximation, or constrained information processing given attention to certain  aspects of a signal.

We consider the behavior of the person's perceived signal strength given $\s$, as the perceived signal strength determines the person's inference from the signal. In Section \ref{subsection:FunctionalForm} below, we take the traditional approach of assuming that the person is a constrained Bayesian: they only receive a noisy estimate of the strength, and they update correctly given the joint distribution of signal strengths and estimates. From these assumptions, we derive the parameterized relationship between signal strength and reaction. Our initial goal in this section, however, is to demonstrate the generality of our main effect given very minimal assumptions on the distribution of $e$ and assuming intuitive (and potentially non-Bayesian) updating rules. That is, rather than deriving updating rules under specific (and likely unrealistic) Bayesian assumptions, we show how our effect obtains under a broad class of updating rules.

\subsubsection*{Reasonable Estimates and Updating Rules}

We start with a minimal set of restrictions on the distribution of estimates given a signal strength, requiring that the estimate be a well-ordered approximation of the true strength:
\begin{assumption2} \label{assumption:e} For each direction $s_d$,\footnote{We allow all statements to potentially condition on $s_d$, but we leave this conditioning implicit to ease notation for the distribution of $e$. That is, $p(e|\mathbb{S})$ is shorthand for $p(e|s_d,\mathbb{S}),$ and so on for related expressions.} estimates are formed such that:
\begin{enumerate}[label=(\alph*),itemsep=0pt,topsep=\parsep]
    \item $e$ is unbiased: $\e[e|\mathbb{S}]=\mathbb{S}$.
    \item $e$ is well-ordered: $\frac{p(e \,|\, \mathbb{S} = \mathsf{S}_2)}{p(e \,|\, \mathbb{S} = \mathsf{S}_1)}\,\text{ strictly increases in $e$ for all $\mathsf{S}_2>\mathsf{S}_1$}$.
    \item $e$ is imperfect: there is no pair $(e,\mathbb{S})$ such that $P(\mathbb{S} | e) = 1$. 
\end{enumerate}    
\end{assumption2}
Part~(a) is effectively a normalization such that the estimate is centered around the correct signal strength. 
Part~(b) assumes the strict monotone likelihood ratio property (MLRP) on estimates. This commonly used property implies that higher estimates are associated with higher levels of $\s$; under Bayesian updating, this implies that posteriors for~$\s$ are monotonic in~$e$ (\cite{M81}). In our case, we impose MLRP to ensure that the distribution of estimates is well-behaved enough to be able to make general statements even in cases of non-Bayesian updating. Part~(c) rules out trivial cases in which $e$ fully reveals the signal strength $\s$. This implies that the set of feasible signal strengths is non-degenerate.

Next, we consider the person's prior perceptions of signal strength. Before observing any information, the person has some subjective expectation $\hat{\mathbb{S}}_0$ of signal strength $\mathbb{S}$. After observing if $s_d$ is positive or negative (but before incorporating the estimate $e$), the person updates this expectation to $\hat{\mathbb{S}}(s_d)$. 
We do not require these expectations to be correct, but we do require the minimal assumption that they be within the feasible set of signal strengths:
\begin{assumption2} $\hat{\mathbb{S}}(s_d)$ is strictly between $\text{min}_{s_m}{\mathbb{S}}(s_d,s_m)$ and $\text{max}_{s_m}{\mathbb{S}}(s_d,s_m)$. 
\label{assumption:priors}
\end{assumption2}

Given the above expectation $\hat{\mathbb{S}}(s_d)$ of $\mathbb{S}$, the person then generates an estimate $e$ following \Cref{assumption:e}. How will the person update given their prior expectation and this new signal? Instead of requiring Bayesian updating, we assume only that the person's posterior expectation of signal strength $\hat{\mathbb{S}}(\hat{s}) = \hat\e[\s | s_d,e ]$ will move from $\hat{\mathbb{S}}(s_d)$ toward $e$:
\begin{assumption2} \label{assumption:uts} For all $\hat s$, the posterior $\hat{\mathbb{S}}(\hat{s})$ is strictly between the prior $\hat{\mathbb{S}}(s_d)$ and estimate~$e$. 
\end{assumption2}
Crucially, given that the estimate is noisy, we assume that the person will not update all the way to $e$. This intuitive property is referred to as ``updating toward the signal'' (UTS) by \textcite{CH12}, who show that it is satisfied in many commonly studied updating environments.\footnote{As \citeauthor{CH12} note, some papers assume UTS directly (e.g., \cite{S86,MH08}), as we do. Note that we assume strict UTS, rather than a weaker version in which $\hat{\mathbb{S}}(s_d)\leq \hat{\mathbb{S}}(\hat{s})\leq e$.} We are, in effect, assuming implicitly that the person is aware that their internal estimate is noisy and therefore shades their signal-strength belief toward their prior. 
Note that we place no further restriction on how much the person updates from a given $e$: for all~$\hat s$, there exists \emph{some} $\alpha\in(0,1)$ such that  $\hat{\mathbb{S}}(\hat{s}) = \alpha e + (1-\alpha) \hat{\mathbb{S}}(s_d)$, but the signal weight $\alpha$ need not be constant and may vary with $e$ (and with $s_d$).

To summarize, the person observes a signal $s=(s_d,s_m)$ containing both directional and magnitude information. A Bayesian would correctly interpret this signal as having strength ${\mathbb{S}}(s)$. In contrast, the person in our model understands $s_d$ but doesn't fully understand $s_m$ and therefore cannot fully resolve ${\mathbb{S}}(s)$. Instead, she forms a reasonable, well-ordered, but noisy estimate $e$ of ${\mathbb{S}}(s)$. She then uses a very general and intuitive updating rule using $\hat{s}=(s_d,e)$ to form her expectation of signal strength $\hat{\mathbb{S}}(\hat{s})$.

\subsubsection*{Overinference and Underinference}

Our primary objective is to study whether a person is over- or underinfering relative to the full (signal-understanding) Bayesian benchmark. While the Bayesian's view of the signal strength is fixed at ${\mathbb{S}}({s})$ given a signal $s$, a person's perceived $\hat{\mathbb{S}}(\hat{s})$ depends on their estimate~$e$, which is stochastic. Consequently, we focus on the expected perception $\e[\hat{\mathbb{S}}(\hat{s})|s]$, and we define over- and underinference in the following natural way.\footnote{Given our focus on inference from signals of varying strengths, we directly define over- and underinference in terms of mean perceived signal strength. Fixing $\pi_0$, this intuitively corresponds to over- and underreaction in beliefs, as belief changes $|\logit(\hat\pi_1(s))-\logit(\pi_0)|$ are generally monotonic in perceived strength $\hat\s(\hat s)$. This connection can fail given the non-linear mapping from signal strength to beliefs, but it will hold to first order (e.g., in a small-noise limit, as in Khaw et al.\ \citeyear{KLW21}, Appendix~G). We can also simply modify Assumptions~\ref{assumption:e}--\ref{assumption:uts} to focus on beliefs (so $e$ is an estimate of the correct $\pi_1$), in which case our results will hold for beliefs. Unless stated otherwise, we assume throughout that belief changes are monotonic in $\hat \s(\hat s)$ with the correct direction.\label{fn:monotonic}}

\begin{definition}
The person \textit{overinfers} from $s$ if $\e[\hat{\mathbb{S}}(\hat{s})|s]>\mathbb{S}(s)$ and \textit{underinfers} from $s$ if $\e[\hat{\mathbb{S}}(\hat{s})|s]<\mathbb{S}(s)$.
\end{definition}

\noindent Our main result is that this person is biased in their perception of signal strength:

\vspace{2mm}
\begin{proposition}[Over- and Underinference]
\label{prop:main}
The person overinfers from weak signals and underinfers from strong signals: there exists a unique switching point $\mathbb{S}^*$ such that they overinfer from $s$ if $\mathbb{S}(s)<\mathbb{S}^*$ and underinfer if $\mathbb{S}(s)>\mathbb{S}^*$.
\end{proposition}
The proof, provided in Appendix~\ref{app:proofs1}, involves expressing $\e[\hat{\mathbb S}(\hat s)|s] - \mathbb{S}(s)$ as an expectation of a single-crossing function $g(e)$ with respect to the conditional distribution $p(e|\mathbb{S}(s))$. Then, using a well-known result (formalized by \cite{K68}, among others) referred to as the variation diminishing property, the fact that $p(e|\mathbb{S}(s))$ satisfies the MLRP (by Assumption~\ref{assumption:e}) implies that $\e[\hat{\mathbb S}(\hat s)|s] - \mathbb{S}(s)$ is single-crossing as well: in particular,  $\e[\hat{\mathbb S}(\hat s)|s]- \mathbb{S}(s)>0$ for small $\mathbb{S}(s)$ and $\e[\hat{\mathbb S}(\hat s)|s]- \mathbb{S}(s)<0$ for large $\mathbb{S}(s),$ with a unique interior switching point $\mathbb{S}^*.$

Although this proof is slightly involved, the results are intuitive. First, consider the extreme case in which the person places no weight on their strength estimate $e$ (because, for example, the estimate is extremely noisy). The person will effectively be fully insensitive to signal strength, such that they expect the same intermediate strength ($\hat{\mathbb{S}}(s_d)$) regardless of actual strength $\mathbb{S}$. This leads to overinference when $\mathbb{S}$ is low and underinference when $\mathbb{S}$ is high. As the weight on $e$ rises, the person will still shrink toward an intermediate strength as compared to a full Bayesian. The resulting partial insensitivity to signal strength leads to overinference from weak signals and underinference from strong signals on average.

Note that under our general assumptions, it is not necessarily the case that a person's expected signal strength $\hat{\mathbb{S}}(\hat{s})$ is monotonic in $e$ or that the amount of over- or underinference $\e[\hat{\mathbb{S}}(\hat{s})|s]-\mathbb{S}(s)$ is monotonic in $\mathbb{S}(s)$.\footnote{For example, if a person updates from their prior $\hat{\mathbb{S}}(s_d)$ strongly toward the estimate $e_1$ but very weakly toward the estimate $e_2=e_1+\epsilon$, the person can have a large drop in $\hat{\mathbb{S}}(\hat{s})$ from a small increase in $e$.}
Appendix~\ref{app:proofs1} provides conditions under which these additional monotonicity results will hold: as long as the weight placed on the estimate does not fall dramatically given small increases in $|e-\hat{\mathbb{S}}(s_d)|$, then $\hat{\mathbb{S}}(\hat{s})$ will be monotonic in $e$; and as long as the weight does not increase dramatically in $e$, then $\e[\hat{\mathbb{S}}(\hat{s})|s]-\mathbb{S}(s)$ will be monotonic in $\mathbb{S}(s)$. 

\subsection{Parametric Example: Updating with Log-Normal Estimates} \label{subsection:FunctionalForm}

In the previous subsection, we showed how a person following a set of intuitive (but potentially non-Bayesian) updating assumptions will overinfer from weak signals and underinfer from strong signals. We now specialize the model to show that a quasi-Bayesian facing log-normal distributions will also update following the predictions in \Cref{prop:main}, with the updating rule taking a particularly simple form that will then guide our experimental analysis.\footnote{The resulting updating rule is similar to one obtained using different foundations (based on \cite{KLW21}) in the previous version of this paper, \textcite{ALT-old}.\nocite{W20,FJ-WP}}

First, we assume that signal strength is log-normally distributed with $\log \mathbb{S} \sim \mathcal{N}(\mu_\mathbb{S}, \sigma_\mathbb{S}^2)$, regardless of direction. A Bayesian's expectation of signal strength after seeing either direction is thus $\hat{\mathbb{S}}(s_d)\equiv \e[{\mathbb{S}}|s_d]=\exp(\mu_\mathbb{S} + \sigma_\mathbb{S}^2/2)$. Next, given a specific strength $\mathbb{S}$, we assume that the person's estimate $e$ is log-normally distributed, $\log e \sim \mathcal{N}(\log \mathbb{S} - \sigma_e^2/2, \sigma_e^2)$. The correction $- \sigma_e^2/2$ ensures that the estimate is centered around the true signal strength: $\e[e|\mathbb{S}]=\mathbb{S}$. 

How will a Bayesian then react to $e$? Using standard results given a log-normal likelihood and conjugate prior, the updating rule for expected signal strength is
\begin{equation}
\underbrace{\hat{\mathbb{S}}(\hat{s})}_{\substack{\text{Posterior} \\ \text{Expectation}}} \! = \,\,\exp\!\bigg[ \underbrace{\!\left(1-\frac{\sigma_\mathbb{S}^2}{\sigma_e^2 + \sigma_\mathbb{S}^2}\right)}_{\substack{ \text{Weight on} \\ \text{Prior}}}\,\, \cdot \!\!\underbrace{\log\hat{\mathbb{S}}(s_d) }_{\substack{ \text{(Log Adjusted)} \\ \text{Prior Expectation}}} + \underbrace{\left(\frac{\sigma_\mathbb{S}^2}{\sigma_e^2 + \sigma_\mathbb{S}^2}\right)}_{\substack{ \text{Weight on} \\ \text{Estimate}}}\, \cdot \,\underbrace{\left(\log e+\sigma_e^2/2\right)}_{\substack{ \text{(Log Adjusted)} \\ \text{Estimate}}} \bigg]. \label{eq:logupdate}
\end{equation}
Intuitively, the Bayesian will take a weighted average of the adjusted prior and estimate in log space, and then exponentiate to form their posterior expectation of strength. The weight on the imperfect strength estimate depends on the relative precision of the estimate versus the prior: as the precision of the estimate rises, the weight on the estimate rises; as the precision of the prior rises, the weight on the estimate falls.

Given this updating rule, people will overinfer from weak signals and underinfer from strong signals on average, with a simple and estimable functional form for the effect. In particular, the expectation of $\hat{\mathbb{S}}(\hat{s})=\hat\s(s_d,e)$ over the distribution of estimates is 
\begin{equation}
\label{eq:ExpSHatAndS}
    \mathbb{E}[\hat{\mathbb{S}}(\hat{s})|s] = k \, \mathbb{S}^\beta, 
\end{equation}
where $\beta \equiv \sigma_\mathbb{S}^2 / (\sigma_\mathbb{S}^2+\sigma_e^2) \in (0,1)$ and $k \equiv \exp(\beta^2 \sigma_e^2 / 2) \hat{\mathbb{S}}(s_d)^{1-\beta}$. Note that $\mathbb{E}[\hat{\mathbb{S}}(\hat{s})|s]\geq \s$ if and only if $\s\leq \s^*\equiv k^\frac{1}{1-\beta}$. That is, as in Proposition \ref{prop:main}, people overinfer from signal strengths below $\s^*$ and underinfer above $\s^*$.

\begin{figure}[htb!]
\caption{Theoretical Predictions of Over- and Underinference by Signal Strength}
\label{fig:theory}
\begin{center}
\vspace{-5.5mm}
    \includegraphics[width=.49\textwidth]{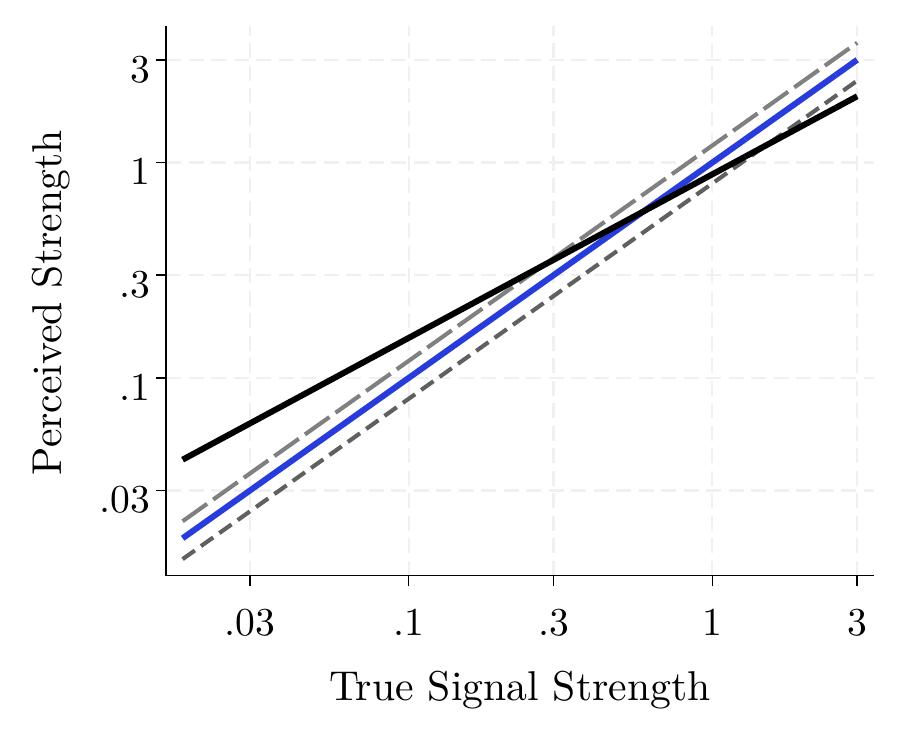}
    \includegraphics[width=.49\textwidth]{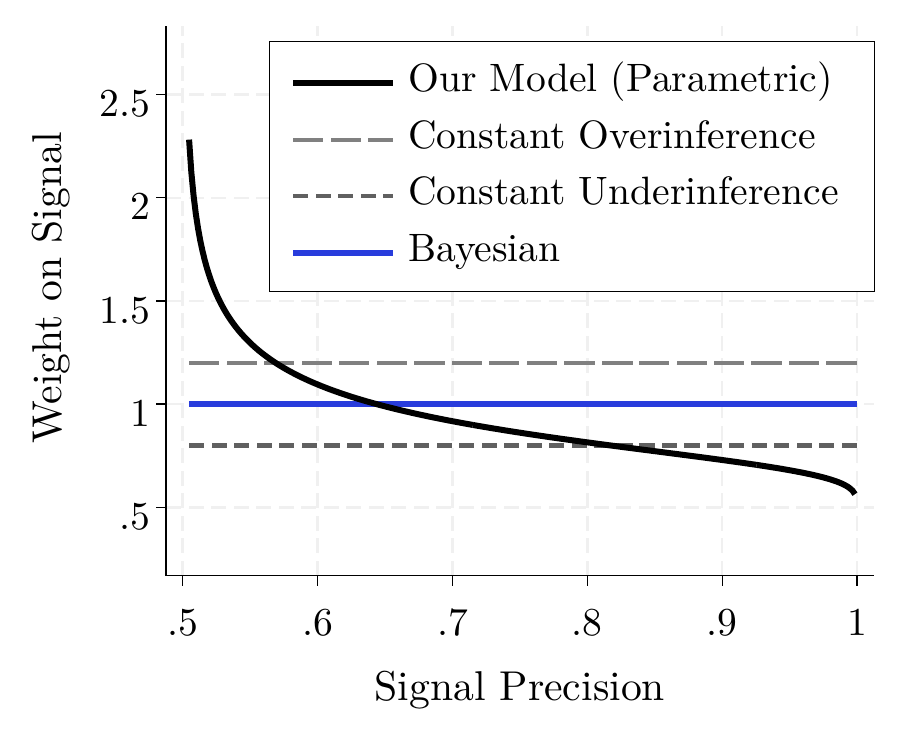}
\end{center}
\begin{threeparttable}
\begin{tablenotes}
\begin{scriptsize}
\vspace{-9.5mm}
\item \textbf{Notes:} These figures provide two representations of the core deviation in our model. Blue lines correspond to Bayesian updating (correct perception of signal strength $\mathbb{S}$), short dashed lines to underinference (with perceived signal strength $0.8 \cdot \mathbb{S}$), long dashed lines to overinference (perceived signal strength $1.2 \cdot \mathbb{S}$), and black lines to the over- and underinference behavior in the parametric version of our model (perceived signal strength $k \cdot \mathbb{S}^\beta$ with $k = 0.88$ and $\beta = 0.76$, as estimated from Study 1a). The left panel plots signal strength perception as a function of signal strength on a log-log scale. The right panel plots the weight put on signals as a function of the true precision. Both figures show that our model predicts overweighting of weak signals and underweighting of strong signals. \par
\end{scriptsize}
\end{tablenotes}
\end{threeparttable}
\end{figure}

The relationship between reaction and signal strength given this setup can be represented and visualized in a number of ways. Taking logs of \eqref{eq:ExpSHatAndS} yields 
\begin{equation}
\log(\mathbb{E}[\hat{\mathbb{S}}(\hat{s})|s]) = \log(k) + \beta \log(\s), \label{eq:loglinear}
\end{equation}
such that there is a log-linear relationship between the expected and true signal strength, with a positive intercept and a muted slope between 0 and 1. The left panel of \Cref{fig:theory} plots this relationship given the parameters $k$ and $\beta$ we estimate from our first experiment (discussed further in \Cref{sec:experiment}). For comparison, we also plot the relationship for a Bayesian (for which they are equal), a person who exhibits constant underinference, and a person who exhibits constant overinference.

%\end{comment}

This relationship can also be represented in terms of the effective weight a person places on a signal with strength $\s$. While a Bayesian observing the full signal will update following~\eqref{bayesianUpdating} using $\s$, a person in our model updates as if the signal strength is, on average, $\hat{w}(\s) \s$ for some weight function $\hat{w}(\s)$. The full Bayesian effectively uses $w(\s)=1$, while for our model,
\begin{equation}
\hat{w}(\s) = k \s^{-(1-\beta)}. \label{eq:weight}
\end{equation} 
This weight is greater than 1 for weak signals and less than 1 for strong signals. Note that $\hat{w}(\s)$ approaches 1 as $\beta\to1$, so the degree of over- and underinference shrinks as the person's estimation process becomes more precise.

Rather than using the relationship in \eqref{eq:weight} directly, we often follow past literature (\cite{B19}) and focus on the relationship between the inference weight and signal \emph{diagnosticity} or \emph{precision} $\rho(s).$ For a symmetric signal (where $p(s|\theta=1)=1-p(s|\theta=0)$), signal precision is $\rho(s)\equiv \max\{p(s|\theta=1),p(s|\theta=0)\}= \text{logit}^{-1}(\s(s)),$ which is a monotonically increasing transformation of strength $\s(s).$\footnote{Signal precision is by definition between $1/2$ and 1. When $\pi_0={1}/{2}$ (as in our first experimental study), the Bayesian posterior after a positive signal is equal to the signal precision, $\pi_1(s)=\rho(s)$.} The qualitative relationship between weight and precision matches the relationship between weight and strength. In particular, the weight $\hat w(s)$ is above~1 for low precisions and below 1 for high precisions:
\begin{equation}
\hat{w}(s) = (\text{logit}(\rho^*))^{-(1-\beta)} \,|\text{logit}(\rho({s}))|^{-(1-\beta)},\label{eq:weight_precision}
\end{equation}
where $\rho^*\equiv\big({1+\exp(-k^{-\frac{1}{1-\beta}})}\big)^{\!-1}$ is the switching point. The right panel of \Cref{fig:theory} plots this relationship, again given parameters $k$ and $\beta$ from the experiment and again as compared to Bayesian updating, underinference, and overinference. We return to these graphs in Section~\ref{sec:experiment}.

\subsection{Relaxing Assumptions} \label{sec:relax}

\subsubsection*{Prior Belief Distortions}

We have assumed to this point that the person starts with a correct prior, $\hat\pi_0=\pi_0$. If the person has an incorrect prior that is observable (and otherwise updates according to the assumptions in \ref{sec:theory-setup}), it is straightforward to correct for the distortion induced by $\hat\pi_0\neq\pi_0$ in our empirical analysis. Rather than estimating perceived signal strength using $|\logit(\hat\pi_1(s))-\logit(\pi_0)|$ from \eqref{bayesianUpdating}, the incorrect prior can be controlled for by using $|\logit(\hat\pi_1(s))-\logit(\hat\pi_0)|$. The person uses their perceived signal strength to update from their prior to their posterior, so perceived strength can be backed out from the posterior and prior, and \Cref{prop:main} continues to provide testable predictions. 
Note that this is true even if the person's prior $\hat\pi_0$ arose after updating from an estimate of the previous-period signal. In this case, even though the person used a noisy estimate and was insensitive to the past signal strength, $\hat\pi_0$  incorporates this uncertainty. See Appendix~\ref{app:proofs2} for details.

This analysis becomes more complicated if the prior is not observed, or if the person uses their prior in a non-standard way (e.g., with base-rate neglect). For example, suppose that an experiment provides a person with both an endowed prior $\pi_0$ and a signal~$s$ simultaneously, and asks for a single updated posterior. In this case, if people are unsure how to use both the prior and the signal, their single answer will reflect estimation uncertainty in how to treat both objects (with no way to separate them). Similarly, if the person states a prior $\hat\pi_0$, but then effectively uses a distorted version of this prior due to base-rate neglect, the posterior will again contain multiple distortions. We discuss in Appendix~\ref{app:proofs2} how these distortions contaminate people's reactions and when they potentially overwhelm our effect: unsurprisingly, it will depend on the relative size of the prior distortion.

This concurrent distortion will make it more challenging to test our core prediction. There are two broad approaches to control for this issue. First, one can focus on uninformative priors $\pi_0=0.5$, where biases like base-rate neglect have no impact. We use this approach in our first experiment. Alternatively, one can vary the prior and then control for potential base-rate neglect using a regression approach following \textcite{G80}. We use such an approach in our other experiments. As shown in Section~\ref{sec:experiment}, our experimental results provide consistent evidence for our core effect, regardless of whether we control for base-rate neglect.

\subsubsection*{Uncertainty About the Direction} 

Our theory is geared to situations in which people know the correct direction to update, but are unclear about the strength. In this case, imperfect estimates lead to insensitivity to strength, which leads to our main effect. We can extend the model to situations in which the person is unsure about both the direction and strength of the signal (such that the person forms an estimate $e$ of \textit{signed} signal strength
$\mathbb{S}_{signed}\equiv\log\!\left(\frac{p(s|\theta=1)}{p(s|\theta=0)}\right)$ and does not observe the direction directly). This version of the model is closely related to that of \textcite{EG-WP}, with the substantive difference being that \citeauthor{EG-WP} work in probability space rather than signal strength space. In this case, as with \textcite{EG-WP}, we predict that insensitivity without directional information generally leads to underinference (see Appendix~\ref{app:proofs2} for details). Intuitively, because people do not know the directional meaning of the signal, they shade their belief updating toward a reaction of zero. That said, our definition of underinference becomes strained in this context, so we are reluctant to make strong statements.\footnote{For example, suppose the correct signed signal strength is 2, but a person perceives it to be -1. Is this an under- or overinference? Alternatively, suppose that half of people perceive the strength as -3 and half as 3. On average, people perceive the signal strength as 0 (and are thus underinferring by our definition), although it could be argued that all are overinferring. In our main model, these issues do not arise because we assume that the person knows the correct updating direction, which we believe is true in our empirical environments.}

\subsection{Multiple People, Limited Attention, and Correlated Estimates}\label{subsec:multiple}

The analysis thus far has focused on the expected reaction of a single person. In this section, we instead consider the average response across different people $i = 1, \dots, N$ (where $N$ should be thought of as large). A natural preliminary way to extend our analysis to this case is to assume that each person understands the direction $s_d$ and generates a mutually independent strength estimate $e_i$. 
That is, people see the same signal and agree on its direction, but there is diversity in people's estimated signal strengths due to different interpretations, models, or perceptions of the problem. Under this assumption, our results (immediately) continue to hold across people. Specifically, define the expectation over people given $s$ as $\e_i[\cdot|s]$ and the person-specific strength perception as $\hat\s_i(\hat{s}_i)$. Then, rather than focusing on expected perceived strength of an individual across potential estimates $\e[\hat\s(\hat{s})|s]$ as in \Cref{prop:main}, the same results hold taking the expectation across people $\e_i[\hat\s_i(\hat{s}_i)|s]$ (see Appendix~\ref{app:proofs3} for a more formal discussion).
Intuitively, under the assumption that estimates are independent across people, there is no formal distinction between taking the expectation with respect to the distribution of estimates and taking a cross-sectional expectation across people.

The assumption of independent estimates is appropriate for some situations. For example, in our naturalistic experiment, we ask people to update their subjective probability of a team winning a basketball game after observing a signal (a made or missed basket) in simple situations. We find that people's perceptions of the strength of a given signal tend to be diverse and smooth, presumably because people have different ways of using their knowledge and experiences to estimate its effect. Similarly, in our more abstract experiments, signals are presented in a computationally challenging form (e.g., a signal with conditional likelihood 202/337) and find similar diversity and smoothness in responses, likely because people have different estimates of the precise value of this number and how to use it to form a posterior.

However, there are also natural situations in which people might form correlated estimates of a given signal's strength. For example, people may have similar simplified models of a given DGP or similar strategies for combining available information to determine a signal's meaning. Similarly, some dimensions of a piece of information may be more salient than others, such that people incorporate similar dimensions in forming their estimate of signal strength. These cases will lead to partially correlated estimates across people and potentially non-smooth multimodal posterior belief distributions, as in \textcite{BCGKS-WP}. 

\subsubsection*{Example: Limited Attention}

To understand correlated estimates more formally, we consider a situation in which the signal's strength component (the second entry in $s=(s_d,\boldsymbol{s}_m)$) has multiple dimensions:   $\boldsymbol{s}_m= (s_{m,1}, \ldots, s_{m,n})$. While a Bayesian uses all components to determine signal strength, people in our model have limited attention, limited processing ability, or attend specifically to certain features of the signal, such that they only appreciate a subset of components. Correlation in estimates will occur if people focus on the same components.

Specifically, we assume that regardless of direction, the components $s_{m,j}$ are independently and identically distributed $\mathcal{N}(\mu_{\s},\sigma^2_m)$, and the true log signal strength is the average of these components:
\begin{equation*}
\log\mathbb{S} = \frac{1}{n}\sum_{j=1}^n s_{m,j}.
\end{equation*}
Consequently, signal strength is log-normally distributed, $\log \mathbb{S} \sim \mathcal{N}(\mu_\mathbb{S}, \sigma^2_\s)$, with $\sigma^2_\s=\sigma^2_m/n$. While a Bayesian uses all $n$ components and can determine $\mathbb{S}$, person $i$ only attends to $n_i\leq n$ of the components, captured in a fixed person-specific vector $\boldsymbol{a}_i \in \{0,1\}^n$, where $a_{i,j}=1$ if the person attends to component $j$. Given this setup, person $i$'s best (log) estimate of $\s$ is
\begin{equation*}
\log e_i = \frac{1}{n_i} \sum_{j=1}^{n} \boldsymbol{1}(a_{i,j}=1) \cdot s_{m,j} - \frac{\sigma_{e,i}^2}{2},
\end{equation*}
where $\sigma_{e,i}^2 = \frac{n-n_i}{n\cdot n_i}\sigma_m^2 = \frac{n-n_i}{n_i}\sigma_\s^2$ (with the term $- \sigma_{e,i}^2/2$ again included so that $\e[e_i|\s]=\mathbb{S}$). The estimate $e_i$ is log-normally distributed conditional on $\s$, $\log e_i \sim \mathcal{N}(\log \mathbb{S} - \sigma^2_{e,i}/2, \sigma^2_{e,i})$. This setting thus maps to the one in \Cref{subsection:FunctionalForm}, with $\sigma^2_\s=\sigma^2_m/n$, and $\sigma^2_{e}$ $\sigma^2_{e,i}=(n-n_i)\sigma^2_m/(n\cdot n_i)$. That is, that model can be microfounded with people who only consider a subset of the full signal.\footnote{The strength sensitivity parameter $\beta$ in \eqref{eq:ExpSHatAndS} becomes $\beta_i=n_i/n$. So fixing $n$, an increase in $n_i$ (e.g., due to greater sophistication) should lead to less-noisy estimates and less insensitivity to true strength for person $i$. An increase in $n$ (e.g., from a more complicated signal) should generate the opposite behavior for all $i$.} Crucially, however, this multi-component model produces correlated updating behavior across people, governed by the overlap in $\boldsymbol{a}_i$ across $i$.

This correlation can create a specific type of violation of \Cref{prop:main}. That result says that \emph{all} signals $s$ of a given strength $\s(s)$ will lead to over- or underinference in the same way on average. But in this setting, the same is not necessarily true. To take an extreme example, if everyone focuses on the same components, then they will have the same estimate $e_i$ for a given signal $s$. While this estimate is random conditional on $\mathbb{S}$ (it is drawn from a distribution with mean $\mathbb{S}$), it is \emph{not} random conditional on the full signal $s$ (since $s$ contains the entries that will be used to determine $\log e_i$). This suggests a simple adjustment under which our results \emph{do} apply: when over- and underinference are defined conditional on $\s$ rather than $s$, then a version of the proposition holds (see Appendix~\ref{app:proofs3}). That is, our results hold when averaging over signals of the same strength.

Finally, as we discuss in Appendix~\ref{app:proofs3}, it is possible to obtain more precise predictions about the correlation in updating behavior under different sets of assumptions about the signal components or attention vectors. If there are few components or all people are drawn to a small set of salient components, people's estimates will be correlated, and we may see multimodality in responses.\footnote{For example, if an abstract problem only includes a few numbers representing the ``prior'' and a ``signal,'' we might see some people focusing on the prior, some on the signal, and some on both. \textcite{BCGKS-WP} provide a richer foundation and set of predictions for this form of behavior arising from bottom-up attention to salient features, which further speaks to instability across problems with the same correct answer.} If people must estimate a probability given a complex DGP and a rich signal, or if the main salient part of a signal is the direction $s_d$ (as may sometimes apply in time-series settings), we might expect more independent strength estimates and smoother distributions of resulting strength perceptions.

\medskip
To summarize, we model an updating environment in which a person knows the directional meaning of signals, but only forms a rough estimate of the exact strength. As this estimate is imperfect, the person shades their perceived strength toward some intermediate value, which leads to overinference from weak signals and underinference from strong signals on average. In the following sections, we test this core prediction for updating in a range of environments. We also predict that the effect will be dampened as a person's estimate becomes more precise. Estimation precision will increase with more thought or sophistication, more experience, or attending to more components in a multi-dimensional problem. While estimation precision is not directly observable, we test this relationship using a variety of proxies in our experiments.

\section{Experimental Evidence} 
\label{sec:experiment}

To test our core prediction that people overinfer from weak signals and underinfer from strong signals, we design and conduct three experiments. Each experiment studies the causal effect of varying signal strengths on participants' updating behavior and, by implication, the level of over- and underinference. The first experiment (``Study 1a'') adapts a classic belief-updating design from \textcite{GHR65}. The next experiment (``Study 1b'') replicates the first experiment and expands the analysis by varying the prior and eliciting a more direct proxy for precision in signal strength estimates. The final study (``Study 2'') uses a novel naturalistic design in which participants predict the win probability of basketball games. Each study was preregistered,\footnote{Study~1a: \url{https://aspredicted.org/ax4wg.pdf}; Study~1b: \url{https://aspredicted.org/8Q4_6Y9}; Study~2: \url{https://aspredicted.org/SYW_QWF}.} and we largely follow the preregistration plans, although some of the estimated results from the first study are placed in Appendix~\ref{app:experiment} to conserve space. We note these cases in the main text.

\subsection{Study 1a: Abstract Updating Experiment}

\label{sec:study1a}

\subsubsection*{Design}

The design of the first experiment follows the broad ``bookbag-and-poker-chips'' (or ``balls-and-urns'') paradigm, which is a benchmark design for measuring underinference and overinference in past literature (\cite{B19}). Participants are told that there are two card decks, each with $N$ cards. One deck is labeled as Green, and the other is labeled as Purple. Each deck is composed of Diamond and Spade cards, with the Green deck having $D_1$ Diamonds and $N - D_1$ Spades and the Purple deck having $D_2$ Diamonds and $N - D_2$ Spades. 

In the main treatment, the computer chooses either the Green or Purple deck with equal probability. Participants do not observe the color. Instead, participants are shown the suit of a single card drawn from the chosen deck. Given this signal, participants are asked to provide a percent chance that the chosen deck is Purple or Green. These probabilities are restricted to be between 0 and 100 percent and must sum to 100. In addition to the main treatment, there are treatments with multiple draws of cards, elicitation of willingness-to-pay for drawing cards, and where the signal precision is unknown. The timing of the treatments and more details are described in \Cref{app:study1a-timing}. Screenshots of the experimental interface are contained in the Supplementary Appendix. 

The relative proportion of suits in each deck determines the signal strength of observing a card. For example, if the Purple deck contains a large proportion $\rho_1=D_1/N$ of Diamonds, while the Green deck contains a very small portion $\rho_2$ of Diamonds, a Diamond card is a strong signal that the chosen deck is Purple. Given our core prediction, we vary these proportions to vary signal strength. Following the literature, we largely focus on symmetric signal structures, in which $\rho \equiv \rho_1 = 1-\rho_2$. We choose 32 possible values of $\rho$ within the range [0.047, 0.495] or [0.505, 0.953]. These values correspond to 16 possible signal strengths ($\mathbb{S} = |\text{logit }\rho|$) in the range $\mathbb{S} \in [0.02, 3.00]$.\footnote{More specifically, we choose whole numbers of cards such that signal strengths would be closest to the following values: $\{0.02, 0.05, 0.10, 0.15, 0.20, 0.30, 0.40, 0.50, 0.75, 1.00, 1.25, 1.50, 1.75, 2.00, 2.50, 3.00\}$.} On each question, we randomized whether the Green deck or Purple deck had more Diamonds or Spades, which suit was chosen, and whether the number of cards in a deck $N$ was 1665 or 337.\footnote{The deck sizes are intentionally large and irregular to (1) allow for a wide range of signal strengths, (2)~remove clear anchor points for people's answers, and (3) induce some uncertainty in mental calculations.}

We use monetary incentives to elicit participants' beliefs, as incentives have been shown to improve decision-making in these settings (e.g., \cite{G92}). We implement a version of the binarized scoring rule (\cite{HO13}) that is easier for participants to comprehend: \textit{paired-uniform scoring} (\cite{VW-WP}).\footnote{In general, binarized scoring rules have been argued to better account for risk aversion and hedging (\cite{ACH18}).} Participants' answers determine the probability that they win a high bonus as opposed to a low bonus.

\subsubsection*{Implementation}

Study 1a was conducted in March 2021. Participants were recruited from the online platform Prolific (\url{prolific.co}). Prolific was designed by social scientists in order to attain more representative samples online; it has been shown to perform well relative to other participant pools (\cite{GRW-WP}). 500 participants completed the experiment and passed the attention check, of whom five were randomly chosen to win bonuses (either a high bonus of \$100, or a low bonus of \$10). 
All participants received a \$3 show-up fee, and the average bonus earnings for the selected participants was \$82. 

Participants play 12 rounds in the main part of the study, in each of which they observe one draw of a card. We elicit 6,000 predictions in this part: 4,036 from one symmetric signal, and 1,964 from one asymmetric signal. To test that participants have a basic understanding of the setting, we randomly make 72 signals fully uninformative.\footnote{That is, both decks have exactly the same composition, so the correct update is to stay at 50 percent. Reassuringly, 96 percent of participants answer exactly 50 percent.} The rest of the study includes an attention check, multiple draws of cards, demand for information, and signals with ambiguous strength. Given space constraints and to emphasize our core results, we largely focus on the main treatment, where people see one symmetric signal and signal strength does not depend on the signal realization. Details for the additional treatments are in \Cref{app:experiment} and our previous working paper (\cite{ALT-old}, or ALT \citeyear{ALT-old}).

\subsubsection*{Main Results}

In the main condition where signals are symmetric, the signal precision from learning the suit of one drawn card is $\rho=\rho_1=\rho_2$. Given that the prior is $1/2$ (both decks are equally likely to be chosen), a Bayesian will place probability $\rho$ that the card was drawn from the deck that contains more of that card's suit.

\begin{figure}[t!]
\caption{Study 1a: Over- and Underinference by Signal Strength}
\begin{center}  
\vspace{-5.5mm}
    \includegraphics[width=.49\textwidth]{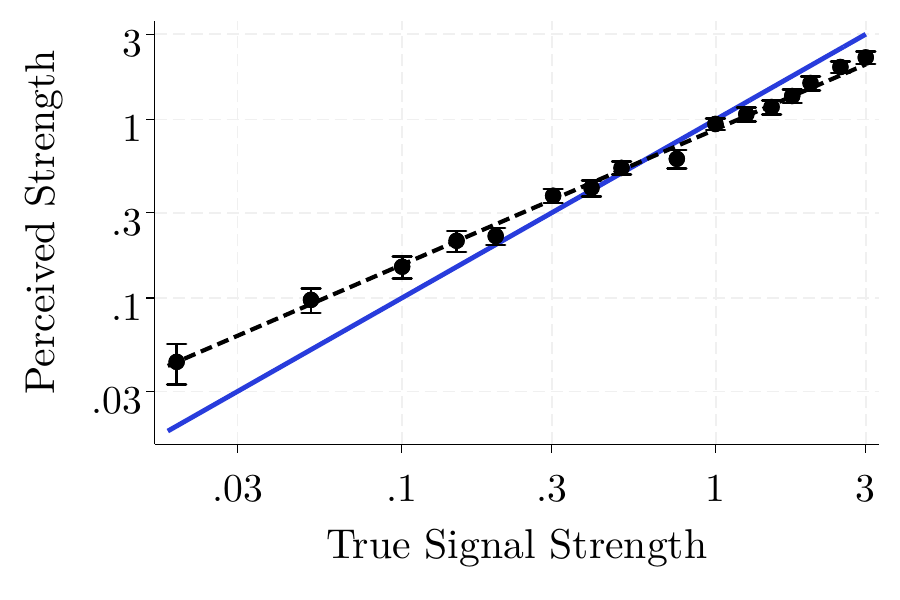}
    \includegraphics[width=.49\textwidth]{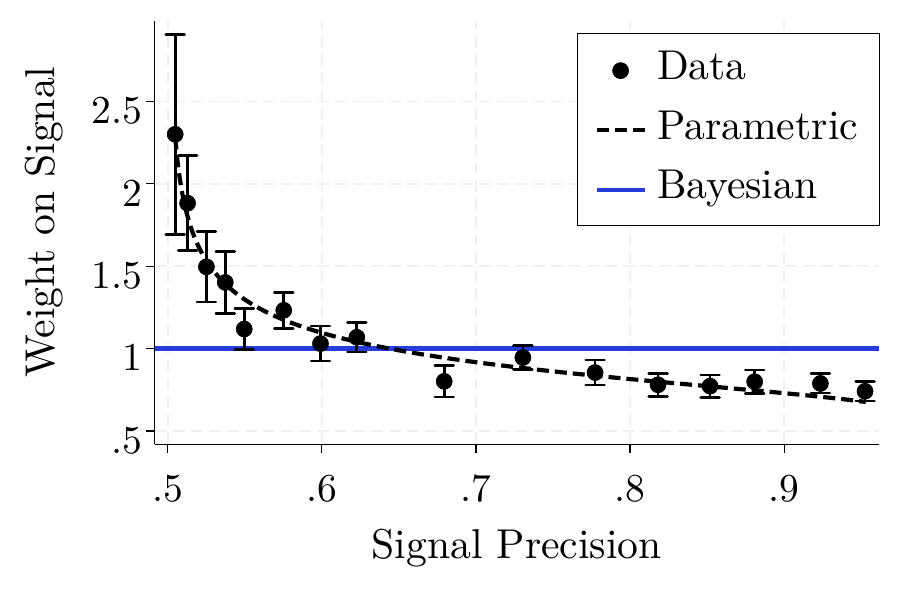}
\end{center}
\begin{threeparttable}
\begin{tablenotes}
\begin{scriptsize}
\vspace{-9.5mm}
\item \textbf{Notes:} 
The left panel plots the perceived signal strength (the logit belief change) as a function of true signal strength on a log-log scale. The right panel plots the average weight participants put on signals relative to a Bayesian for whom the weight is 1. In both panels, black markers plot the data (with 95\% confidence intervals). Observations are winsorized for each signal strength category at the 5\% and 95\% level. Dashed lines fit the data using the power weighting function from equation~\eqref{eq:weight}, estimating parameters using nonlinear least squares. Blue lines indicate Bayesian behavior.  Both panels show that participants overweight weak signals and underweight strong signals.\par
\end{scriptsize}
\end{tablenotes}
\end{threeparttable}
\label{fig:main_effects}
\end{figure}

We used \Cref{fig:theory} in the theoretical section to visually represent our core predictions under the log-normal parameterization of our model. \Cref{fig:main_effects} presents the same graphs with the addition of the actual data from the experiment, where we back out participants' perception of signal strength from their posterior (given a fixed prior of $1/2$). We compare our estimates for each condition (black circles) and the fitted predictions of the parameterized model (dashed lines, described below) with Bayesian updating (blue lines). The left panel shows that participants' behavior is not purely random: they qualitatively understand that stronger signals are in fact stronger, as average perceived signal strength rises monotonically with true strength. But this relationship is \emph{quantitatively} muted, so participants systematically overinfer from weak signals and underinfer from strong signals. As in the parameterized model, this relationship between true and perceived signal strength is close to linear in logs.

The right panel presents the same information in a different way, showing that people are effectively overweighting weak signals and underweighting strong signals, with a shape that again largely hews to the predictions of the parameterized model. For very weak signals, participants are acting as if signals are more than twice as strong as they truly are; for very strong signals, they are acting as if signals are roughly 2/3 as strong as they truly are.

The parametric curves in \Cref{fig:main_effects} are obtained by estimating the model parameters $k$ and $\beta$ from equation~\eqref{eq:weight}, $\hat{w}(\mathbb{S}) = k \cdot \mathbb{S}^{\,-(1-\beta)},$  using nonlinear least squares. 
The estimated value for $k$ is 0.88 (s.e.\ 0.02) and for $\beta$ is 0.76 (s.e.\ 0.03). The value of $\beta$ is statistically significantly less than one ($p<0.001$), as predicted. These values correspond to an estimate for the switching point $\rho^*$ of 0.64 (s.e.\ 0.01).\footnote{Equivalently, people are updating as if the distribution of strengths is such that $\s^* = \logit(0.64) = 0.58$.} All standard errors are clustered by participant.

Experiments using this paradigm have been run many times in the past, largely focusing on higher signal strengths. In Appendix \Cref{fig:benjamin_comp}, we compare our estimates to the many studies discussed in \textcite{B19}. Our results line up with previous studies' estimates for these higher signal strengths. In particular, we match the literature in finding underinference for signals with precision at or above 2/3. It is only for signals with precision below 0.6 that we see overinference, and this is a range that the past literature had not explored.

To explore our main results more formally in a consistent way across studies, \Cref{tab:expt_results} presents the results of regressions for the weight on the signal, $\hat w(\s)$, on a constant and the true signal strength $\s$. Bayes' rule would predict a constant of 1 and slope of 0 on $\s$ in such a regression, while our theory predicts a constant above 1 (indicating overinference for very weak signals) and a slope below 0 (indicating that people are partially insensitive to signal strength, and switch to underinference for strong signals).\footnote{Both our model and \Cref{fig:main_effects} suggest a nonlinear  relationship between weight and strength. We thus see the linear specification in \Cref{tab:expt_results} as providing a clean hypothesis test of the key effect predicted by our theory, rather than identifying model parameters directly (which we do separately via nonlinear least squares).} Column~(1) confirms the relationship suggested by \Cref{fig:main_effects} for Study 1a: the constant is above 1, the slope is below~0, and both effects are precisely estimated and strongly significant. 

\begin{table}[t!]
\centering
\begin{threeparttable}%[tb!]
\centering
\onehalfspacing
\begin{small}
\caption{The Effects of Signal Strength on Over- and Underinference\\[-3pt]}
\begin{tabular}{l !{\color{gray!0}\vrule width 0.0pt}
cc!{\color{gray!0}\vrule width 0.0pt} c!{\color{gray!0}\vrule width 0.0pt} 
cc!{\color{gray!0}\vrule width 0.0pt} 
cc}
\toprule
  \hphantom{Dep.\ Var.: Weight on Signa}
  &\multicolumn{1}{c}{Bayes} 
  &\multicolumn{1}{c!{\color{gray!0}\vrule width 0.0pt}}{Theory} 
  &\multicolumn{1}{c!{\color{gray!0}\vrule width 0.0pt}}{Study 1a} 
  & \multicolumn{2}{c!{\color{gray!0}\vrule width 0.0pt}}{Study 1b} 
  & \multicolumn{2}{c}{Study 2} \\

  \cmidrule(lr){2-2} \cmidrule(lr){3-3} \cmidrule(lr){4-4} \cmidrule(lr){5-6} \cmidrule(lr){7-8}

\multicolumn{2}{l}{\small Dep.\ Var.: Weight on Signal}
  &\multicolumn{1}{c!{\color{gray!0}\vrule width 0.0pt}}{}
  &\multicolumn{1}{c!{\color{gray!0}\vrule width 0.0pt}}{(1)}
  &\multicolumn{1}{c}{(2)}
  &\multicolumn{1}{c!{\color{gray!0}\vrule width 0.0pt}}{(3)}
  &\multicolumn{1}{c}{(4)}
  &\multicolumn{1}{c}{(5)}\\[.1em]

\midrule

Constant         &  1 & $>1$   & 1.420   & 2.180 & 2.182 & 1.706 & 1.700 \\
\seline 
&&   & (0.030) & (0.049) & (0.048) & (0.024) & (0.025)  \\
Signal Strength    & 0 & $<0$   & -0.308  & -0.957 & -0.958 & -2.078  & -2.060 \\
\seline
                     && & (0.031) & (0.065) & (0.065) & (0.111)  & (0.112) \\
\rule{0pt}{0.35ex}Weight on Prior  & 1 &  &   &  &  0.980&  & 0.976\\
\seline
                          &&  &   &  &   (0.013)& &  (0.009)\\
                    
\rule{0pt}{1ex}Participant FE    &&                &      Yes&      Yes&      Yes&     Yes &      Yes\\
Round FE                     && &      Yes&      Yes&      Yes&      Yes&      Yes \\[.1em]
\midrule
Observations                 && &     3964&     7500&     7500&     8000 &     8000\\
\(R^{2}\)                    && &     0.23&     0.16&     0.17&     0.24 &     0.24\\
$p$-val.: $\text{Const.}=1$                   && &     <0.001&     <0.001&     <0.001&     <0.001 &     <0.001\\
$p$-val.: $\text{Slope}=0$                    && &     <0.001&     <0.001&     <0.001&     <0.001 &     <0.001\\ 
\bottomrule
%\multicolumn{8}{l}{\footnotesize Standard errors in parentheses}\\
\end{tabular}

\label{tab:expt_results}
\end{small}
\begin{tablenotes}[para,flushleft]
\singlespacing
\vspace{-3.85mm}
\begin{scriptsize}
\textbf{Notes:} Columns (1)--(5) show OLS estimates with standard errors in parentheses clustered by participant. The dependent variable is the weight put on the signal compared to a Bayesian, defined following \Cref{sec:theory} as $\hat w(\s) = \hat\s/\s$, where the perceived strength $\hat\s$ is estimated for a given observation from the logit change in beliefs. Weights greater than 1 correspond to overinference; weights less than 1 correspond to underinference. Our theory predicts a constant of $>1$ (indicating overinference as signals become very weak) and a coefficient on signal strength of $<0$ (indicating relatively more underinference as signals become stronger). Columns (3) and (5) control for weight on prior following equation~\eqref{eq:brn_est}, and a coefficient below 1 indicates base-rate neglect. \par
\end{scriptsize}
\end{tablenotes}
\end{threeparttable}
\end{table}

\subsubsection*{Heterogeneity}

The theory assumes that people use a randomly drawn estimate of signal strength to form their beliefs. Consequently, it predicts our main effect occurs \textit{on average}, but also that there will be heterogeneity: some people will overreact and some will underreact to any given signal. 
Appendix \Cref{fig:individual-cdfs} plots the raw cumulative distribution and  probability density functions at the individual level for strong and weak signals. Nearly everyone updates in the right direction, and the distributions are centered in accordance with our main effect. But given the non-trivial spread in the distributions, there is clear heterogeneity in perceived signal strength and associated updating behavior.

The model also makes the prediction that the core effect will be larger as a person's estimate of signal strength becomes less precise. Naturally, we cannot observe the precision of a person's internal estimates of strength, and therefore must rely on a set of proxies. To estimate heterogeneity in treatment effects, we then interact these proxies with the signal strength in regressions for the weight placed on the signal, with results presented in \Cref{tab:expt_results_heterogeneity}.

\begin{table}[t!]
\centering
\begin{threeparttable}%[tb!]
\centering
\onehalfspacing
\begin{small}
\caption{Heterogeneity in Treatment Effects\\[-3pt]}
\begin{tabular}{l ccc!{\color{gray!0}\vrule width 0.0pt} ccc}
\toprule
  &\multicolumn{3}{c!{\color{gray!0}\vrule width 0.0pt}}{Study 1a} & \multicolumn{3}{c}{Study 1b} \\

% &\multicolumn{4}{c!{\color{gray!0}\vrule width 0.0pt}}{(Abstract)} & \multicolumn{3}{c}{(Vary Priors)} \\

\cmidrule(lr){2-4} \cmidrule(lr){5-7} 

\multicolumn{1}{l}{\small Dep.\ Var.: Weight on Signal}
  &\multicolumn{1}{c}{(1)}
  &\multicolumn{1}{c}{(2)}
  &\multicolumn{1}{c!{\color{gray!0}\vrule width 0.0pt}}{(3)}
  &\multicolumn{1}{c}{(4)}
  &\multicolumn{1}{c}{(5)}
  &\multicolumn{1}{c}{(6)} \\[.1em]

\midrule

Constant   &   1.395&    1.416&  1.421&  2.147 &   2.183 & 2.181 \\
\seline
                              &  (0.021)&  (0.030)& (0.030)&   (0.037) &  (0.048) & (0.048) \\
Strength          &   0.118&    -0.072&  -0.175&    -0.002 &   -0.698 &   -0.756\\
\seline
                              &  (0.036)&  (0.009)&  (0.028)& (0.070) &  (0.089) &  (0.120)\\    
\hspace{3mm}  Strength $\times$ Noise        
& -0.383   & & & -0.433 & &  \\
\seline
& (0.036)   & & & (0.046) & &  \\               
\hspace{3mm}  Strength $\times$ Inexperience   
& & -0.042  & & & -0.037 &  \\
\seline
& & (0.009)  & & & (0.012) &  \\              
\hspace{3mm}  Strength $\times$ CRT Incorrect       
& & & -0.102 & & &  \\
\seline
& & & (0.028) & & &  \\          
\hspace{3mm}  Strength $\times$ Uncertainty
& & & & & & -0.542 \\
\seline
& & & & & & (0.288) \\   
\rule{0pt}{0.45ex}Base-Rate Neglect        &   &    & & 0.980 &  0.980  & 0.980  \\
\seline
                              &  &  & &  (0.013) &  (0.013) &  (0.013)\\    
\rule{0pt}{1ex}Participant FE                    &      Yes&          Yes& Yes &   Yes &      Yes  &      Yes\\
Round FE                      &      Yes&          Yes& Yes &   Yes &      Yes  &      Yes\\[.1em]
\midrule
Observations                  &     3964&         3964&    3964&     7500 &     7500 &     7500\\
\(R^{2}\)                     &     0.28&     0.24&      0.23&    0.20 &     0.17 &     0.17\\
\bottomrule
%\multicolumn{6}{l}{\footnotesize Standard errors in parentheses}\\
\end{tabular}

\label{tab:expt_results_heterogeneity}
\end{small}
\begin{tablenotes}[para,flushleft]
\singlespacing
\vspace{-3.85mm}
\begin{scriptsize}
\textbf{Notes:} OLS, with standard errors in parentheses clustered at the participant level. The dependent variable is the weight put on the signal compared to a Bayesian, as in \Cref{tab:expt_results}. Weights greater than 1 correspond to overinference; weights less than 1 correspond to underinference. Study 1b controls for weight on prior, where $\alpha < 1$ indicates base-rate neglect. Noise is defined as the SD of weights on other questions. CRT incorrect ranges from 0 to 3. Inexperience equals the number of rounds remaining in the main experiment. Uncertainty ranges from 0 to 1 and is described in \Cref{sec:study1b}. We do not include the estimation precision proxies as separate regressors, as they are absorbed by either the participant or round fixed effects included in all regressions. \par
\end{scriptsize}
%\vspace{2mm}
\end{tablenotes}
\end{threeparttable}
\end{table}

Our first proxy for (im)precision uses the standard deviation in a person's implied signal weights across the experiment. Intuitively, a person whose estimates have very high precision will have low variance in these weights, since most weights will be around 1; meanwhile, a person whose estimates have low precision will have high variance in weights.\footnote{There is a small endogeneity issue in using the same observation to measure a person's reaction and also to calculate a person's weight variance across choices. As a result, we relate a person's reaction in one decision to the standard deviation in their weights for all \textit{other decisions} on similar problems.} As shown in column (1) of \Cref{tab:expt_results_heterogeneity}, our effect is stronger (i.e., the interaction term is negative) for people who have higher standard deviation of their weights on other questions. 

Our second proxy for estimation precision is task experience: if people become better at understanding and estimating signal strength as they get more practice, then they will be less precise earlier in the experiment (and our core effect will be stronger). As shown in column~(2) of Table~\ref{tab:expt_results_heterogeneity}, consistent with this idea, people overweight weak signals and underweight strong signals by more in earlier rounds of the experiment. 

In addition to these two proxies, we also preregistered correlating our effects with performance on a three-item cognitive reflection test (CRT; \cite{F05}). In column (3) of Table~\ref{tab:expt_results_heterogeneity}, we find that people with lower CRT scores show the core effect significantly more.\footnote{We also preregistered looking at an additional heterogeneity by self-reported news consumption, and indeed find that less experience with news consumption is correlated with our core effect (see ALT \citeyear{ALT-old}).}

\subsubsection*{Extensions: Asymmetric, Multiple, and Ambiguous Signals}

The main treatment of the experiment focuses on how people respond to one symmetric signal with a deterministic signal strength. The experiment included additional treatments in which we relax each of these features. We report some key takeaways here. 

First, we consider asymmetric signals such that one deck has a similar share of Spades and Diamonds, but the other deck does not. We find that our main results continue to hold for these asymmetric signals, and as predicted by our theory, the more complicated problem leads to a stronger effect. Using the same nonlinear least squares estimation as above, we estimate a value for $k$ of 0.84 (s.e.\ 0.03) and for $\beta$ of 0.56 (s.e.\ 0.04), with a similar estimated switching point $\rho^*$ of 0.66 (s.e.\ 0.01). 

Second, in Appendix \Cref{fig:multiple}, we replicate the finding in \textcite{TK92} that people react less to multiple signals than a single signal with the same overall strength. The reduction in reaction to multiple signals is essentially constant for all strengths, so this effect is orthogonal to our main effect. Third, we consider ambiguous signals by telling participants that the share of suits in each deck is equal to one of two possible values (high or low). Our main effect continues to hold, and results suggest that people first estimate each possible signal strength, and then average these estimates, to form their overall expected strength (see ALT \citeyear{ALT-old}).

\subsubsection*{Other Concerns}

Finally, we consider a set of alternative hypotheses for our results that are unrelated to over- or underinference. First, it is possible that participants simply dislike stating ``50 percent'' (or a number close to it), even when signals are not informative. However, 96\% of participants who see a completely uninformative signal say exactly 50 percent, suggesting that there is no aversion to this answer. We additionally vary the prior from 50\% in the replication study below, finding no change in our main results.\footnote{Results are consistent in a treatment that uses a non-round prior (33.$\overline{3}$\%), further suggesting that a preference for stating the closest round number above/below the prior given a weak signal is not a key driver.} 
Second, it is possible that participants' behavior is driven by reactions to particular components of the experiment, for example being influenced by the relative salience of the first deck or the second deck (or the Green and Purple color), positive or negative signal, the suit of the signal, or the particular deck size. However, we find a tightly estimated null effect of color and suit asymmetry. For the deck size (which was randomly either 1,665 or 337), the estimated switching point $\rho^*$ does change, but sensitivity parameters $\beta$ are statistically indistinguishable.\footnote{People's average estimate of the first (Green) deck is 0.503 (s.e.\ 0.002), which suggests no preference for the first (Green) deck and no differential distortion of up or down signals. For suits, participants' average signal weight of a Diamond is 1.156 (s.e.\ 0.031) and 1.134 (s.e.\ 0.035) for a Spade. Similarly, deck size does not impact the core sensitivity parameter $\beta$: for the larger deck, $\beta = 0.75$ (s.e.\ 0.03), and for the smaller deck, $\beta = 0.75$ (s.e.\ 0.04). However, there is a difference in $\rho^*$: the larger deck is 0.68 (s.e.\ 0.02) and the smaller 0.61 (s.e.\ 0.01). Although speculative, one explanation for the shift in $\rho^*$ is that participants put some weight on the \textit{differences} between numbers of cards, and don't only focus on the ratios.} 

\subsection{Study 1b: Follow-Up Experiment}
\label{sec:study1b}
\subsubsection*{Design}

To probe the robustness of the results from Study 1a, we run a follow-up in Study~1b using the same general design, but now considering asymmetric prior beliefs. Given our focus on the robustness of the main results, Study 1b drops the additional treatments in the original study and focuses on the reaction to a single symmetric signal. 

To allow for asymmetric prior beliefs, we vary the probability that the first (Green) deck is chosen. In the original study, the chosen deck is picked randomly from 2 decks, such that the likelihood of picking the Green deck is 1/2. In order to vary this prior likelihood in a salient way, Study 1b includes treatments in which there are 2, 3, or 4 decks, and each deck is chosen with equal probability (1/2, 1/3, or 1/4, respectively). As in the original study, the first deck is Green and has $D_1$ Diamonds and $N - D_1$ Spades. The other decks are different shades of Blue and have identical compositions of $N - D_1$ Diamonds and $D_1$ Spades. Given this setup, the signal strength matches that of the original study, but the person's prior that the Green deck is chosen is either 1/2, 1/3, or 1/4.\footnote{Another way to vary the prior would have been to continue to use two decks, but tell participants that the Green deck would be chosen with some specific probability. We instead chose the multi-deck design as it makes the change in the prior more clear and, from our perspective, easier to understand.} After the suit of the drawn card is shown to the participant, we elicit the probability that each deck was chosen. Our analysis considers the stated probability for the Green deck as the belief outcome of interest.

\subsubsection*{Implementation}

Study 1b was conducted in March 2024. Participants were again recruited from Prolific. As preregistered, 500 participants completed the experiment and passed an attention check. Ten participants were randomly chosen to win bonuses. If they won the high bonus, they received \$50; if not, they received no bonus. All participants received a \$3.60 show-up fee, and the average bonus earnings for the selected participants was \$35. Participants played 15 rounds in the study, in each of which they received one draw of a card. The experiment involved three blocks of five rounds. Each block gave participants a different prior, in which the Green deck, as above, had either a 1/2, 1/3, or 1/4 probability of being chosen.

\subsubsection*{Results}

\begin{figure}[t!]
\caption{Study 1b: Over- and Underinference by Signal Strength}
\begin{center}
%\textbf{(b)} {Study 1b}
%\vspace{-2mm}
%\rule{4cm}{0.4pt}
\vspace{-5.5mm}
    \includegraphics[width=.49\textwidth]{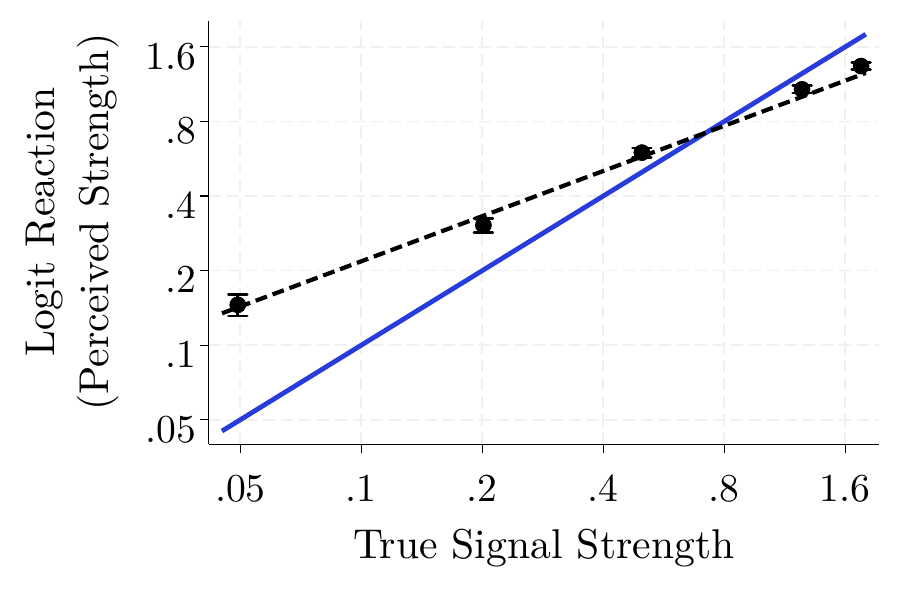}
    \includegraphics[width=.49\textwidth]{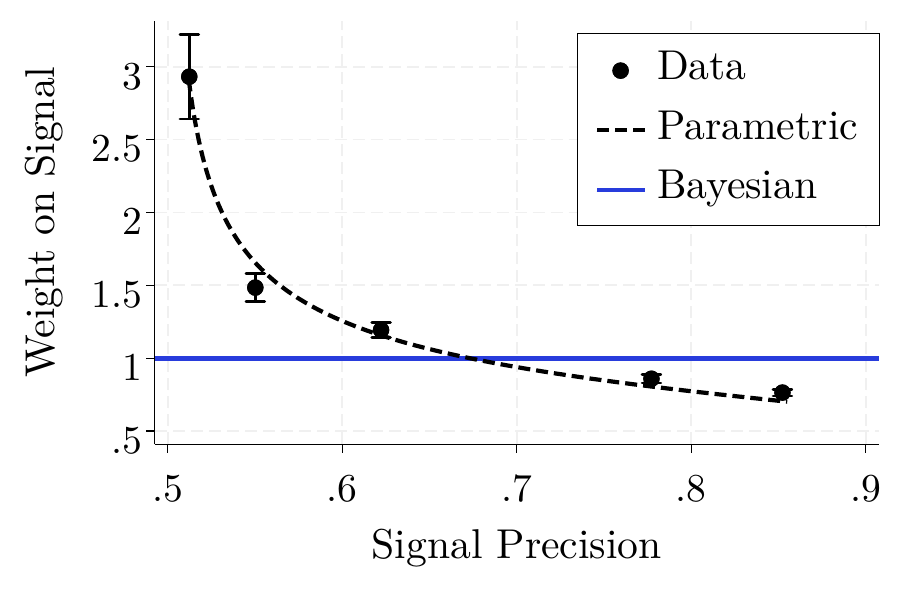}
\end{center}
\begin{threeparttable}
\begin{tablenotes}
\begin{scriptsize}
\vspace{-9.5mm}
\item \textbf{Notes:} The left panel plots the logit belief change (equal to perceived strength in our theory) as a function of true signal strength on a log-log scale. The right panel plots the average implied weight participants put on signals relative to a Bayesian for whom the weight is 1. In both panels, black markers plot the data (with 95\% confidence intervals). Observations are winsorized for each category of signal strength and prior at the 5\% and 95\% level. Dashed lines fit the data using the power weighting function from equation~\eqref{eq:weight}, estimating parameters using nonlinear least squares. Blue lines indicate Bayesian behavior.  Both panels show that participants overweight weak signals and underweight strong signals. \par
\end{scriptsize}
\end{tablenotes}
\end{threeparttable}
\label{fig:main_effects_2}
\end{figure}

We first visually present the main results in \Cref{fig:main_effects_2}, which replicates \Cref{fig:main_effects} using the new data from Study~1b. We find that the broad patterns in the logit belief changes (in the left panel) and resulting implied signal weights are very similar to those from Study~1a, even when allowing for asymmetric priors.

Next, we replicate the regression from Study~1a for the implied weight on the signal, with results shown in column~(2) of \Cref{tab:expt_results}. Since $\hat w(\s)$ is measured from the logit belief change, this analysis implicitly assumes that people correctly incorporate the prior probability. But as discussed in Section~\ref{sec:relax}, these results may be contaminated if people do not appreciate or misweight the prior (as is true with base-rate neglect). As such, column~(3) estimates and controls for the effect of misweighted priors. In particular, it includes the additional regressor $\frac{\logit\hspace{1pt}\pi_0}{\logit\hspace{1pt}\pi_1-\logit\hspace{1pt}\pi_0}$ in the regression, which (omitting the error term and fixed effects) is now
\begin{equation}
    \hat w(\s(s)) = \underbrace{\gamma_0 + \gamma_1\cdot\s(s) \vphantom{\frac{\logit\hspace{1pt}\pi_0}{\logit\hspace{1pt}\pi_1-\logit\hspace{1pt}\pi_0}}}_{\text{Previous Terms}} + \underbrace{(\alpha-1)\cdot\frac{\logit\hspace{1pt}\pi_0}{\logit\hspace{1pt}\pi_1-\logit\hspace{1pt}\pi_0}}_{\text{Base-Rate Neglect Term}}, \label{eq:brn_est}
\end{equation}
and an estimated $\alpha<1$ represents base-rate neglect.\footnote{The regressor's denominator $\logit\hspace{1pt}\pi_1-\logit\hspace{1pt}\pi_0$ is included to make $\alpha$ here match its typical interpretation in a \textcite{G80} regression. The typical \citeauthor{G80} regression is $\logit\hspace{1pt}\hat\pi_1 = \alpha\,\logit\hspace{1pt} \pi_0 + \gamma\, (\logit\hspace{1pt}\pi_1-\logit\hspace{1pt}\pi_0),$ or $\logit\hspace{1pt}\hat\pi_1- \logit\hspace{1pt}\pi_0= (\alpha-1)\,\logit\hspace{1pt} \pi_0 + \gamma\, (\logit\hspace{1pt}\pi_1-\logit\hspace{1pt}\pi_0).$ Our regression sets $\gamma=\gamma_0+\gamma_1\s(s)$ and uses   $\hat w(s) = \frac{\logit\hspace{1pt}\hat\pi_1- \logit\hspace{1pt}\pi_0}{\logit\hspace{1pt}\pi_1- \logit\hspace{1pt}\pi_0}$ as the outcome variable. So dividing both sides of the \citeauthor{G80} equation by $\logit\hspace{1pt}\pi_1-\logit\hspace{1pt}\pi_0$, we obtain equation~\eqref{eq:brn_est}, with $\alpha$ having the same interpretation as in \citeauthor{G80}'s case. Intuitively, base-rate neglect matters more for the estimated weight the greater the distance of $\pi_0$ from 0.5 (the regressor's numerator) relative to the signed signal strength (its denominator).\label{fn:brn}} Controlling for misweighted priors does not affect our main results: the estimated constant and slope on strength in columns~(2) and~(3) are close to identical, and we continue to strongly reject the Bayesian null in the manner predicted by our theory. If anything, comparing the columns for this study with column~(1) for Study~1a, asymmetric priors seem to strengthen our effects, potentially because signal-strength estimation is more challenging when there are more decks.\footnote{This is at best a tentative conclusion: the estimates across the two studies may not be directly comparable, as Study~1b uses a more limited set of signal strengths.}

We also analyze the effects of asymmetric priors in a specification that follows the \citeauthor{G80} (\citeyear{G80}) regression approach more directly:
\begin{align}
    \text{logit}\hspace{1pt}\hat\pi_1(s) = \alpha \cdot \text{logit}\hspace{1pt}\pi_0 \pm \gamma \cdot \mathbb{S}(s). \label{eq:grether_reg}
\end{align}
To allow for differences in inference in response to signals of different strengths, we estimate~\eqref{eq:grether_reg} separately for each signal strength $\s(s)$. The results are presented in Appendix~\Cref{tab:expt_results_grether}, and they align with those in \Cref{tab:expt_results}, albeit with different interpretation for the strength coefficient $\gamma$. We find that participants significantly overweight weak signals ($\hat\gamma>1$) and underweight strong signals ($\hat\gamma<1$). We find that there is significant base-rate neglect for strong signals but none for weak signals, indicating that the modest estimates for overall base-rate neglect may partly reflect the inclusion of the weak-signal treatments.\footnote{The table also presents a set of additional analyses. Column~(1) considers only the $\pi_0=0.5$ treatment, finding our usual results. Column~(2) replicates the analysis for all priors, imposing $\alpha=1$, and finds slightly stronger results. Column~(3) allows for separate $\gamma$ across strengths but sets $\alpha$ to be constant, with similar results and mild base-rate neglect. Column~(4) presents the full set of $\alpha$ and $\gamma$ estimates described in the text.}

In the rightmost columns of  \Cref{tab:expt_results_heterogeneity}, we examine heterogeneity in our main treatment effect by interacting signal strength with proxies for estimation precision (or imprecision), controlling for base-rate neglect as in equation~\eqref{eq:brn_est}. Column~(4) considers the same noise proxy from Study 1a, finding again that people who have higher variance in weights exhibit stronger effects. Column~(5) considers how our effect correlates with task experience, again finding that people exhibit a stronger effect earlier in the experiment.

We also elicit one additional measure to proxy for participants' estimation precision, based on the elicitation procedure used by \textcite{EG-WP}. In particular, we ask people to answer ``How certain are you that the optimal guess is somewhere between $x-1$\% and $x+1$\%?'' on a scale from 0 to 100. We ask people this question three separate times during the experiment, and average their answers to get an additional measure of a person's estimation precision: intuitively, a person with low precision will report higher subjective uncertainty than a person with high precision (as shown by \cite{EG-WP}, and \cite{EGO-WP}). Column (6) suggests that this new proxy of cognitive uncertainty is also associated with our effect in the direction predicted by the theory: people with more stated uncertainty about their answer seem to exhibit our core effect more strongly.

Finally, we again estimate $k$ and $\beta$ from equation~\eqref{eq:weight}. The estimated value for $k$ is 0.89 (s.e.\ 0.02) and for $\beta$ is 0.61 (s.e.\ 0.02). The value of $\beta$ is statistically significantly less than one (p-value $<0.001$). These values correspond to an estimate for $\rho^*$ of 0.68 (s.e.\ 0.01). Allowing for base-rate neglect in the model gives an estimate of 0.94 for the weight on prior, and leads to little change in the other estimates ($k = 0.87$ and $\beta = 0.69$).

\subsection{Study 2: Naturalistic Experiment}
\label{sec:nba-experiment}

\subsubsection*{Overview}

The benefit of the abstract data-generating process in Studies 1a and 1b is that it is cleanly and fully defined. This constrained structure allows for straightforward manipulation of signal strength and calculation of a precise Bayesian benchmark, which is a key reason this paradigm is so widely used. But one possible concern is that this abstract, numerically-oriented environment is unnatural for most people, more closely mirroring a math exam than a real-life updating situation. If people solve abstract inference problems differently than more naturalistic problems, our results might not generalize to real-life behavior. 

Given this concern, our next experiment attempts to study updating behavior in a more naturalistic environment. In particular, we analyze how NBA basketball fans update their beliefs that a team wins a game given information that they make or miss a shot in different situations. This environment provides an experimental parallel to one of the observational-data settings considered in \Cref{sec:markets}. We choose to focus on it here because while the DGP is naturalistic and complex, fans intuitively understand this process well and can easily make reasonable predictions. A made basket is almost always a positive signal, and a missed basket is a negative signal, but the exact strength of this signal is unclear.

This last feature also represents the main challenge in analyzing a naturalistic setting: not having exact knowledge of the signal strength would seem to make it difficult to test for over- and underreaction. Crucially, however, this environment is one where we can obtain credible estimates of the correct probabilities in different situations using historical game data. We do so using 
an online win-probability calculator from Inpredictable, a sports analytics site that provides estimates for different game situations.\footnote{Our estimates are taken from \url{https://stats.inpredictable.com/nba/wpCalc.php}. This calculator takes as input the current score differential, time remaining, and which team has possession, and outputs a win probability based on historical data. To check whether this calculator gives reasonable estimates, we also created our own simple calculator based on more, or fewer, years of data; the estimates from our versions of the calculator and the online calculator are extremely similar. We tie our hands by using this third-party tool.} To provide participants signals with varying strength, we vary the game situation. As detailed below, the key source of signal-strength variation across scenarios is similar to the one we later use in our analysis of sports betting data: the timing of the event. NBA basketball games have four quarters, and a basket made in the fourth quarter is a stronger signal of the game's winner than a basket made in the first quarter. Our core prediction, therefore, is that people will overreact to made or missed shots in early quarters (when signals are weaker), and underreact to made or missed shots in late quarters (when signals are stronger).

\subsubsection*{Design}

Participants are told that they will see a variety of simple scenarios in an NBA  game (which include the score differential, time remaining, and which team has possession) between two unnamed teams (e.g., Team A and Team B), and that their goal is to estimate the probability that each team wins the basketball game in that scenario. Participants are sequentially given four sets of scenarios, with each scenario set starting with 2:40 left in one of the four quarters and the time decreasing by 10-15 seconds after each event. The order of the sets is random.

Within each scenario set, the person is first given a \textit{base scenario}. They are then told the actual calculated probability of the base scenario, such that all participants have the same prior. To provide variation in priors within a quarter, we randomize whether the lead in the base scenario is 1 or 5. Then, the participants are told the outcome of the next possession. This signal is equally likely to be good news for the team on offense (a made 2-point basket) or good news for the team on defense (a missed basket that leads to the defensive team getting possession). They are then asked for the probability that a given team will win after observing this event. We again elicit beliefs using the paired-uniform scoring version of the binarized scoring rule (\cite{VW-WP}; \cite{HO13}), with participants' answers determining the probability that they win a high bonus rather than a low bonus.\footnote{Our study instructions include the following: ``We have used a model based on a database of regular-season NBA games with several years of play-by-play data to estimate the likelihoods of each team winning in these scenarios. The closer your answer is to the likelihood, the more likely you are to win the \$50 bonus.''}

\begin{figure}[t!]
\caption{Study 2: Example of an Information Page Participants See}
\label{fig:nba_screenshot}
\centering
%\begin{center}  
%\vspace{0.5mm}
\fbox{
\parbox[c]{.89\textwidth}{
\centering
\vspace{1.5pt}
\includegraphics[trim=0in 7.05in 0in 0in, clip, width=.88\textwidth]{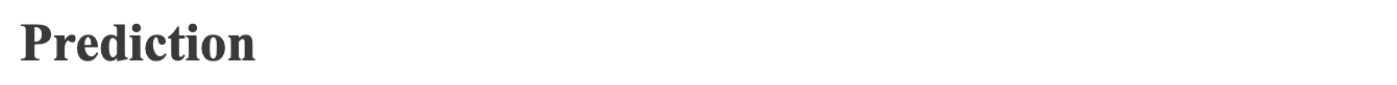} 

\includegraphics[trim=0in 5.815in 0in 0.875in, clip, width=.88\textwidth]{screenshots_nba/screenshot_nba-q3_alt_shorter.pdf} 

\includegraphics[trim=0in 2.875in 0in 2.05in, clip, width=.88\textwidth]{screenshots_nba/screenshot_nba-q3_alt_shorter.pdf} 

\includegraphics[trim=0in 2.2375in 0in 4.975in, clip, width=.88\textwidth]{screenshots_nba/screenshot_nba-q3_alt_shorter.pdf} 

\includegraphics[trim=0in 1.625in 0in 5.6in, clip, width=.88\textwidth]{screenshots_nba/screenshot_nba-q3_alt_shorter.pdf} 

\includegraphics[trim=0in 0in 0in 7.025in, clip, width=.88\textwidth]{screenshots_nba/screenshot_nba-q3_alt_shorter.pdf} 
\vspace{6pt}
}
}
%\end{center}
%\vspace{-11mm}
\end{figure}

After the person enters their answers, they go through this process for 3 more consecutive possessions in the same quarter. For each subsequent possession within this scenario set, they see the sequence of previous events within the quarter, as well as the answers they entered. After completing a scenario set, they move on to the next scenario set in a different quarter, where they are again told a base scenario and shown a series of signals. \Cref{fig:nba_screenshot} shows a screenshot with an example of the page participants see after the base scenario and one event, and a full set of screenshots of the study pages are again in the Supplementary Appendix.

We identify our core effect by exploiting variation in signal strength across these scenarios. Interestingly, the empirical variation in signal strength in these scenarios is driven almost entirely from variation in the amount of time left rather than the event or score differential.\footnote{Fixing time and initial score difference, our events (made and missed 2-point shots) have similar strengths, as NBA teams average close to 1 point per possession. 
Similarly, fixing time, baskets have surprisingly similar strengths given different initial score differences. Intuitively, while a basket shifts probability when tied more than when up by 10 (say 50\% to 60\% vs.\ 90\% to 93\%), these have virtually the same signal strength $\s$ given the different base probabilities. Quantitatively, using past game data and regressing estimated strength on time remaining yields an $R^2$ of about 55\%, and adding the score margin only improves this to 57\%.\label{fn:strength}} This motivates us to group our estimates by quarter when visually presenting our results below, to see whether we indeed observe overreaction on average in response to the weak signals in early quarters, and underreaction given the strong signals in late quarters.

\subsubsection*{Implementation}

Study 3 was conducted in April 2024. Participants were recruited from Prolific from a sample of Americans who reported that they were basketball fans. As preregistered, 500 participants completed the experiment, passed an attention check, and stated that they followed the NBA. Ten participants were randomly chosen to win bonuses. If they won the high bonus, they received \$50; if not, they received no bonus. All participants received a \$2.50 show-up fee and the average bonus earnings for the selected participants was \$25. Participants played 16~rounds in the study (4~possessions in a given quarter's scenario set, and 4~quarters).

\subsubsection*{Results}

To study the relationship between participant's perceived signal strength and the true signal strength, we first back out a participant's perceived signal strength from their beliefs before and after an event as follows:
\begin{align}\label{eq:basketball_update}
    \underbrace{\text{logit}(\hat{\pi}_{t+1}(s_{t+1}))}_{\substack{ \text{Logit of} \\ \text{Guess}}} = \underbrace{\text{logit}(\hat{\pi}_{t})}_{\substack{ \text{Logit of} \\ \text{Prior}}} \hspace{4mm} \underbrace{\vphantom{(\pi_t)} \pm}_{\substack{ \text{Signal} \\ \text{Direction}}} \hspace{4mm}  \underbrace{\hat{\mathbb{S}}(s_{t+1})}_{\substack{ \text{Perceived} \\ \text{Signal Strength}}}.
\end{align}
For the base scenario ($t=0$) in a given set, we set the prior $\hat{\pi}_0$ to the calculator-estimated~${\pi}_0$, as we give this win probability to the participant at the beginning of a set. We then provide signals (events) $s_{t+1}$ and elicit $\hat{\pi}_{t+1}(s_{t+1})$ for each $t=0,1,2,3$, backing out $\hat{\mathbb{S}}(s_{t+1})$ from $\hat{\pi}_{t+1}(s_{t+1})$ and their previous $\hat{\pi}_{t}$ (which they still see onscreen).\footnote{One benefit of giving a sequence of signals is our ability to observe the previous $\hat{\pi}_{t}$ in backing out $\hat{\mathbb{S}}(s_{t+1})$.}  We back out true signal strength $\mathbb{S}(s_{t+1})$ in a similar manner, but using the calculator's estimated $\pi_{t+1}(s_{t+1})$ after each signal. Following the previous studies, we then compare $\hat{\mathbb{S}}(s_{t+1})$ to $\mathbb{S}(s_{t+1})$.

\begin{figure}[t!]
\caption{Study 2: Over- and Underinference by Quarter}
\begin{center}  
\vspace{-5.5mm}
    \includegraphics[width=.49\textwidth]{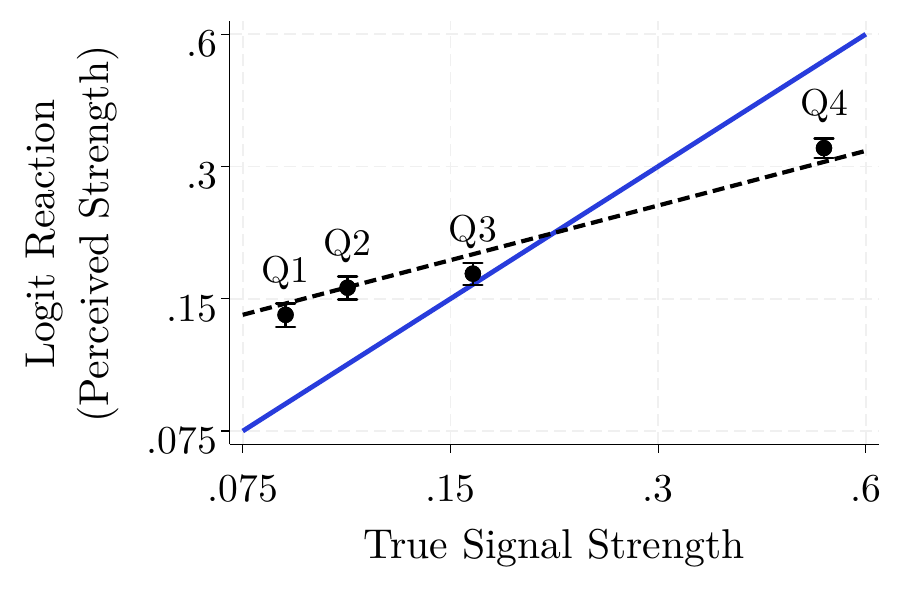}
    \includegraphics[width=.49\textwidth]{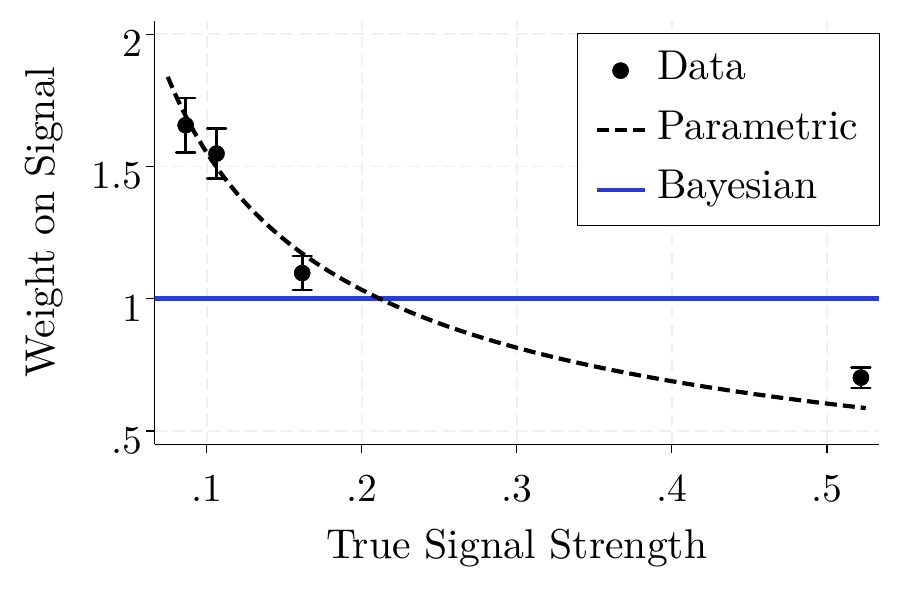}
\end{center}
\begin{threeparttable}
\begin{tablenotes}
\begin{scriptsize}
\vspace{-9.5mm}
\item \textbf{Notes:} The left panel plots the logit belief change (our measure of perceived strength, as in \eqref{eq:basketball_update}) as a function of true signal strength on a log-log scale, where true signal strength is based on the \texttt{inpredictable.com} win-probability calculator. The right panel plots the average weight participants put on signals relative to a Bayesian for whom the weight is 1, also against true signal strength (rather than precision, since signals are not necessarily symmetric).  In both panels, black markers plot the data (with 95\% confidence intervals), averaged by quarter; signal strengths increase in each quarter. Observations are winsorized for each category of quarter and score differential at the 5\% and 95\% level. Dashed lines fit the data using the power weighting function from equation~\eqref{eq:weight}, estimating parameters using nonlinear least squares. Blue lines indicate Bayesian behavior based on the calculator's average change. Both panels show that participants overweight weak signals (in earlier quarters of the game) and underweight strong signals (in the fourth quarter of the game). \par
\end{scriptsize}
\end{tablenotes}
\end{threeparttable}
\label{fig:main_effects_nba}
\end{figure}

We visually present our main results in \Cref{fig:main_effects_nba}, averaging perceived and true signal strength across all events in each quarter. The left panel shows that as in the previous studies, the relationship between perceived and true signal strength is approximately linear in logs, with a positive intercept and a muted slope. The dots are ordered by quarter from left to right: the first quarter has the lowest true signal strength, the fourth quarter has the highest, and participants understand this ordering. But while average perceived signal strength does rise over quarters, participants are insensitive to how much true signal strength is increasing, such that they overreact early and underreact late (switching around the third quarter). This can also be seen in the right panel, which plots the implied weights placed on events by quarter. People weight first-quarter events by about 1.6 times as much as the win-probability estimates suggest, and weight fourth-quarter events by less than 2/3 as much.

We then conduct regressions for the estimated signal weights as in the previous studies, with results shown in the last two columns of \Cref{tab:expt_results}. As usual, we estimate these regressions at the individual observation level, and thus do not group by quarter for this analysis. In column~(4), we regress $\hat w(\s)$ only on a constant and $\s$, implicitly assuming that people correctly incorporate their prior beliefs. We find the same qualitative patterns as in the abstract experiments. Quantitatively, we see greater insensitivity to signal strength than in the abstract studies, possibly because this is a more complex environment. In column~(5), we allow for misweighting prior beliefs, with estimation proceeding from equation~\eqref{eq:brn_est}. We find modest base-rate neglect, but minimal change in our main coefficients of interest. 

We also run a Grether-style regression in Appendix \Cref{tab:expt_results_nba}, again following equation~\eqref{eq:grether_reg} and now estimated separately for each quarter. As before, we find that participants overinfer from events in the first half, underinfer from events in the second half, and exhibit modest base-rate neglect overall. This modest overall base-rate neglect may be because the sequential setting makes the prior belief more salient than in some other contexts, leading participants to internalize their prior.  
But mimicking the results in Study 1b, the last column of that table shows that priors are appropriately weighted for weak signals (in early quarters), but that there is statistically significant base-rate neglect for stronger signals (in later quarters). 

Finally, we estimate $k$ and $\beta$ from equation~\eqref{eq:weight}, again using nonlinear least squares. The estimated value for $k$ is 0.40 (s.e.\ 0.02) and for $\beta$ is 0.41 (s.e.\ 0.02). The value of $\beta$ is statistically significantly less than one (p-value $<0.001$). Allowing for base-rate neglect in the model gives an estimate of 0.976 (s.e.\ 0.06) for the weight on prior, and leads to little change in the other estimates ($k = 0.42$ and $\beta = 0.44$).

\subsection{Discussion}

Across our three experiments, we find robust evidence that people overinfer from weak signals and underinfer from strong signals. Our findings hold in both abstract decision problems (Studies 1a and 1b) and naturalistic ones (Study 2), as well as with fixed symmetric priors (Study 1a), exogenously varied asymmetric priors (Study 1b), and endogenous priors based on previous belief-updating questions (Study 2). While prior-weighting biases like base-rate neglect can theoretically contaminate our predictions of overreaction and underreaction, in our data they have little impact on our main estimated effect. 

We find that these observed patterns of over- and underinference are consistent with people understanding the direction they should update their beliefs, but only imperfectly estimating the strength of the signals they receive. Our heterogeneity analyses provide suggestive evidence of this as well: greater answer precision, subjective confidence, task experience, and cognitive reflection are all correlated with greater sensitivity to signal strength and belief-updating patterns that are closer to Bayes' rule.

\section{Evidence from Finance and Sports Betting}
\label{sec:markets}

To build on our experimental evidence and test our theory in relevant observational settings, 
we now consider evidence from a set of sports betting markets and financial markets. Departing from the lab setting comes with multiple costs: (1) it is generally infeasible for us to estimate the true conditional probability of an outcome or true signal informativeness, as we no longer have knowledge of the true DGP (as we did in Studies~1a and 1b) nor the full information set available to participants over time (as we did in Study 2); and (2)~measuring subjective beliefs and perceived signal informativeness is also less straightforward. To overcome these issues, we apply new theoretical tools that allow us to proxy for signal informativeness and test updating behavior, given a set of beliefs data. We then choose a set of markets from which to measure price-implied beliefs: we consider the prices of different bets (with payouts of either \$0 or \$1) with known terminal dates. By considering price movements across informativeness regimes, we test whether the patterns of over- and underinference documented in the experiments apply in these real-world settings. We first describe our theoretical approach in more detail, before turning to our empirical data and results.

\subsection{Conceptual Framework and Approach}
\label{sec:mrsubsec}

Our conceptual framework for testing the behavior of beliefs builds closely on \textcite{AR21} (AR \citeyear{AR21}) and \textcite{AL-WP} (AL \citeyear{AL-WP}). Whereas \Cref{sec:theory} provided a model of over- and underinference from signals, our goal here is different: rather than a full alternative model of inference, we aim to characterize the Bayesian null in a way that allows for empirically implementable hypothesis tests. But while our starting point is someone who updates according to Bayes' rule, our tests are designed such that rejections are consistent with over- or underinference and therefore speak to the patterns predicted from \Cref{sec:theory}. We also build on that section's notation where appropriate, generalizing it to a dynamic setting with arbitrary signal structures.

Time is discrete, $t=0,1,2,\ldots,T$, and there is again a binary state $\theta \in \{0,1\}$. 
Each period, a person observes a signal $s_t$ from arbitrary distribution $p(s_t\,|\,\theta, H_{t-1}),$ where $H_t\equiv \{s_\tau\}_{\tau=1}^t$ is the history of signal realizations. The person's prior belief in state 1 is denoted by $\pi_0$, and their belief at time $t$ given the DGP (i.e., their prior and $p(\cdot)$) and history $H_t$ is $\pi_t(H_t)$, or $\pi_t$ for short. The \emph{belief stream} $\bm{\pi}$ refers to the collection of the person's beliefs over time.

While we cannot directly test for overinference vs.\ underinference without knowledge of the DGP, keeping track of the following two objects will allow for well-motivated indirect tests. First, define the \textit{movement} of a belief stream from period $t_{1}$ to $t_{2}>t_{1}$ as the sum of squared changes of beliefs over these periods: 
\begin{equation*}
m_{t_{1},t_{2}}(\bm{\pi} )\equiv \sum\nolimits_{\tau =t_{1}}^{t_{2}-1}(\pi
_{\tau +1}-\pi _{\tau })^{2}.
\end{equation*}%
Then, defining the \textit{uncertainty} of belief at
period $t$ as $u_{t}(\bm{\pi} )\equiv (1-\pi _{t})\pi _{t}$, we define \textit{uncertainty reduction} from
period $t_{1}$ to period $t_{2}>t_{1}$ as:
\begin{equation*}
r_{t_{1},t_{2}}(\bm{\pi} )\equiv \sum\nolimits_{\tau =t_{1}}^{t_{2}-1}(u_{\tau
}(\bm{\pi} )-u_{\tau +1}(\bm{\pi} ))=u_{t_{1}}(\bm{\pi} )-u_{t_{2}}(\bm{\pi} ).
\end{equation*}%
For each variable, we define the
concomitant random variable in capital letters (e.g., $M_{t_{1},t_{2}}$).

Our null model will be that the person fully understands the meaning of all signals and updates according to Bayes' rule. Under this null, beliefs satisfy $\pi_t(H_t) = \e_t[\theta] \equiv \mathbb{E}[\theta\,|\,H_t]$ for all $H_t$, where $\mathbb{E}$ is the expectation under the true (\emph{physical}) measure. 

\subsubsection*{The Equality of Movement and Uncertainty Reduction}

As in AR (\citeyear{AR21}), the martingale property of beliefs under the null implies that, \textit{regardless of the DGP},  expected Bayesian belief movement from any period $t_{1}$ to period $t_{2}$ must equal expected uncertainty reduction:

\vspace{2mm}

\begin{proposition}[Movement and Uncertainty Reduction]
\label{Prop_mr_substreams} Assume $\pi_t(H_t) = \mathbb{E}_t[\theta]$. For \textit{any} DGP and for any periods $t_{1}$ and $t_{2}$, 
$\mathbb{E}_{t_1}[M_{t_{1},t_{2}}]=\mathbb{E}_{t_1}[R_{t_{1},t_{2}}].$
\end{proposition}
This result formalizes the ``correct'' amount of belief volatility (or movement) under rationality, without the need to know the true unobservable DGP. (We provide a brief review of the proof in Appendix~\ref{app:proofmr}.) One can then follow AR (\citeyear{AR21}) to use this as the basis for a statistical test for Bayesian updating: given a set of belief streams, one can calculate  the difference between movement and uncertainty reduction (which they call ``excess movement'') and then apply a means test to see if the average difference is statistically different from zero. If so, one can reject --- with a certain confidence level --- that the beliefs arose from Bayesian updating. 

The result thus provides a testable link between belief movement, uncertainty reduction, and signal strength: when we observe a Bayesian person's beliefs moving, this must (on average) mean that she is receiving informative signals and reducing her uncertainty accordingly.\footnote{Formally, note from \eqref{bayesianUpdating} that for any $\pi_t$, belief movement $(\pi_{t+1}(\s(s_{t+1}))-\pi_t)^2$ is increasing in signal strength $\s(s_{t+1})$. So if we are in a regime with high signal strength ex ante, $\e_t[M_{t,t+1}]$ will be high, and by \Cref{Prop_mr_substreams}, so will $\e_t[R_{t,t+1}]$. We will verify that both of these increase with our informativeness proxy.} Crucially, this test (1) is valid regardless of the DGP, and (2) can be applied to arbitrary belief substreams (from period $t_1$ to $t_2$), as \Cref{Prop_mr_substreams} applies ex ante in all cases. Thus, given some ex ante known and observable sorting variable related to signal strength, we can test whether excess movement is related to signal strength. We will use \emph{time to resolution} ($T-t$) as our separating variable, and we discuss its relation to signal strength~--- and the relation of excess movement to over- and underinference --- below.

\subsubsection*{Excess Movement and Over- and Underinference}

We now consider what kinds of non-Bayesian behavior generate different violations of the equality in \Cref{Prop_mr_substreams}. Most importantly for us, there is a natural positive connection between excess movement and overinference: people who overinfer are intuitively changing their beliefs ``too much'' relative to the informativeness of signals on average, generating $\mathbb{E}_{t_1}[M_{t_{1},t_{2}}-R_{t_1,t_2}]>0$. The opposite is true in the case of underinference.  

AR (\citeyear{AR21}) formalize this connection. First, in a two-period environment, a person with a correct prior who overinfers from signals will exhibit a positive excess movement statistic, while a person who underinfers will exhibit a negative statistic.
Second, they show that the same relationship holds over many periods in a symmetric binary-signal environment, despite the complication that the person's prior may not be correct in later periods.\footnote{Specifically, the paper considers a specification of over- or underinference equivalent to eq.~\eqref{eq:grether_reg}, in which $\text{logit}(\hat{\pi}_{t+1})=\text{logit}(\hat{\pi}_{t})\pm\gamma\s(s_{t+1})$. Their Proposition 6 states that a person with $\hat\pi_t=\pi_t$ and $\gamma>1$ will have $\e[M_{t,t+1}]>\e[R_{t,t+1}]$ (and the opposite if $\gamma<1)$. Proposition 7 states that, given a DGP with a constant signal strength and $\pi_0=1/2$, a person with $\hat\pi_0=1/2$ and $\gamma>1$ will have $\e[M_{t_1,t_2}]>\e[R_{t_1,t_2}]$ given any history $H_{t_0}$ (and the opposite if $\gamma<1$). One quarter of a basketball game very roughly approximates such a binary symmetric environment, to take an example (see footnote~\ref{fn:strength}).} We suspect that the same relationship between inference and excess movement applies quite generally, but it is difficult to characterize other DGPs analytically. We therefore turn to simulations to verify that the same intuitive relationships hold under our updating model in environments that more closely map to our empirical setting. We also use our simulations for further verification that our time-based measure of signal strength is a good proxy in this setting. 

As a caveat, we note that while overinference (underinference) generates positive (negative) excess movement both analytically and in simulations, there could be other drivers of excess movement that we cannot rule out in observational beliefs data: we only observe overall reactions.\footnote{Base-rate neglect, for example, tends to generate positive excess movement (AR \citeyear{AR21}). Another bias, probability weighting, effectively matches the results from constant underinference. In fact, given a prior of 50\%, the classic symmetric functional form for probability weighting from \textcite{gonzalez1999shape} is exactly equivalent to a person who constantly underinfers from all signals.} But given that the patterns we observe in the data end up aligning closely with the predictions of our model, we view the data as providing supporting evidence alongside our experimental results (in which we can isolate inference behavior more directly).

\subsubsection*{Simulated Belief Streams} 

We now consider patterns in movement and uncertainty reduction for a person who updates according to our model in \Cref{sec:theory} --- as well as a person who exhibits constant over- or underinference --- when forming beliefs about the outcome of a sporting event or the future level of the stock market in simulated data. Empirically, these settings feature similar random-walk-like DGPs with signals (points scored, daily returns) received in each period, with the aggregate of that information determining the final state. To transparently model such situations in our simulated economy, we consider a simple random-walk-like DGP in which there are two ``teams'' representing the two states, exactly one team scores in each of $T$ periods, each team has equal probability of scoring in each period, and the final state is which team has the highest score after the final period. For example, if a team is leading by one score with two periods left, they have a 75\% chance of being the final winner because they win if they score in one of the final two periods.

\begin{figure}[t!]
\caption{Simulated Movement and Uncertainty Reduction Over Time For Different Models}
\label{fig:simulations}
\begin{center}
\vspace{-8.5mm}
    \centerline{\includegraphics[width=.995\textwidth]{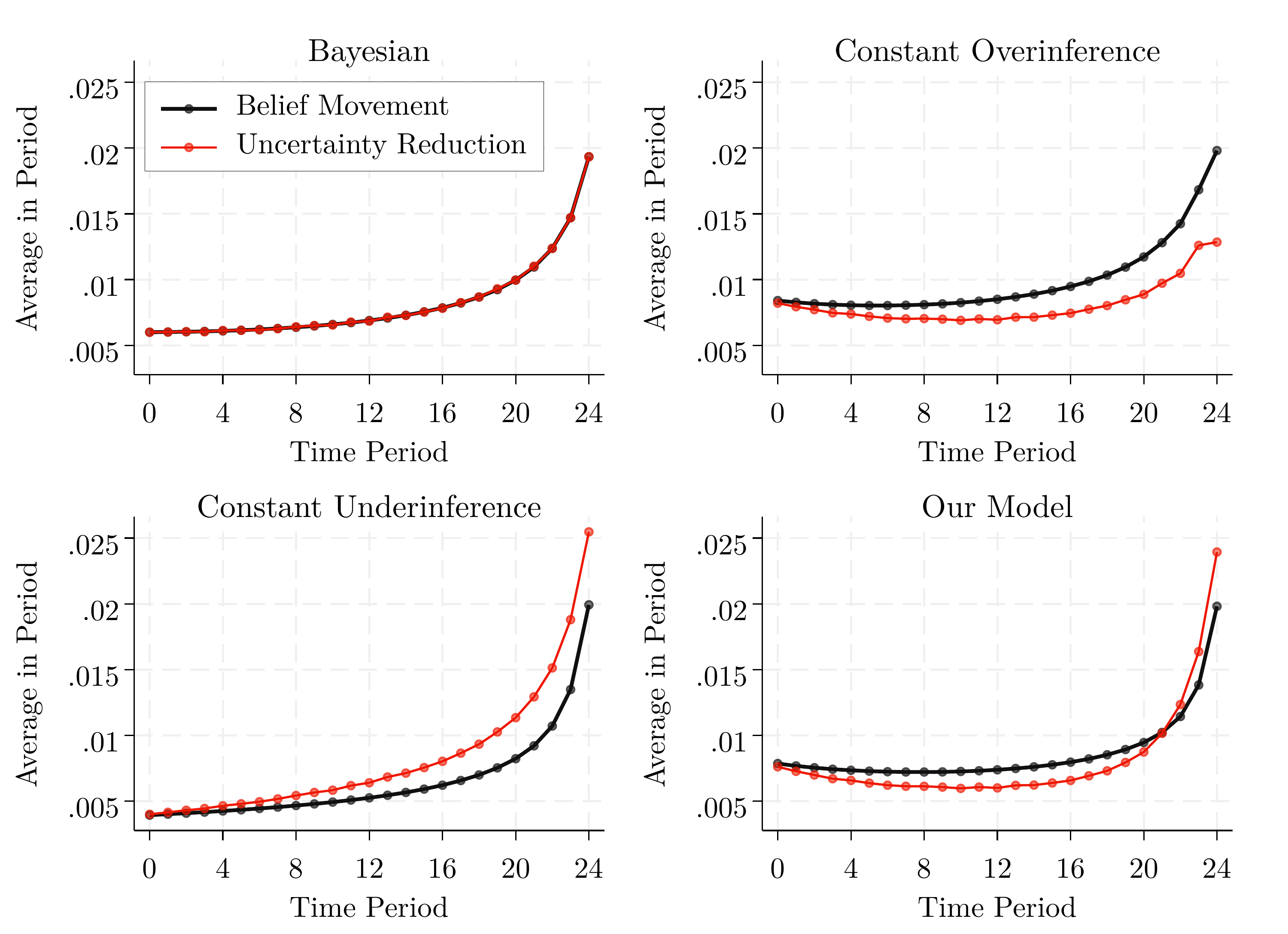}}
    %height=.6\textheight]{combo_mrAll4Sim24.png}}
\end{center}
\begin{threeparttable}
\begin{tablenotes}
\begin{scriptsize}
\vspace{-13mm}
\item \textbf{Notes:} This figure shows the average belief movement (thicker black line) and uncertainty reduction (thinner red line) statistics over time for four different models, averaged over 1~million simulations of the game-like DGP discussed in the text with $T=27$. We drop the first and last period, as they always have zero excess movement given this DGP, and plot the remaining 24 periods. The updating models are (1) Bayesian updating (correct perception of signal strength $\mathbb{S}$), (2) underinference (with perceived signal strength $0.8 \cdot \mathbb{S}$), (3) overinference (with perceived signal strength $1.2 \cdot \mathbb{S}$), and (4) our model (perceived signal strength $k \cdot \mathbb{S}^\beta$ with $k = 0.88$ and $\beta = 0.76$). For Bayesian updating, these statistics are always equal. For underinference, movement is always less than uncertainty reduction, and the opposite is true for overinference. For our model, movement is greater than uncertainty reduction in early time periods (when signals are generally weak) and lower in later time periods close to resolution (when signals are generally strong). \par
\end{scriptsize}
\end{tablenotes}
\end{threeparttable}
\end{figure}

We conduct one million simulations of this DGP, and we present average results by time period in \Cref{fig:simulations}. The top-left panel shows the expected movement and uncertainty reduction statistics over time for a Bayesian. First, following Proposition \ref{Prop_mr_substreams}, the statistics must be equal at each period. Next, note that both statistics are rising as the resolution of the game approaches. Initial periods always contain very little information, while the later periods sometimes convey no information (because one team has an insurmountable lead) and sometimes convey strong information (because the scores are close). Overall, though, signal strength rises over time, and average movement and uncertainty reduction increase accordingly. This is intuitive theoretically.\footnote{For example, for option prices, the Black--Scholes model predicts that the sensitivity of an option price to the same change in the underlying price (i.e., option \emph{delta}) decreases exponentially with time to maturity, and the same applies (in fact more strongly) for the option spreads used to construct option-implied beliefs. That is, the same underlying price change rationally generates a bigger change in beliefs about the option payoff closer to maturity. Our simulated random walk is in fact a discrete-time approximation of a Black-Scholes economy, but the same logic will hold in practically any option-pricing model beyond this one.} And as we show shortly, it also applies in all of our empirical settings; importantly, we consider settings where all uncertainty will be resolved by some fixed end period, and as a result strength increases closer to that expiration.

What do the statistics look like for people who over- or underinfer from signals? Following intuition and the theoretical results from simpler DGPs, overinference (top-right panel) leads to positive excess movement in every period, while the opposite is true for underinference (bottom-left panel). The bottom-right panel displays the results for our model, with parameters estimated from our Study~1a. In the early periods, average signal strength is low, leading to overinference, which in turn generates excess movement. In later periods, the amount of information revealed is higher, leading to underinference. Belief movement increases, but not in line with the increase in uncertainty reduction. There is therefore a switching period at which average movement crosses below uncertainty reduction. This switching is in effect the signature pattern for our model, as it does not occur under Bayesian updating or when there is universal overinference or underinference. We now proceed to test whether the same patterns hold empirically.

\subsection{Sports Betting Data}

\subsubsection*{Data Description}

We start with data on sports betting. Our data comes from Betfair, which operates a large prediction market in which individuals are matched on an exchange to make opposing financial bets about the outcome of a sporting event. We observe time-stamped transaction prices for a contract in which one party pays another party a set amount given a particular realized outcome of the game (e.g., Team A beats Team B). Prices are quoted as fractional odds; for example, a transaction for the Team~A contract might occur at 3/1 odds, meaning the person buying one unit of it will receive \$4 if Team A wins and lose \$1 if Team~A loses. These odds can then be normalized to obtain an implied probability (in this case, 1/4). As in a standard centralized exchange, contract prices (and implied beliefs) change with supply and demand.

These are the same data as used in AR (\citeyear{AR21}), and we use the same sample (2006--2014) and similar data filters as in that paper. In particular, we focus on markets for five major sports (soccer, basketball, baseball, ice hockey, and American football), and we consider only contracts over the final winner of the game (omitting more exotic contracts, such as which team will be winning at the midpoint or number of goals scored). If there are multiple contracts (e.g., one paying off if Team A wins, another if Team B wins), we use the contract for which the starting beliefs are closest to 0.5. We use observations only when the game is being played. To remove high-frequency noise, we follow AR (\citeyear{AR21}) and keep only the first transaction in a given minute increment. We also drop trades with less than 1\% of the overall average transacted amount. Finally, we attempt to have similar timing in events by dropping (less common) events in a category for which the timing of the game is different (such as WNBA games, in which the game time is shorter than the NBA). We are left with over 5~million transaction prices from about 260,000 sporting events over the sample.

Given our focus in this section on equilibrium bet-price data, we follow the literature that interprets these prices as ``market beliefs.''\footnote{The interpretation of market prices as averages of individual beliefs has been studied in a range of work. In standard Bayesian settings with complete markets, this interpretation is straightforward (see AL \citeyear{AL-WP}). With heterogeneity, \textcite{G-WP} and \textcite{WZ-WP} show the interpretation is valid when traders have log utility and trade statically (cf.\ \cite{M06}). But with speculative trading, prices often react more to new information than individual beliefs (\cite{MP22}).} A test based on Proposition~\ref{Prop_mr_substreams} can thus be viewed as a test of the joint null that market prices may be interpreted as beliefs and that these beliefs are Bayesian. But while this might affect the interpretation of full-sample excess movement tests, it poses less of a problem for our purposes. We are {fixing} the environment (i.e., the particular betting market in question) and comparing excess movement as one varies the signal strength (proxied by time to maturity) within this environment. If we assume that the mapping from individual to market beliefs does not change systematically within a stream as one moves closer to maturity, our findings are at minimum directionally informative about both individual and market-level reactions to information across signal-strength regimes.

\subsubsection*{Graphs of Movement and Uncertainty Reduction}

\begin{figure}[t!]
\caption{Movement and Uncertainty Reduction Over Time for Sports Betting Data}
\label{fig:sports}
\begin{center}
\vspace{-8.5mm}
    \centerline{\includegraphics[width=.9975\textwidth]{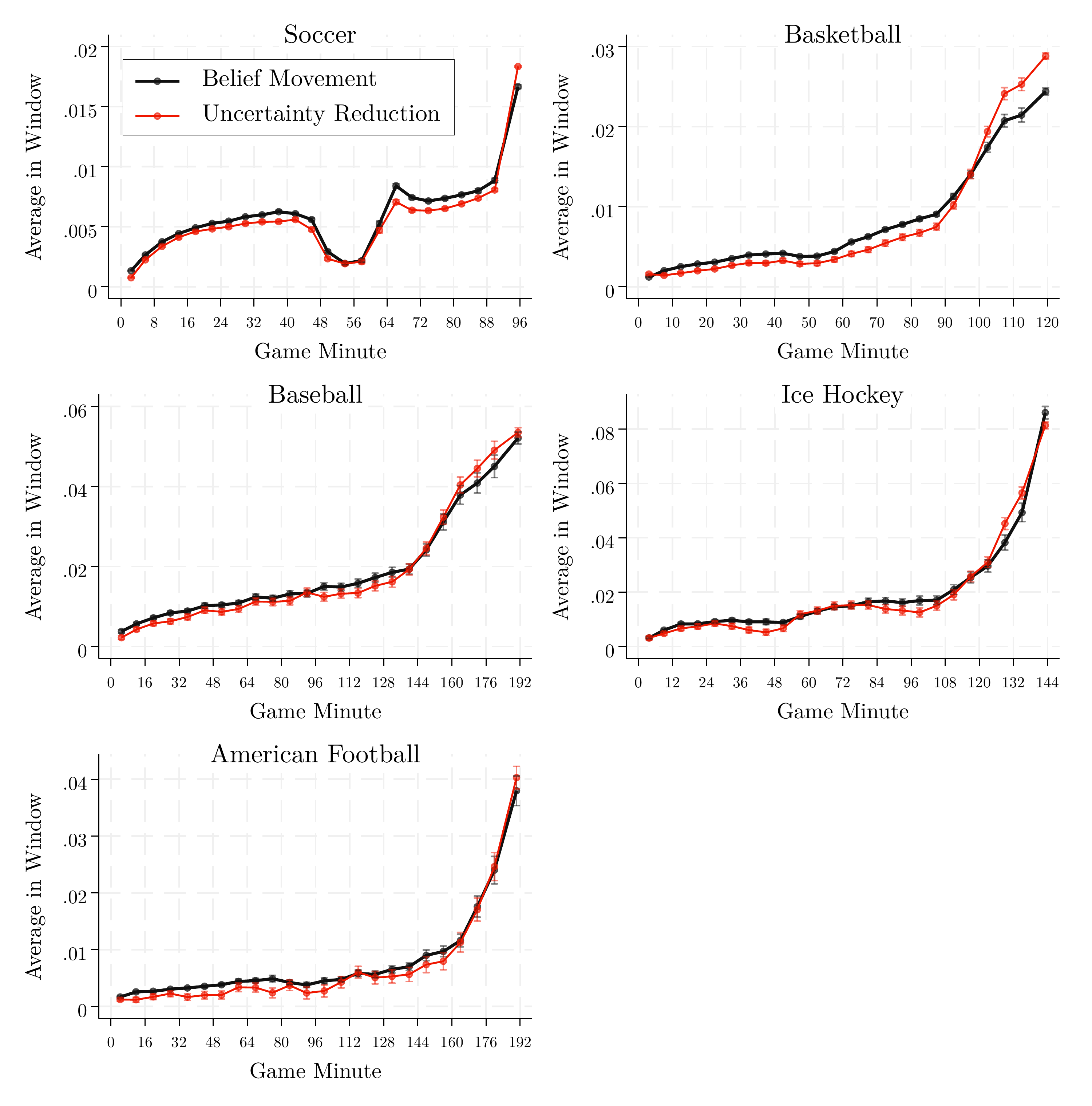}}
\end{center}
\begin{threeparttable}
\begin{tablenotes}
\begin{scriptsize}
\vspace{-15.5mm}
\item \textbf{Notes:} This figure shows average belief movement (thicker black line) and uncertainty reduction (thinner red line) statistics over time for the beliefs implied by betting prices for five different sports (with 95\% confidence intervals). Minute~0 is the beginning of the game, and the last minute is the end of the game. Each estimated point is the summed movement or uncertainty reduction within one of 24 equal-length time windows, averaged over all the games in the sample. In each case, movement is greater than uncertainty reduction in early time periods (when signals are generally weak), and it is lower than uncertainty reduction close to the end of the game (when signals are generally strong), as predicted by the model.\par
\end{scriptsize}
\end{tablenotes}
\end{threeparttable}
\end{figure}

\Cref{fig:sports} shows average movement and uncertainty reduction (as well as confidence intervals) across time for each sport. Observations occur in continuous time and therefore must be aggregated in some way. Our data contain observations in clock time (``1:31pm'') rather than game time (``4:50 through the third quarter''); we therefore consider average movement and uncertainty reduction for observations within 24 time windows, each of which corresponds to $1/24$ the length of an average game.\footnote{For example, as the average basketball game lasts around 132 minutes, basketball games are broken into 24 chunks of 5.5 minutes. The final chunk then includes all observations that occur after 132 minutes. Results are similar if we use different numbers of chunks. Separately, in constructing confidence intervals for this figure (but not for the regressions), we assume observations are uncorrelated across contracts.} 
As in the simulations, average movement and uncertainty reduction are generally increasing over time (with the exception of mid-period breaks). As discussed in \Cref{sec:nba-experiment}, signals in basketball games increase in strength strongly over time; the increase in both movement and uncertainty reduction over time shows  that the same pattern applies for all sports.\footnote{This follows unless markets completely misunderstand directional changes in signal strength (thinking stronger signals are weaker), seemingly counter to all available evidence (e.g., \cite{CR13}).} 

The relative patterns of the two series, though, follow the predictions of our model of over- and underinference. Early in games for each sport, movement is greater than uncertainty reduction, and for each sport there is a time at which movement drops below uncertainty reduction. For four of the five sports, movement then continues to be lower than uncertainty reduction after this time (for hockey, movement stays lower than uncertainty until the final period). The market accordingly appears to overreact to the less-informative signals at the beginning of a game, and underreact to the more-informative signals at the end of a game. Interestingly, for basketball (in the top-right panel), excess movement switches from positive to negative around the end of the third quarter, precisely mirroring the switch from over- to underinference observed in our experimental basketball setting in Study~2 (see \Cref{fig:main_effects_nba}).

\subsubsection*{Statistical Tests}

Are the patterns in the figures statistically meaningful? To answer this question, we require a test to determine if there is overreaction (captured by expected movement being greater than uncertainty reduction) when signals are weak (captured by low uncertainty reduction), and underreaction when signals are strong. But we cannot simply sort observations by realized uncertainty reduction (or some ex post proxy for signal strength) and then test how excess movement changes across this sort. Instead, \Cref{Prop_mr_substreams} tells us that we must test whether \emph{expected} movement $\e_t[M_{t,t+1}]$ equals \emph{expected} uncertainty reduction $\e_t[M_{t,t+1}]$ ex ante. We must therefore consider an ex ante sort variable, and analyze the relationship between \emph{average} movement and uncertainty reduction across settings with different strength. 

As we have seen, time to resolution is a strong such ex ante variable separating low (early) from high (late) signal-strength periods. We therefore, for each sport, regress average movement in each time window on average uncertainty reduction in the same time window. Under the null of Bayesian updating, the constant will be equal to 0 and the slope coefficient equal to 1, as average movement should be equal to average uncertainty reduction in every period. However, for a person who updates according to our model, average movement will be higher than average uncertainty reduction when reduction is low, but lower than uncertainty reduction when reduction is high, such that the constant will be positive and the slope coefficient will be less than one.

The results for these regressions are shown in the first five columns of \Cref{tab:mrRegResults}. Given that the regressions use calculated averages (collapsing to 24 observations per regression), we bootstrap standard errors by resampling events (games) with replacement and recalculating averages and regression coefficients 10,000 times  (though OLS standard errors are very similar). For each sport, the constant and slope coefficients are highly statistically significantly different from the Bayesian benchmark in the direction predicted by the theory: we consistently observe evidence consistent with overinference from weak signals (when average uncertainty reduction is low) and underinference from strong signals (when reduction is high). 

To understand the magnitude of the estimates, note that beliefs moving 3 percentage points up and then 3 points down would produce movement of 0.0018 (close to the average constant coefficient) and no uncertainty reduction. Given this average constant, the average slope coefficient then implies that movement will cross uncertainty reduction when both are around 0.014, which occurs before the end of the game for all the sports in \Cref{fig:sports}. 

%\begin{center}
\begin{table*}[t!]
\centering
\begin{threeparttable}%[t!]
\onehalfspacing
\begin{small}
\caption{Regressions of Average Movement on Average Uncertainty Reduction\\[-3pt]}
\begin{tabular}{l*{8}{c}}
\toprule
  \hspace{-2pt}\raisebox{-7pt}[\height][\depth]{Dep. Var.:}   &\multicolumn{5}{c}{Sports}  &\multicolumn{2}{c}{Finance} \\ \cmidrule(lr){2-6} \cmidrule(lr){7-8}
{Movement} &\multicolumn{1}{c}{Soccer}&\multicolumn{1}{c}{Basketball}&\multicolumn{1}{c}{Baseball}&\multicolumn{1}{c}{Hockey}&\multicolumn{1}{c}{Football} &\multicolumn{1}{c}{Raw} &\multicolumn{1}{c}{Risk-Adj.}\\
\midrule
Constant              &          0.0009&      0.0018&       0.0026&       0.0018&       0.0015&       0.0065&       0.0060\\
     \seline                 &        (0.0001)&    (0.0001)&     (0.0002)&     (0.0002)&     (0.0002)&     (0.0003)&     (0.0003)\\
     
Uncert.\ Red.\       &           0.918&       0.806&        0.889&        0.945&        0.912&        0.680&        0.733\\
     \seline                 &         (0.005)&     (0.008)&      (0.013)&      (0.013)&      (0.027)&      (0.040)&      (0.041)\\
\midrule
%$\sqrt{\text{Constant}}$&          0.037&        0.043&        0.049&       0.036&        0.039&        0.067&       0.068\\
$R^2$                   &          0.977&        0.985&        0.995&       0.976&        0.995&        0.944&       0.941\\
Time Chunks      &             24&           24&           24&          24&           24&           24&          24\\
Events                  &          175,026&        48,430&        16,536&       19,445&        3,212&          955&         955\\
Belief Obs.\            &      4,589,289&      867,567&      166,346&     109,751&       86,193&       58,864&      58,864\\
$p$-val: $\!\text{Const} =0$     &         <0.001&       <0.001&       <0.001&      <0.001&        <0.001&       <0.001&      <0.001\\
$p$-val: $\!\text{Slope} =1$     &          <0.001&      <0.001&       <0.001&      <0.001&        0.007&       <0.001&      <0.001\\
\bottomrule
\end{tabular}

\label{tab:mrRegResults}
\end{small}
\begin{tablenotes}[para,flushleft]
\singlespacing
\vspace{-3.85mm}
\begin{scriptsize}
\textbf{Notes:} This table presents the results from OLS regressions of average movement in each of the 24 time periods on average uncertainty reduction in that time period, for five sports and the options data (both raw and risk-adjusted, as described in the text). Bootstrapped standard errors, in parentheses, are calculated by resampling events with replacement and recalculating averages and regressions 10,000 times. Bayesian updating predicts constant $=0$ and slope $=1$, while our theory predicts constant $> 0$ (corresponding to overinference for very weak signals) and slope $<1$ (relatively more underinference for stronger signals). \par
\end{scriptsize}
%\vspace{5mm}
\end{tablenotes}
\end{threeparttable}
\end{table*}
%\end{center}

A potential concern when testing for a one-to-one slope is measurement error in the regressor and resulting attenuation bias. This would be a meaningful concern if, instead of following \Cref{Prop_mr_substreams}, we regressed period-by-period realized $m_{t,t+1}$ on $r_{t,t+1}$.\footnote{For a Bayesian, $\mathbb{E}_t[M_{t,t+1}]=\mathbb{E}_t[R_{t,t+1}]$, but $r_{t,t+1}$ is equal to that expectation plus a mean-zero error.} But because we take averages over thousands of belief changes at a given time horizon, we are able to estimate expected uncertainty reduction at that time plus a tiny error term. In our case, the estimated variance of the error term at each period is more than 100,000 times smaller than the estimated variance of the regressor, so any resulting attenuation bias is negligible.\footnote{By averaging the movement and uncertainty reduction statistics over time chunks, we do face the subjective question of how many chunks to use. We show in Appendix Tables~\ref{tab:mrRegResults12}--\ref{tab:mrRegResults36} (with accompanying figures) that estimated slopes change slightly (differently across sports) when using 12 or 36 chunks, but $p$-values remain highly significant in all cases, with the exceptions of hockey with 12 chunks and football with~36.} Note that the $R^2$ values in all cases are very close to 1:  average movement and uncertainty reduction move very closely together, but with a muted slope.

\subsection{Index Options Data}\label{sec:findata}

\subsubsection*{Data Description}

The sports betting data provide a useful lab for studying beliefs in an incentivized setting similar to the one in Study~1b of our experiment. We now consider whether similar patterns apply to a large-scale financial market, where beliefs are expressed over outcomes of first-order macro importance. In particular, we consider options on the S\&P 500 index, which are effectively bets on the value of the market index as of a fixed future expiration date.\footnote{An option contract specifies an expiration date $T$ and strike price $K$, which together with the realized value of the S\&P ($V_T$) determine the payoff to the buyer of the contract. If $V_T>K$, then the holder of a call option receives $\$V_T-K$; otherwise they receive \$0. They pay $c_t$ for the option upfront, and the seller receives the negative of the buyer's payoff. (For a put option, the holder instead receives $\max(K-V_T,0)$.)} We use the OptionMetrics database to obtain option price quotes for S\&P index options traded on the Chicago Board Options Exchange (CBOE), which is the largest U.S.\ exchange. We observe the best posted bid and ask quotes at the end of every day for each strike price and expiration date, and we take the average of these two and use this as our end-of-day price.

These are the same data as used in AL (\citeyear{AL-WP}), and we use the same sample (1996--2018) and similar filters as in that paper. As our list of filters is somewhat long, we relegate details to \Cref{app:details}. After filters, we are left with over 4~million option prices corresponding to about 955 option expiration dates (the analogue of a single event in the sports betting data) and 5,500 trading dates. To ensure that prices are liquid, AL (\citeyear{AL-WP}) consider options with expiration date at most one year away from the trading date. For our purposes, we cut off the analysis at 100 trading days from expiration (in calendar time, roughly 4.5 months).\footnote{This somewhat arbitrary choice is largely so that our movement and uncertainty reduction figures are easy readable, and results continue to hold when using longer-horizon options.}

\subsubsection*{Converting Option Prices to Market-Implied Beliefs}

On any given trading date $t$, there are prices for a range of S\&P options with the same expiration date $T$. They differ only in their strike prices $K$ (for a call option, the minimum S\&P index value at which the option will obtain a positive payoff at expiration). Using minimal assumptions (following \cite{BL78}), the set of option prices for such a $(t,T)$ pair can be translated into a market-implied (or \emph{risk-neutral}) probability distribution over the future S\&P price on the option expiration date. Intuitively, by buying a set of options, one can construct a strategy that pays off \$1 if, say, the S\&P is between 5,500 and 6,000 on September 30, and \$0 otherwise. The market price of constructing such a binary bet can be read as an option-implied probability that the S\&P will indeed be in this range.

Unlike in the case of sports betting data, index options have payoffs that are tied (by construction) to the value of aggregate wealth. Option prices therefore reflect risk aversion in addition to subjective probability assessments about the future index value.
This is the main complication in using option-implied probability distributions: they do not, in general, correspond to any notion of aggregate subjective beliefs. (They are equivalent to subjective beliefs only in the case of risk neutrality over the index value, which motivates referring to them as risk-neutral beliefs.) For example, suppose that there are two possible date-$T$ macroeconomic states that are perceived by investors as equally likely. If investors value a marginal dollar in the ``bad'' state (when the market is low) more than in the ``good'' state (when the market is high), they will be willing to pay more for the option that pays off in that state. If these risk preferences are not taken into account, one will (falsely) conclude that investors believe that the bad state is more likely.

Addressing this issue is the main theoretical task taken up in AL (\citeyear{AL-WP}). That paper shows that under certain assumptions, one can place a \emph{bound} on excess movement in risk-neutral (RN) beliefs under the null that underlying subjective beliefs are rational. The bound is tight in the space of possible DGPs --- that is, one can construct a DGP under which it holds exactly --- but it is not necessarily tight under the true real-world DGP.\footnote{While the bound is sufficient for the full-stream tests considered in that paper, it might not be here: we wish to understand how ``true'' excess movement evolves with signal informativeness within a stream.}  
We therefore provide two sets of results in the current analysis, (1) using the raw (unadjusted) RN beliefs, and alternatively (2)~translating these beliefs to a set of physical (subjective, risk-adjusted) beliefs under an assumption on risk aversion. For (2), we consider many possible assumptions in translating from risk-neutral to physical beliefs, detailed in Appendix~\ref{app:details}. While the dozens of possible assumptions and parameterizations affect the physical belief estimated for a given risk-neutral belief, it turns out their effect on our movement and uncertainty-reduction statistics is so small as to be nearly indetectable.\footnote{The brief intuition is that risk aversion is unlikely to be changing meaningfully from day to day. But the main point of interest for this analysis is that the basic patterns found in the experimental data and in the sports betting data are also observed in the finance data, regardless of the RN beliefs correction used.} We thus report results here under our main translation, which assumes a representative investor with power utility over the terminal index value. 
For robustness, we present estimates under a wide range of alternative parameterizations in Appendix~\Cref{fig:financealt}, finding that these choices have little effect on our results.

To implement our measurement of RN beliefs empirically, we need a set of discrete possible outcomes as of date $T$. In particular, we must partition the set of possible date-$T$ index values into discrete ranges (in the example above, the single range considered was from 5,500 to 6,000). To maintain the same set of possible outcomes across different expiration dates, we set these states to correspond to ranges for the log excess return on the S\&P 500 from the first observable option trading date to the expiration date. In particular, we define 10 potential return outcomes~$\theta_j$, each of which (aside from the two tails) corresponds to a five-percentage-point range for the S\&P's log return in excess of the risk-free rate: state $\theta_1$ is realized if the S\&P's log excess return is below -0.2 (i.e., roughly -20\%) from date 0 to date~$T$; state $\theta_2$ is realized if the log excess return is in the range $(-0.2,-0.15]$ (between -20\% and -15\%); $\theta_3$ if $(-0.15,-0.10]$; and so on, up to $\theta_{9}=(0.15,0.2]$ and $\theta_{10}$ above 0.2.

We then use options to measure RN beliefs $\pi_{t,j}^*$ for each state $j$ over trading days $t=0,1,\ldots,99.$\footnote{Full details on how we construct the risk-neutral belief distribution are again provided in \Cref{app:details}. Given $\pi_{t,j}^*$, we then also calculate the corresponding risk-adjusted physical belief $\pi_{t,j}$ using the power-utility risk adjustment described above, and all the calculations for movement and uncertainty reduction are then duplicated for these adjusted beliefs.} For example, $\pi_{t,3}^*$ is the option-implied belief, as of $t$, that the S\&P's excess return from 0 to $T$ will end up being between -15\% and -10\%. At $T=100$,  we assign probability 1 to the actual realized return state. %, which is observed at expiration on that day.
Note that unlike for the sports betting data, we no longer have only two possible states. Instead, we are using the full \emph{histogram} of beliefs over 10 possible return outcomes. This departs from AL (\citeyear{AL-WP}), where the histogram is converted into a set of binarized conditional beliefs. We keep the full histogram here in order to minimize the potential effects of noisy prices, which AL (\citeyear{AL-WP}) show can induce meaningful measurement error in the binarized statistics.\footnote{They also provide and estimate a correction for this error on the binarized statistics (which are used in that paper given their theoretical setting). We show in Appendix~\Cref{fig:financeBinary} that using their noise-corrected, binarized risk-neutral beliefs, we observe very similar patterns in movement and uncertainty reduction as with our histogram data. The Appendix also includes figures and tables showing that results are unchanged with different numbers of time windows, as well as in subsamples of the option data (post-2000 and post-2010).}

So for each trading day and expiration date, we calculate the belief movement and uncertainty reduction statistics for each state's risk-neutral belief ($m_{t,t+1,j,T}$ and $r_{t,t+1,j,T}$), and we then average the resulting statistics across all 10 states ($\overline m_{t,t+1,T}=\frac{1}{10}\sum_{j=1}^{10} m_{t,t+1,T}$, and similarly for $\overline r_{t,t+1,T}$). Given that each belief has an interpretation as a binary belief over whether the given state will be realized or not, \Cref{Prop_mr_substreams} still applies to these aggregated statistics (see AR \citeyear{AR21}, Proposition~3). Finally, as before, after calculating these values, we then break the data into 24 equal-length time windows sorted by trading days to expiration. Within each such chunk (for trading days $t_1$ through $t_2$), we calculate average overall movement and uncertainty reduction over all events $T$ (e.g., $\frac{1}{955}\sum_{T=1}^{955} \overline m_{t_1,t_2,T}$) as our empirical measures of $\e[M_{t_1,t_2}]$ and $\e[R_{t_1,t_2}]$.

\subsubsection*{Graphs of Movement and Uncertainty Reduction}

\begin{figure}[tb!]
\caption{Movement and Uncertainty Reduction Over Time for Finance Data}
\label{fig:finance}
\begin{center}
\vspace{-7.5mm}
    \centerline{\includegraphics[width=.995\textwidth]{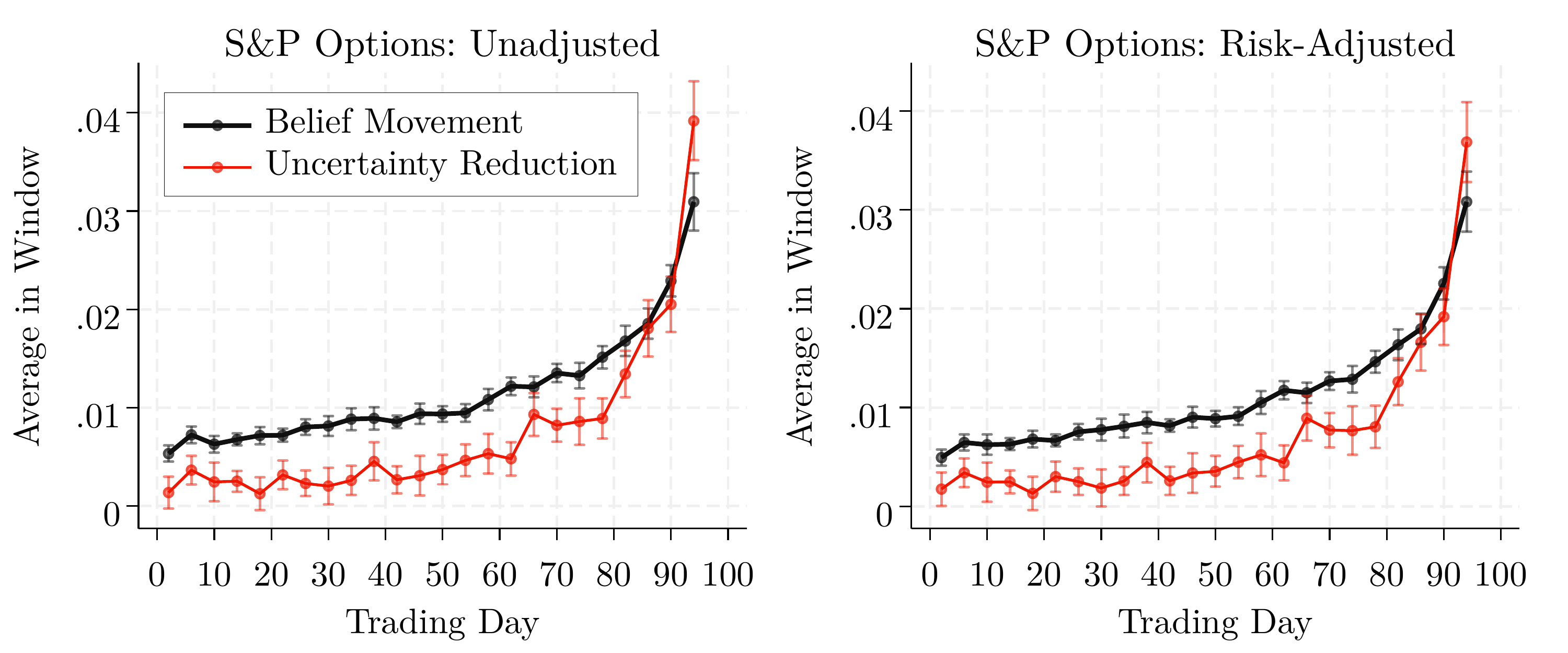}}
    %{combo_mrFinanceWithAll.png}}
\end{center}
\begin{threeparttable}
\begin{tablenotes}
\begin{scriptsize}
\vspace{-12.75mm}
\item \textbf{Notes:} This figure shows the average belief movement (thicker black line) and uncertainty reduction (thinner red line) statistics over time for the beliefs implied by option data (with 95\% confidence intervals). Trading day~0 corresponds to 100 days from expiration, and the last trading day is the expiration date. Each estimated point is the summed movement or uncertainty reduction within a 4-trading-day window, averaged over all option-expiration dates in the sample (1996--2018). The left panel uses the unadjusted (risk-neutral) beliefs implied by options. The right panel uses a risk adjustment described in the text. Movement is greater than uncertainty reduction far from expiration (when signals are generally weak), and it is lower than uncertainty reduction close to expiration (when signals are generally strong), as predicted by the model. \par
\end{scriptsize}
\end{tablenotes}
\end{threeparttable}
\end{figure}

\Cref{fig:finance} shows average movement and uncertainty reduction over time in the options data, analogous to \Cref{fig:sports}. Date 0 is again 100 trading days from expiration, while date 100 is expiration. The left panel shows the average movement and uncertainty reduction statistics for the raw risk-neutral beliefs, and the right panel shows the statistics for physical beliefs obtained under the main risk adjustment procedure. 
In both cases, movement is consistently above uncertainty reduction relatively far from expiration, when signals are only very weakly informative and uncertainty reduction is statistically indistinguishable from zero. Uncertainty reduction increases dramatically closer to expiration (when market movements are more informative regarding the true index value at the expiration date). And while option-implied belief movement increases alongside uncertainty reduction, it appears to do so less than one for one, with uncertainty reduction crossing above movement roughly~10 days from expiration.

The patterns observed in this high-stakes financial market are similar to those in the sports betting data for many sports plotted in \Cref{fig:sports}. They are also similar to the simulated results from our theoretical framework plotted in the bottom-right panel of \Cref{fig:simulations}. Recall that these simulations are parameterized using the estimates from our experimental data, so the theory accordingly helps unify the evidence obtained in both the lab and real-world data.

\subsubsection*{Statistical Tests}

We conclude this analysis by conducting the same formal tests as in the previous case, regressing average movement on average uncertainty reduction in each time window. The results are shown in the final two columns of \Cref{tab:mrRegResults}. For both the raw and risk-adjusted data, the estimated slope and constant are again highly statistically significantly different from the Bayesian benchmark in the direction predicted by our theory. The positive constant again indicates overreaction when signal informativeness (uncertainty reduction) is low, as movement is significantly positive in these cases; meanwhile, the slope being less than one (and numerically nearly identical to the estimated slope in the sports betting data) indicates underreaction for high enough levels of signal informativeness. The market therefore appears to over- and underreact in the way predicted for individuals modeled in \Cref{sec:theory}. More broadly, the consistent results from the lab and from observational data indicate a key determinant for this updating behavior that applies across settings.

\section{Discussion and Conclusion}
\label{sec:discussion}

We provide evidence that people overinfer from weak signals and underinfer from strong signals. We demonstrate this phenomenon using three tightly controlled experiments and using a new empirical method applied to betting and financial markets. In each setting, beliefs appear to move in the correct direction and shift more when signals are stronger. But perceptions of signal strengths appear consistently anchored toward some intermediate level; in other words, people act as if they are {partially} insensitive to the objective signal strength, leading on average to overinference from weak signals, underinference from strong signals, and corresponding over- and underreaction in beliefs. This partial insensitivity to signal strength is well captured by a model in which a person understands the directional meaning of a signal but is less certain about the strength of the information. These findings help unify seemingly contradictory results in past literature on inference behavior. Naturally, we view this as one of many possible reasons why people may react to information in a non-Bayesian manner, and we see several potential paths for future research.

Although we find similar main results in our abstract experiments and our more naturalistic experiment, it is worth understanding better whether participants facing math-test-like experimental environments use different decision-making processes and heuristics than they do in naturalistic environments. Using abstract environments has huge benefits --- better control and mapping to abstract models --- but may come at the cost of only observing a very particular class of behaviors. For example, our theory and results suggest that estimates of base-rate neglect might depend on whether the experiment uses abstract ``endowed'' priors (telling people that their prior should be 80\%) or uses a naturalistic ``internalized'' prior (a person genuinely believing 80\% given previous experience and updating). Relatedly, our work suggests that some standard results (like underinference, in our case) may be limited to classic parameters (like strong signals) used in experiments (an insight also seen in work like \cite{mcgranaghan2024distinguishing} and \cite{blavatskyy2023common}). Using past parameters and experimental designs that ``work'' has large benefits --- the ability to manipulate effects and benchmark results --- but may again limit external validity.

Second, these results suggest future directions to study the demand for news in the real world. There has been a shift in news provision and consumption away from traditional news outlets and toward other platforms (\cite{LG22}), despite concerns about platforms' low-quality news and misinformation (\cite{AG17}). One potential explanation is that people respond to news in general, but are insufficiently sensitive to the quality of the information source. In our abstract environment in Study 1a, we in fact find some suggestive evidence for this: we ask people to decide how many signals to purchase (related to \cite{AL18}), and find that people purchase too many weak signals and too few strong signals relative to the instrumental value of the information (Appendix Figure~\ref{fig:wtp}). It would be valuable to empirically understand whether these effects generalize outside the lab in a way that might help explain the prevalence of lower-quality news sources.

Finally, while our results speak most directly to inference behavior, we see potential connections to the behavior of forecasts (stated expectations, rather than beliefs) at different horizons. As noted in the introduction, \textcite{AKLMT-WP} and \textcite{FLP-WP} find evidence for overreaction to news in a set of experimental forecasting tasks, as well as a selection of survey data. But there is evidence in recent work that such overreaction decreases with the persistence of the given series in a range of settings, in many cases switching to underreaction as a series approaches unit-root persistence.\footnote{Among others, \textcite{RH11} and \textcite{AKLMT-WP} provide evidence in the lab, and \textcite{BGMS20} provide evidence in survey data. Using both options and stock-return surveys, \textcite{GGL-WP} show evidence for overreaction in forecasts of the future equity premium, which is a moderately persistent series. For the Treasury yield curve, \textcite{W-WP} and \textcite{FNS-WP} show evidence for effective underreaction (e.g., positive coefficients in regressions of survey-based forecast errors on forecast revisions) for the short-horizon interest rate, which is a very persistent series with an annualized autocorrelation of above 0.9. \textcite{G19} provides a review and further discussion.} We conjecture that a model in which the conditional mean (rather than $\pi_t$) is the object of interest may help speak to these patterns: if forecasters understand a signal's directional impact on the conditional mean but do not perfectly understand how much it should change, then stronger news will take the form of more-persistent shocks, potentially generating the observed patterns of over- and underreaction. Given the importance of forecast behavior for macroeconomic and financial-market contexts outside of the ones we consider here, it would be useful to explore this connection both theoretically and empirically. We leave these, and other potential applications of our findings, for future work. 
\vspace{15mm}

\newpage

\vspace{1cm}
%\nocite{*}
 \AtNextBibliography{\small}
{\singlespace\printbibliography} %[check=uncited]

\newpage 

\appendix

\renewcommand{\thefigure}{A\arabic{figure}}
\setcounter{figure}{0}
\renewcommand{\thetable}{A\arabic{table}}
\setcounter{table}{0}
\renewcommand*{\theHtable}{\thetable}

\newpage

\section*{\LARGE Appendix}
%\vspace{1cm}

\renewcommand{\theequation}{A-\arabic{equation}}
\setcounter{equation}{0}

\section{Proofs and Additional Theoretical Results} \label{app:proofs}

\subsection{Proofs for Section~\ref{sec:theory-setup}} \label{app:proofs1}

\subsubsection*{Proof of Proposition~\ref{prop:main}} 
Fix an arbitrary direction $s_d$. Given \Cref{assumption:uts}, we can write  $\hat{\mathbb{S}}(\hat{s}) = \alpha(e) e + (1-\alpha(e)) \hat{\mathbb{S}}(s_d)$ for some $\alpha(e)\in(0,1)$, where $\alpha(\cdot)$ may depend on $s_d$. We want to characterize
\begin{align}
    \e[\hat{\mathbb S}(\hat s)|s] - \mathbb{S}(s) &= \e\!\left[(1-\alpha(e)) \hat{\mathbb{S}}(s_d) + \alpha(e)e \,\middle|\, s\right] - \mathbb{S}(s). \label{eq:diff1}
\end{align}

For notational convenience, assume a continuous space of estimates $e$ (in which $p(e|s)$ is a probability density function with support $E\subseteq\mathbb{R}$).\footnote{The steps in the proof carry through for discrete~$e$ when replacing integrals with sums and adjusting straightforwardly (though tediously) for discontinuities.} From \Cref{assumption:e}, 
    $\e[e|s] = \mathbb{S}(s),$ so 
   %\begin{align*}
   $\mathbb{S}(s) = {\displaystyle \int_E e \,p(e|s)\,de},$
   %\end{align*}
   with $p(e|s)$ non-degenerate. Using this in \eqref{eq:diff1}, 
   \begin{align}
    \e[\hat{\mathbb S}(\hat s)|s] - \mathbb{S}(s) &= \left[\int_E\left((1-\alpha(e)) \hat{\mathbb{S}}(s_d) + \alpha(e)e\right)p(e|s)\,de\right] - \mathbb{S}(s) \notag \\[.35em]
    &= \int_E (1-\alpha(e)) ( \hat{\mathbb{S}}(s_d) -e)\, p(e|s)\,de. \label{eq:diff2}
    \end{align}
    Denote $g(e)\equiv (1-\alpha(e)) ( \hat{\mathbb{S}}(s_d) -e).$ For the first term in $g(e),$  \Cref{assumption:uts} gives that $1-\alpha(e)>0$. For the second term, $ \hat{\mathbb{S}}(s_d)-e$ crosses 0 exactly once for $e\in \mathbb{R}$: it is positive for $e< \hat{\mathbb{S}}(s_d)$, and negative for $e> \hat{\mathbb{S}}(s_d)$. 
    We thus have that
       \begin{align}
    \e[\hat{\mathbb S}(\hat s)|s] - \mathbb{S}(s) 
    &= \int g(e)\, p(e|s)\,de = \e[g(e) | s] = \e[g(e)|s_d,\mathbb{S}], \label{eq:diff3}
    \end{align}
    where $g(e)$ is strictly single-crossing from above and where $p(e|s)=p(e|s_d,\mathbb{S})$ has the strict MLRP in $\mathbb{S}$, from \Cref{assumption:e}(b). By the variation diminishing property of \textcite{K68},\footnote{See \citeauthor{K68}'s Theorem 3.1 of Chapter 5, or \citeauthor{G01} (\citeyear{G01}, Proposition 16) for a textbook reference based on the generalization of \citeauthor{A02} (\citeyear{A02}, Theorem 2). These results are typically stated for a function that is single-crossing from below (SCB); in our case, one can define the SCB function $\tilde g(e)\equiv - g(e)$ and then take  $ \mathbb{S}(s) - \e[\hat{\mathbb S}(\hat s)|s] = \int \tilde g(e)\, p(e|s)\,de$, and all the statements carry through with signs changed appropriately. Note also that \eqref{eq:diff3} can be restated, suppressing dependence on the arbitrary and fixed $s_d$, as $\e[\hat{\mathbb S}(\hat s)|\mathbb{S}] - \mathbb{S}=\e[g(e)|\mathbb{S}]$, and it is this expression to which we apply \citeauthor{K68}'s result.} the expectation of a strictly single-crossing function with respect to an MLRP distribution is also strictly single-crossing, with the same arrangement of signs as the function (here, positive and then negative). That is, if  $\e[\hat{\mathbb S}(\hat s)|s] - \mathbb{S}(s)=0$ at $\mathbb{S}(s)=\mathbb{S}^*,$ then there is overreaction ($\e[\hat{\mathbb S}(\hat s)|s] - \mathbb{S}(s)>0$) for $\mathbb{S}(s)<\mathbb{S}^*$ and underreaction ($\e[\hat{\mathbb S}(\hat s)|s] - \mathbb{S}(s)<0$) for $\mathbb{S}(s)>\mathbb{S}^*$. 
    
    Further, this switching point $\mathbb{S}^*$ must exist and lie within the range of feasible values $\mathbb{S}(s)\in[\text{min}_{s_m}{\mathbb{S}}(s_d,s_m),\text{max}_{s_m}{\mathbb{S}}(s_d,s_m)]$. To see this, consider the case $\mathbb{S}(s)=\text{min}_{s_m}{\mathbb{S}}(s_d,s_m)$. By \Cref{assumption:priors}, $\hat{\mathbb{S}}(s_d)>\mathbb{S}(s)$ in this case, while $\e[e|\mathbb{S}]=\mathbb{S}(s)$ by \Cref{assumption:e}. Thus $  \e[\hat{\mathbb S}(\hat s)|s] = \e[\alpha(e) e + (1-\alpha(e)) \hat{\mathbb{S}}(s_d)|s]$ must satisfy    $\mathbb{S}(s)<\e[\hat{\mathbb S}(\hat s)|s]<\hat{\mathbb{S}}(s_d)$ by \Cref{assumption:uts}, where the lower bound obtains from $\alpha(e)\to1$ and $\e[e|\mathbb{S}]=\mathbb{S}(s)$ (and the upper bound obtains from $\alpha(e)\to0$). Thus $\e[\hat{\mathbb S}(\hat s)|s]-\mathbb{S}(s)>0$ at this minimal $\mathbb{S}(s).$ The same argument gives that  $\e[\hat{\mathbb S}(\hat s)|s]-\mathbb{S}(s)<0$ at the maximal $\mathbb{S}(s)$. The intermediate value theorem then gives that there is such a switching point $\mathbb{S}^*\in(\text{min}_{s_m}{\mathbb{S}}(s_d,s_m),\text{max}_{s_m}{\mathbb{S}}(s_d,s_m))$ at which $\e[\hat{\mathbb S}(\hat s)|s] - \mathbb{S}(s)= \e[\hat{\mathbb S}(\hat s)|s] - \mathbb{S}^*=0$, and the single-crossing result above guarantees its uniqueness, completing the proof.    
%     \begin{align*}
%    \e[\hat{\mathbb S}(\hat s)|s] - \mathbb{S}(s) &= \e[g_s(e) h(e,s)],
%    \end{align*}
%    where $g_s(e)$ is strictly single crossing, and 
\hfill $\square$

%\vspace{2mm}
%\noindent\emph{Derivation of monotonicity results}. 
\subsubsection*{Derivation of Monotonicity Results}
As at the end of \Cref{sec:theory-setup}, under Assumptions~\ref{assumption:e}--\ref{assumption:uts}, it is not necessarily the case that a person's expected signal strength $\hat{\mathbb{S}}(\hat{s})$ is monotonic in $e$ or that the amount of over- or underreaction  
$\e[\hat{\mathbb{S}}(\hat{s})|s]-\mathbb{S}(s)$ is monotonic in $\mathbb{S}(s)$. For conditions under which these additional monotonicity results hold, we again use \Cref{assumption:uts} to write  $\hat{\mathbb{S}}(\hat{s}) = \alpha(e) e + (1-\alpha(e)) \hat{\mathbb{S}}(s_d)$ for some $\alpha(e)\in(0,1)$, and for simplicity assume that $\alpha(e)$ is continuously differentiable (as are other relevant functions of $e$ or $\mathbb{S}$ considered below).

Using this representation, a necessary and sufficient condition for $\hat{\mathbb{S}}(\hat{s})$ to be (strictly) monotonically increasing in $e$ is that 
\begin{align}
    \alpha'(e)\!\left(e-\hat{\mathbb{S}}(s_d)\right) + \alpha(e) > 0. \label{eq:monoton1}
\end{align}
For $e>\hat{\mathbb{S}}(s_d),$ this requires that the weight on the estimate, $\alpha(e)$, not fall dramatically given small increases in $e$. For $e<\hat{\mathbb{S}}(s_d),$ the weight on the estimate must not fall dramatically given small decreases in $e$. Taken together, $\hat{\mathbb{S}}(\hat{s})$ will be monotonic in $e$ as long as the weight on the estimate does not fall dramatically given small increases in $|e-\hat{\mathbb{S}}(s_d)|$ (i.e.,  as $e$ moves further from the default $\hat{\mathbb{S}}(s_d)$), as stated in the text. Note that one simple sufficient condition for \eqref{eq:monoton1} is the constant-weighting case ($\alpha(e) = \alpha$), since in this case $\alpha'(e)=0$ and the condition reduces to $\alpha(e)>0$, which is guaranteed by \Cref{assumption:uts}.

Meanwhile, for $\e[\hat{\mathbb{S}}(\hat{s})|s]-\mathbb{S}(s)$ to be (strictly) monotonically decreasing in $\mathbb{S}(s)$, we must have that 
\begin{align*}
\frac{d\left(\e\!\left[\hat{\mathbb{S}}(\hat{s}) \,\middle|\, s\right]\right)}{d\mathbb{S}} - 1 <0.
\end{align*}
 Fix a direction $s_d$, so that conditioning on $s$ is equivalent to conditioning on $\mathbb{S}$. Since $\e[\hat{\mathbb S}(\hat s)|s]  = {\displaystyle \int\left((1-\alpha(e)) \hat{\mathbb{S}}(s_d) + \alpha(e)e\right)p(e|s)\,de}$, the above condition requires
\begin{align*}
    \int\left((1-\alpha(e)) \hat{\mathbb{S}}(s_d) + \alpha(e)e\right)\frac{\partial p(e|s)}{\partial \mathbb{S}}\,de &<1 \\[.35em]
    \Longleftrightarrow \,\,\,  \int\left((1-\alpha(e)) \hat{\mathbb{S}}(s_d) + \alpha(e)e\right)\frac{\frac{\partial p(e|s)}{\partial \mathbb{S}}}{p(e|s)}p(e|s)\,de &<1 \\[.35em]
    \Longleftrightarrow\,\,\,\e\!\left[\left((1-\alpha(e)) \hat{\mathbb{S}}(s_d) + \alpha(e)e\right)\frac{\frac{\partial p(e|s)}{\partial \mathbb{S}}}{p(e|s)} \,\middle|\, s  \right] &<1,
\end{align*}
or equivalently that
\begin{align}
    \text{Cov}_s\!\left(\alpha(e)(e-\hat{\mathbb{S}}(s_d)),\frac{\frac{\partial p(e|s)}{\partial \mathbb{S}}}{p(e|s)} \right)  < 1-\e[\hat{\mathbb{S}}(\hat{s})|s]\e\!\left[\frac{\frac{\partial p(e|s)}{\partial \mathbb{S}}}{p(e|s)} \,\middle|\, s\right], \label{eq:monoton2}
\end{align}
where $\text{Cov}_s(\cdot,\cdot)$ is the covariance conditional on $s$. Note further that
\begin{align*}
    \e\!\left[\frac{\frac{\partial p(e|s)}{\partial \mathbb{S}}}{p(e|s)} \,\middle|\, s\right] = \int \frac{\frac{\partial p(e|s)}{\partial \mathbb{S}}}{p(e|s)} p(e|s)\,de = \int \frac{\partial p(e|s)}{\partial \mathbb{S}} de = 0,
\end{align*}
since the density must integrate to 1 for all $\mathbb{S}$. The monotonicity condition in \eqref{eq:monoton2} can therefore be simplified to
\begin{align}
    \text{Cov}_s\!\left(\alpha(e)(e-\hat{\mathbb{S}}(s_d)),\frac{\frac{\partial p(e|s)}{\partial \mathbb{S}}}{p(e|s)} \right)  < 1. \label{eq:monoton3}
\end{align}
By \Cref{assumption:e}(b), $\frac{\frac{\partial p(e|s)}{\partial \mathbb{S}}}{p(e|s)}$ increases in $e$. Monotonicity in the degree of over-/underreaction in $\mathbb{S}$ therefore requires that $\alpha(e)$ not increase dramatically with $e$, as stated in the text, so that the covariance on the left side of~\eqref{eq:monoton2} is less than 1. One can verify that this condition is again immediately satisfied in the constant-weighting case. 
%\hfill $\square$

\subsection{Additional Discussion for Section~\ref{subsection:FunctionalForm}}
\label{app:FunctionalForm}

This appendix briefly discusses the mapping between the general environment in Section~\ref{sec:theory-setup} and the log-normal environment in \ref{subsection:FunctionalForm}. First, it is straightforward to verify that Assumptions~\ref{assumption:e} and \ref{assumption:priors} are satisfied in the log-normal environment. Assumption~\ref{assumption:uts} is slightly more complex. This assumption requires that $\hat{\mathbb{S}}(s_d)<\hat{\mathbb{S}}(\hat{s})<e$ when $e>\hat{\mathbb{S}}(s_d)$, and  $\hat{\mathbb{S}}(s_d)>\hat{\mathbb{S}}(\hat{s})>e$ when $e<\hat{\mathbb{S}}(s_d)$. Given the updating rule in \eqref{eq:logupdate}, this requires $\exp(\sigma^2_\s/2)<\frac{e}{\hat{\mathbb{S}}(s_d)}$ when $e>\hat{\mathbb{S}}(s_d)$, and $\exp(\sigma^2_e/2)<\frac{\hat{\mathbb{S}}(s_d)}{e}$ when $e<\hat{\mathbb{S}}(s_d)$.\footnote{These conditions will hold for most draws of $e$ given reasonably small variances. More formally, these conditions are satisfied almost surely in a small-noise limit in which $\sigma^2_e/\sigma^2_\s$ is fixed while $\sigma^2_e, \sigma^2_\s\to0$. A similar limit is considered, for example, in \citeauthor{KLW21} (\citeyear{KLW21}, Section~4 and Appendix~G).} In this case, %But they are also satisfied under more reasonable and intuitive parameterizations: as long as $\sigma^2_\s<2$ and $\sigma^2_e<2$, then the required conditions hold, and 
the posterior is strictly between the prior and estimate. Alternatively, to guarantee that Assumption~\ref{assumption:uts} holds for all $e$, one could drop the unbiasedness requirement of Assumption~\ref{assumption:e}(a) (i.e., $\e[e|\mathbb{S}]=\mathbb{S}$, which is an unimportant normalization) and assume $\log e \sim \mathcal{N}(\log \mathbb{S}, \sigma_e^2)$, in which case Assumption~\ref{assumption:uts} will always hold.

However, even if Assumption~\ref{assumption:uts} is not guaranteed to hold in this log-normal setting, this is unimportant for our main results on over- and underreaction. This is demonstrated in equation~\eqref{eq:ExpSHatAndS}, which shows that the conclusions in \Cref{prop:main} apply regardless, and the person accordingly overreacts to weak signals and underreacts to strong signals (with resulting switching point $\s^*$ discussed in the text) in this log-normal environment. We accordingly do not focus on the conditions under which the primitive assumptions hold; what is important is that the main results continue to hold in this setting.

\subsection{Proofs and Discussion for Section~\ref{sec:relax}}
\label{app:proofs2}

\subsubsection*{Prior Distortions: Incorrect Priors, Uncertain Priors, and Base-Rate Neglect} 

In the case of an incorrect prior belief $\hat\pi_0$ discussed at the beginning of Section~\ref{sec:relax}, we can calculate the belief change $|\logit(\hat\pi_1(s))-\logit(\hat\pi_0)|$ when $\hat\pi_0$ is observed. Perceived signal strength still follows the predictions in \Cref{prop:main}.
Under the maintained assumption that belief changes are monotonic in perceived signal strength (see \cref{fn:monotonic}), the overreaction to weak signals and underreaction to strong signals in \Cref{prop:main} will thus continue to be reflected in the belief change $|\logit(\hat\pi_1(s))-\logit(\hat\pi_0)|$.

We now consider the case in which the correct prior is uncertain. We can model this by adding a pre-period $t=-1$, and we assume that the person entered this previous period with a prior $\hat\pi_{-1}$ known with certainty, then observed a signal $s_0$ (with known direction $s_{d_0}$) and used a strength estimate $e_0$ to form $\hat\s_0(\hat s_0)$ and $\hat\pi_0(\hat s_0)$ following Bayes' rule given distributions $p(\s_0 | s_{d_0})$ and $p(e_0 | s_{d_0},\s_0)$.\footnote{As in \Cref{subsection:FunctionalForm}, we continue to assume quasi-Bayesian updating. This allows us to formalize the statement in the text that $\hat\pi_0$ incorporates all uncertainty about past signals. In the more general case considered in \Cref{sec:theory-setup}, the statement that the person overreacts to weak signals and underreacts to strong signals in period~1 is almost tautological: as long as the belief change continues to be monotonic in perceived signal strength, and perceived signal strength in period~1 is formed following Assumptions~\ref{assumption:e}--\ref{assumption:uts}, then the results hold immediately.} This post-estimation prior is then the center of a non-degenerate distribution for the correct prior $\pi_0(s_0)$, representing a situation with uncertainty over this correct prior. The person then observes $s_1$ and updates to $\hat\pi_1(\hat s_1)$ as before (again following Bayes' rule), with $s_1$ independent of $s_0$ conditional on $\theta$, and with $e_0$ and $e_1$ depending only on $s_0$ and $s_1$, respectively.\footnote{Note that this setup does not depend on the specific timing of periods $0$ and 1; this notation simply formalizes the idea that the correct prior is formed from some signal (like information provided in an experiment) separate from the additional piece of information in signal $s_1$.}

With this setup, applying Bayes' rule twice, the posterior given $\hat s_0$ and $\hat s_1$ is
\begin{align*}
    \text{logit}(\hat\pi_1) &= \text{logit}(\hat\pi_{-1}) + \log\!\left(\frac{p(\hat s_0|\theta=1)}{p(\hat s_0|\theta=0)}\right) +\log\!\left(\frac{p(\hat s_1|\hat s_0,\theta=1)}{p(\hat s_1|\hat s_0,\theta=0)}\right) \\[.35em]
    &=  \text{logit}(\hat\pi_0(\hat s_0)) + \log\!\left(\frac{p(\hat s_1|\hat s_0,\theta=1)}{p(\hat s_1|\hat s_0,\theta=0)}\right).
\end{align*}
Note that $p(\hat s_1|\hat s_0,\theta)=p(\hat s_1 | \theta)$, since $s_0$ and $s_1$ are independent conditional on $\theta$, and $e_0$ and $e_1$ depend only on $s_0$ and $s_1$, respectively. Therefore, 
\begin{equation*}
      \text{logit}(\hat\pi_1) =   \text{logit}(\hat\pi_0(\hat s_0)) + \log\!\left(\frac{p(\hat s_1|\theta=1)}{p(\hat s_1| \theta=0)}\right).
\end{equation*}
The belief update in period 1, $ |\text{logit}(\hat\pi_1)-  \text{logit}(\hat\pi_0(\hat s_0))|,$ accordingly depends on perceived signal strength $\left\lvert\log\!\left(\frac{p(\hat s_1|\theta=1)}{p(\hat s_1| \theta=0)}\right)\right\rvert$ exactly as was the case before, with the previous period's estimate (or multiple previous periods' estimates) affecting only $\hat\pi_0$. Under the assumption that $\left\lvert\log\!\left(\frac{p(\hat s_1|\theta=1)}{p(\hat s_1| \theta=0)}\right)\right\rvert$ is monotonic in $\hat\s_1(\hat s_1) = \hat\e[\s_1|s_{d_1},e_1]$ (again as in footnote~\ref{fn:monotonic}), all our results therefore carry through to this case. 

In the case that the previously formed prior is unobserved, though, we cannot calculate $ |\text{logit}(\hat\pi_1)-  \text{logit}(\hat\pi_0(\hat s_0))|$ directly. Instead, we again use $ |\text{logit}(\hat\pi_1)-  \text{logit}(\pi_0)|$ as our proxy for reaction. This measure now includes both perceived signal strength and the prior distortion:
\begin{equation}
    |\text{logit}(\hat\pi_1)-  \text{logit}(\pi_0)| = |\text{logit}(\hat\pi_0(\hat s_0))-  \text{logit}(\pi_0)| \pm \left\lvert\log\!\left(\frac{p(\hat s_1|\theta=1)}{p(\hat s_1| \theta=0)}\right)\right\rvert. \label{eq:prior_distorted}
\end{equation}
There are thus two cases to consider. (1) If the expected prior distortion in the first term has the same sign as the signal direction, then $|\text{logit}(\hat\pi_1)-  \text{logit}(\pi_0)|$ will overstate the degree of overreaction in the perceived signal strength $\hat\s(\hat s)$, and there may appear to be overreaction even to strong signals. This will apply, for example, if the correct prior is much lower than 0.5, but people do not use this correct prior and instead shade toward a default uninformative prior of 0.5. This will push up the apparent reaction to a positive signal. (2) If the expected prior distortion has the opposite sign as the signal direction, then $|\text{logit}(\hat\pi_1)-  \text{logit}(\pi_0)|$ will understate the degree of overreaction in the perceived signal strength $\hat\s(\hat s)$, and there may appear to be underreaction (or incorrectly signed reactions) even to weak signals. Intuitively, the prior distortion offsets the signal reaction in this case. We should expect these issues to matter less when the prior estimation is more precise than the signal strength estimation, or when the default prior (often 0.5) is close to the correct prior.

The same analysis applies to the case with base-rate neglect, which will simply move the effective prior $\hat\pi_0$ in \eqref{eq:prior_distorted} toward the person's default prior (which, in this binary-state setting, is again often modeled as the uninformative prior of 0.5). Cases in which the correct prior is equal to or close to 0.5 will therefore have little to no role for such base-rate neglect. More generally, we expect our results to hold within a range of priors around $\pi_0=0.5.$ (Based on our experimental results, this range appears reasonably wide.) For correct priors close to 0 or 1, meanwhile, given strong enough base-rate neglect, this can offset our main effect according to situations (1) and (2) as described in the preceding paragraph. 

We discuss and control for the effects of base-rate neglect in additional detail in \Cref{sec:study1b}, which presents the results of an experiment with priors different from 0.5. In particular, as shown in \cref{eq:brn_est} and discussed in footnote~\ref{fn:brn}, the measured signal weight $\hat w(s)$ (which we estimate as $\frac{\logit\hspace{1pt}\hat\pi_1-\logit\hspace{1pt}\pi_0}{\logit\hspace{1pt}\pi_1-\logit\hspace{1pt}\pi_0}$) will be distorted by base-rate neglect to the extent that $\logit\hspace{1pt}\pi_0$ (the distance of the prior from 0.5) is high relative to $\logit\hspace{1pt}\pi_1-\logit\hspace{1pt}\pi_0$ (the true signed signal strength), though of course this only matters to the degree that the person engages in strong base-rate neglect.

\subsubsection*{Uncertainty About the Direction}

Following the discussion in the text, we now assume that the person forms an estimate $e$ of $\s_{signed}$, with that estimate satisfying Assumption~\ref{assumption:e} with respect to $\s_{signed}$. In place of Assumption~\ref{assumption:priors}, we assume that the default value (the person's subjective prior) is $\hat\s_{0,signed}=0.$ This effectively assumes a symmetric signal strength distribution where $\e[\s_{signed}]=0.$\footnote{We note that the prediction of underreaction does not necessarily apply in asymmetric cases, which can lead to strange situations. While the expected change in beliefs is always equal to 0 (at least for a Bayesian), due to the non-linear transformation from signal strength to belief changes, it is not necessarily the case that the signal strength has a mean of zero. If one removes the assumption of symmetry, we can say only that there is underreaction for sufficiently extreme signed strengths, but we cannot necessarily make statements across all signal strengths. This analysis is, however, not the main focus given the settings we seek to describe.} Similar to Assumption~\ref{assumption:uts}, we assume that the posterior $\hat\s_{signed}(\hat s)$ is strictly between 0 and the estimate $e$. Given this, it is immediate that on average, there is underreaction in perceived signed strength: $\e[|\hat\s_{signed}|\mid s] = a(s)|\s_{signed}|<|\s_{signed}|,$ where $a(s)\in(0,1)$. Note that this definition of underreaction is in terms of the absolute perceived signed strength relative to the absolute true signed strength. The interpretation of this result as ``underreaction'' becomes more strained when the signs of $\hat\s_{signed}$ and $\s_{signed}$ are different, as discussed in the main text.

\subsection{Proofs and Discussion for Section~\ref{subsec:multiple}}
\label{app:proofs3}

\subsubsection*{Independent Estimates}
We formally define the cross-sectional expectation as $\e_i[X_i|s]=\lim_{N\to\infty}\frac{1}{N}\sum_{i=1}^N X_i$ for any measurable $X_i$ (whose distribution implicitly depends on $s$). Under the assumptions that the estimates $e_i$ are independent across people, there is no formal distinction between taking the expectation with respect to the distribution of estimates (as we did previously) and taking the cross-sectional expectation across people. Therefore, \Cref{prop:main} continues to apply, in the sense that there exists a unique switching point $\s^*$ such that there is overreaction on average ($\e_i[\hat\s_i(\hat{s}_i)|s]>\mathbb{S}(s)$) if $\s(s)<\s^*$, and there is underreaction on average ($\e_i[\hat\s_i(\hat{s}_i)|s]<\mathbb{S}(s)$) if $\s(s)>\s^*$. %$\hfill\square$

\subsubsection*{Correlated Estimates and Over-/Underreaction Conditional on $\s$}
We consider the case with multi-dimensional signals and perfect correlation in estimates given identical attention vectors $\boldsymbol{a}_i$ for all $i$. In this case, as stated in the text, \Cref{prop:main} holds under the following new definition: there is overreaction if $\e[\hat\s_i(\hat{s}_i)|\s]>\mathbb{S}$, and underreaction if $\e[\hat\s_i(\hat{s}_i)|\s]<\mathbb{S}$. 

Since $s_{m,j}$ are i.i.d.\ over components $j$ and exchangeable, $\e[s_{m,j}|\s]=\log\s$. (By comparison,  conditional on $s$, $s_{m,j}$ is known, so a given person's $\hat{\s}(s)$ in that case could potentially differ across $s$ for the same $\s$, invalidating our results. This motivates our conditioning on $\s$ here.) 
Similarly, $e_i$ is log-normally distributed conditional on $\s$, $\log e_i \sim \mathcal{N}(\log \mathbb{S} - \sigma^2_{e,i}/2, \sigma^2_{e,i})$, where this (and the expression for $\sigma_{e,i}^2$ provided in the text) follow from standard characterizations of a multivariate normal distribution along with some algebra. Therefore, conditioning on $\s$, $\e[\hat\s_i(\hat{s}_i)|\s] = k \s^\beta$, where $k$ and $\beta$ are as given in the text. This is exactly as in \eqref{eq:ExpSHatAndS}, and we conclude that \Cref{prop:main} applies using the above definition of over- and underreaction. Further, since we have assumed the extreme case of perfectly correlated estimates, this will also apply when considering the expected cross-sectional expectation $\e[\e_i[\hat\s_i(\hat{s}_i)|s]|\s]$ given that $\hat\s_i(\hat{s}_i)$ is identical across $i$ for a given $s$. %\hfill$\square$

\subsubsection*{Predictions on Correlation Behavior}
As stated in the text, one can make further statements about the correlation in updating behavior across people under additional assumptions about the signal components and person-specific vectors $\boldsymbol{a}_i$. For example, if the components are ordered by salience, then it is natural to assume that $\boldsymbol{a}_i$ is such that $a_{i,j}=1$ for $j\leq n_i$ and $a_{i,j}=0$ for $n_i<j\leq n$ (i.e., person $i$ pays attention to the first $n_i$ components, and the only difference across people is in how large $n_i$ is). In this case, the following expression holds for the ex ante correlation between estimates for any two people $i$ and $i'$, ordered such that $0<n_i\leq n_{i'}$:
\begin{equation*}
    \textnormal{Corr}(e_i,e_{i'}) = \sqrt{\frac{n_i}{n_{i'}}}\in(0,1].
\end{equation*}
This expression follows from the fact that $\text{Cov}(e_i,e_{i'})=\text{Var}(s_{m,j})/\max(n_i,n_{i'})$, while $\text{Var}(e_i)=\text{Var}(s_{m,j})/n_i$ and $\text{Var}(e_{i'})=\text{Var}(s_{m,j})/n_{i'}$.

The above expression can be generalized to cases where the components are not salience-ordered. In these cases, the correlation will simply scale down as one decreases the overlap in the entries of $\boldsymbol{a}_i$ and $\boldsymbol{a}_{i'}$ that are equal to 1. For example, if there are two components and two types of people, with type 1 attentive only to component 1 and type 2 attentive only to component 2, then there will be multimodal estimates, with perfect correlation across people within type and none across types. With high-dimensional vectors in which the components are not ordered according to salience (i.e., cases where people's attention vectors are varied), estimates will be closer to the independent case, and we will see smoother distributions of resulting strength perceptions. 

\subsection{Proofs for Section~\ref{sec:mrsubsec}} \label{app:proofmr}

\subsubsection*{Proof of Proposition~\ref{Prop_mr_substreams}} 
Here, we provide a brief restatement of the proof of Proposition~1 of AR (\citeyear{AR21}) for completeness. Since $\pi_t = \pi_t(H_t) = \mathbb{E}_t[\theta]$, by the law of iterated expectations (LIE), beliefs are a martingale: $\pi_t = \mathbb{E}_t[\pi_{t+1}].$ Therefore, for arbitrary $t_1$,
\begin{align*}
    \e_{t_1}[M_{t_1,t_1+1}-R_{t_1,t_1+1}] &= \e_{t_1}[(\pi_{t_1+1}-\pi_{t_1})^{2}-(\pi_{t_1}(1-\pi_{t_1})-\pi_{t_1+1}(1-\pi_{t_1+1}))] \\
    &= \e_{t_1}[(2\pi_{t_1}-1)(\pi_{t_1}-\pi_{t_1+1})] \\
    &= (2\pi_{t_1}-1)(\e_{t_1}[\pi_{t_1}-\pi_{t_1+1}]) = 0,
\end{align*}
where the first line uses the definition of movement and uncertainty reduction, the second line simplifies, and the last line rearranges and uses the martingale property of beliefs. Similarly, $\e_{t_1+\tau}[M_{t_1+\tau,t_1+\tau+1}-R_{t_1+\tau,t_1+\tau+1}]=0$ for any $\tau\geq0$, and therefore by the LIE, $\e_{t_1}[M_{t_1+\tau,t_1+\tau+1}-R_{t_1+\tau,t_1+\tau+1}]=0$. So summing all these terms from $t_1$ to arbitrary $t_2>t_1$, 
\begin{align*}
    \e_{t_1}[M_{t_1,t_2}-R_{t_1,t_2}] = \sum_{\tau=0}^{t_2-t_1 - 1} \e_{t_1}[M_{t_1+\tau,t_1+\tau+1}-R_{t_1+\tau,t_1+\tau+1}] = 0,
\end{align*}
as stated. \hfill $\square$

\clearpage

\section{Additional Details and Estimation Results}
\label{app:experiment}

\subsection{Experimental Studies}

\subsection*{Study 1a}

\subsubsection*{Timing Details}
\label{app:study1a-timing}
Participants saw the following five treatment blocks: (1) one symmetric signal, (2) one asymmetric signal, (3) three symmetric signals, (4) demand for information, (5) uncertain signals. Details of each are in the subsequent subsections. The ordering of when they saw each treatment block was as follows:

\vspace{5mm}
\noindent
\renewcommand\arraystretch{1.2}
\begin{tabular}{l l l l}  
\toprule
\textbf{Rounds} \hspace{9mm} & \textbf{Treatment Block} \hspace{21mm} & \textbf{Frequency} \hspace{10mm} & \textbf{Observations} \\
\midrule
   1--12  &  One symmetric signal  &  67 percent  &  4,036  \\
   1--12  &  One asymmetric signal  &  33 percent  &  1,964  \\
   13  &  Attention check  &  100 percent  &  500  \\
   14--18  &  Three symmetric signals &  100 percent  &  2,500  \\
   19--23  &  Demand for information  &  100 percent  &  2,500  \\
   24--25  &  One uncertain signal  &  100 percent  &  1,000  \\
\bottomrule
\end{tabular}
\renewcommand\arraystretch{1}
\vspace{5mm}

Questions within each treatment block were randomized for each participant. The ordering of treatment blocks (besides ``one symmetric'' and ``one asymmetric'') were fixed for ease of participant comprehension. For instance, participants do not see the ``demand for information'' treatment until they have played rounds in which they inferred from one signal and from multiple signals. The uncertain-signals treatment comes after the demand-for-information treatment because they do not reflect the signals that participants would purchase. 

\newpage

\subsubsection*{Additional Results Discussed in the Text}

\begin{figure}[htb!]
\caption{Comparison to Literature}
\begin{center}  
\vspace{-3mm}
    \includegraphics[width=.99\textwidth]{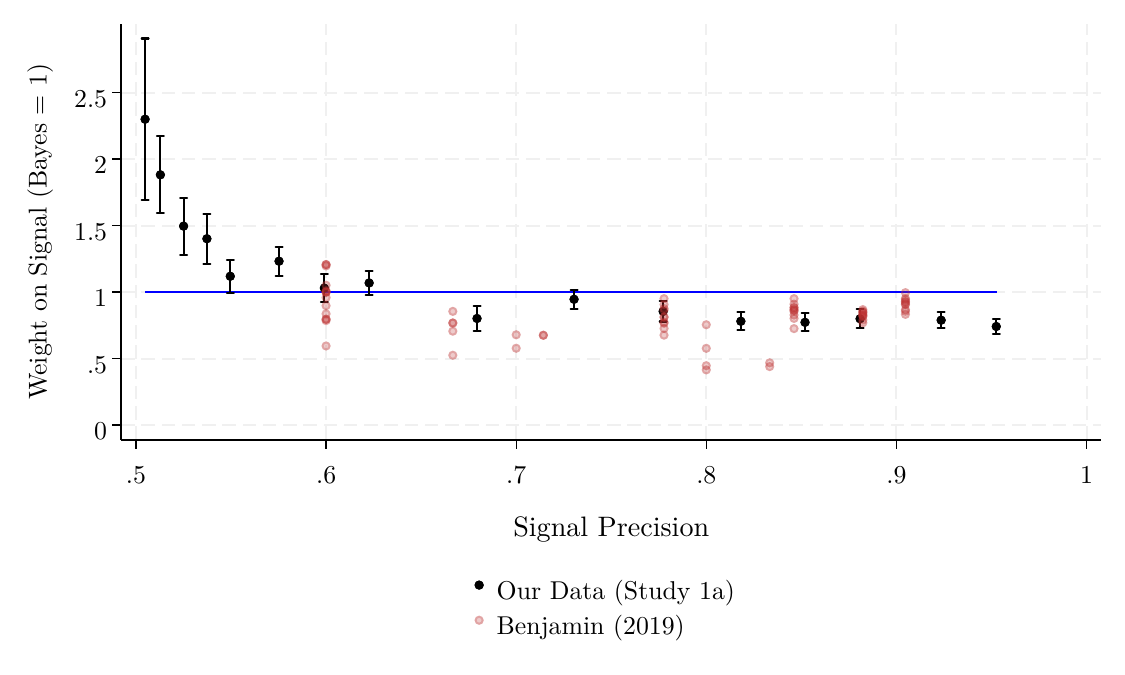}
\end{center}
\begin{threeparttable}
\begin{tablenotes}
\begin{scriptsize}
\vspace{-8mm}
\item \textbf{Notes:} This figure shows the weight put on signals of different precisions, where weight is defined relative to a Bayesian (whose weight of 1 is in the blue line) as in the main text. Black circles correspond to data from our Study 1a with 95\% confidence intervals (as plotted in \Cref{fig:main_effects}). Red circles correspond to data from \textcite{B19}. We use the data from his supplementary files, restricting to the 70 studies in which participants update from one binary signal when the prior is 0.5 and signal precision is symmetric. Note that most papers include multiple studies. %\par 
\end{scriptsize}
\end{tablenotes}
\end{threeparttable}
\label{fig:benjamin_comp}
\end{figure}

\begin{figure}[htb!]
\caption{Heterogeneity in Inference at the Individual Level}
\begin{center}
\vspace{-5mm}
    \includegraphics[height=.42\textheight]{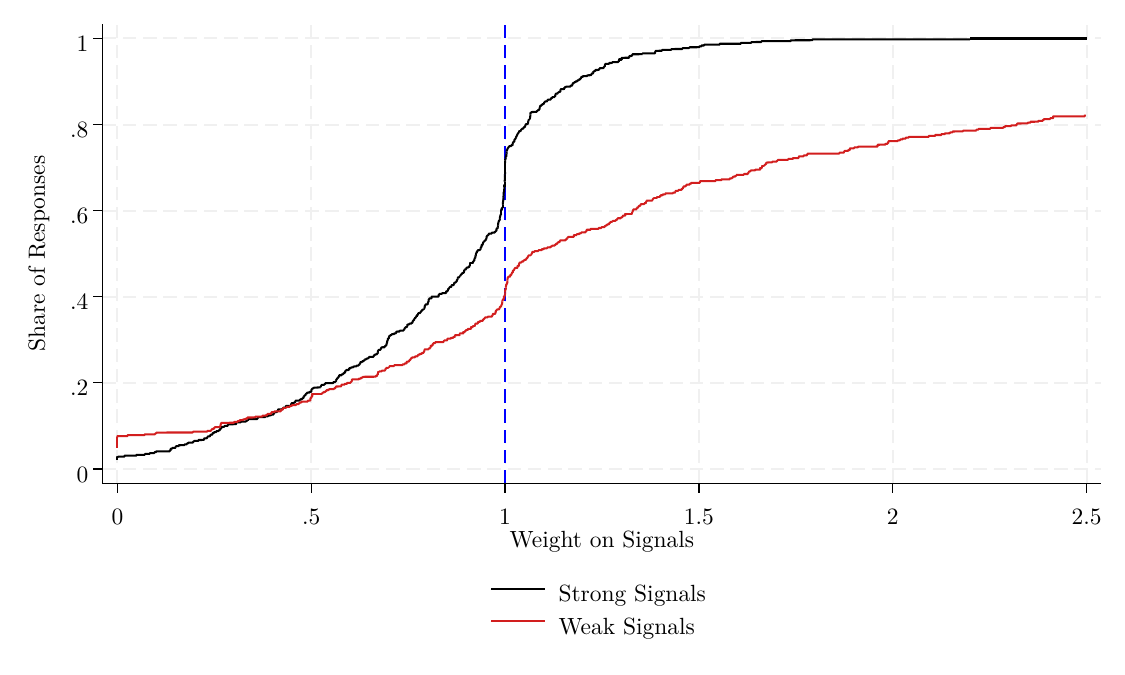}
    \includegraphics[height=.42\textheight]{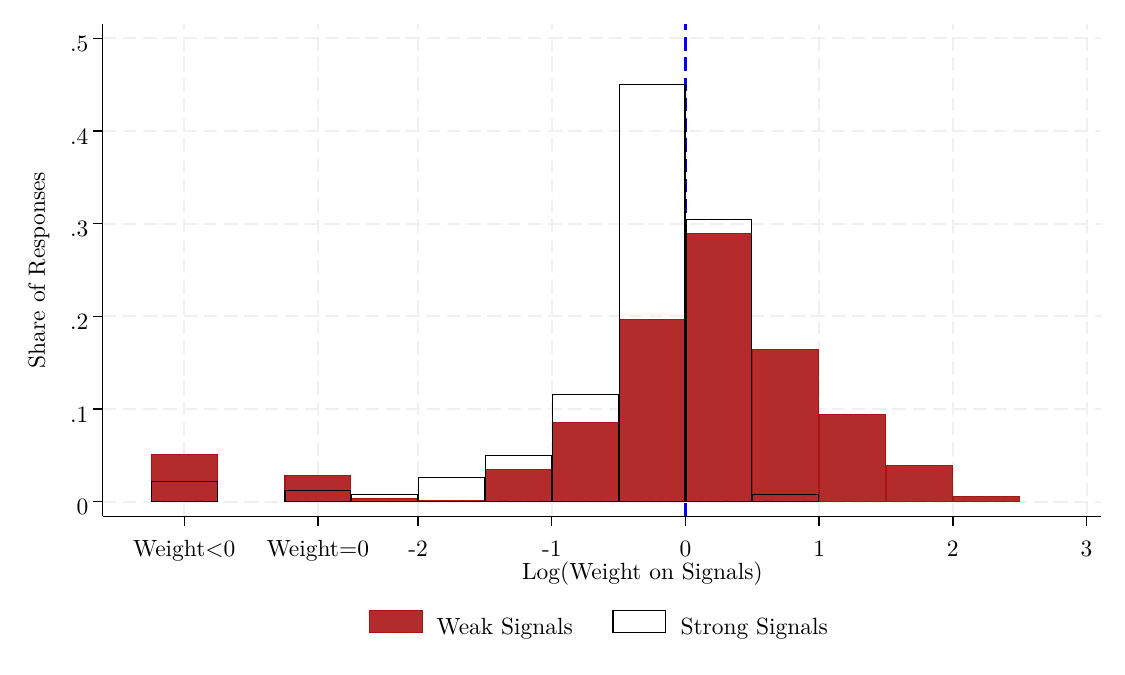}
\end{center}
\begin{threeparttable}
\begin{tablenotes}
\begin{scriptsize}
\vspace{-10mm}
\item \textbf{Notes:} This figure shows how much weight is put on weak and strong signals at the individual level, where weight is defined relative to a Bayesian as in the main text. The top panel shows the CDF of individuals' weights on strong and weak signals. The vertical line at 1 represents Bayesian updating. The bottom panel shows the PDF of the log of individuals' weights on strong and weak signals. The vertical line at 0 represents Bayesian updating. Participants with nonpositive weight are separated out. Weak signals have precision $p < 0.6$ and strong signals have precision $p > 0.7$. Observations are winsorized, for each signal strength, at the 5\% level. 
\vspace{-11mm}
\end{scriptsize}
\end{tablenotes}
\end{threeparttable}
\label{fig:individual-cdfs}
\end{figure}

\begin{figure}[htb!]
\caption{Over- and Underinference by Number and Strength of Signals}
\begin{center}
\vspace{-4mm}
    \includegraphics[width=\textwidth]{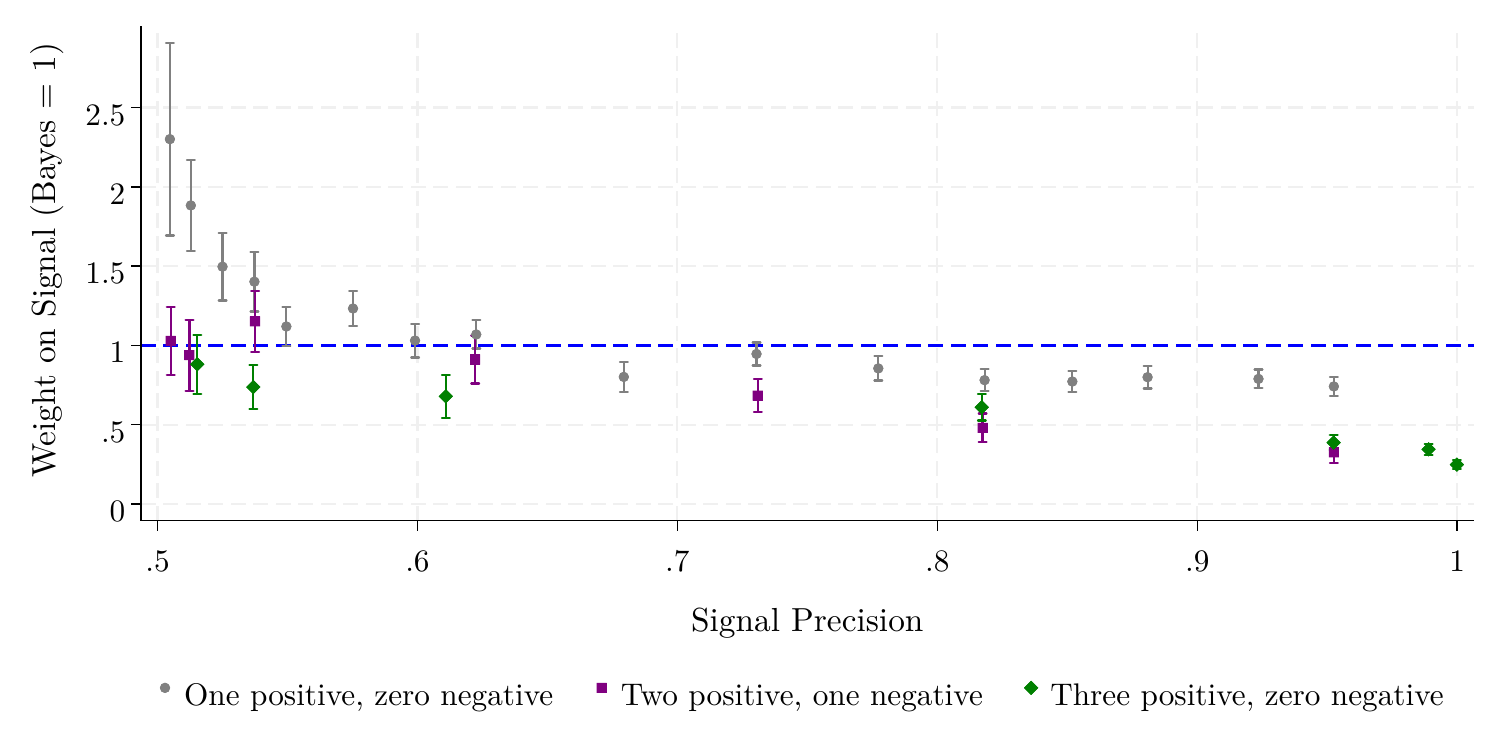}
\end{center}
\begin{threeparttable}
\begin{tablenotes}
\begin{scriptsize}
\vspace{-7mm}
\item \textbf{Notes:} This figure plots the average weight participants put on signals relative to a Bayesian (indicated by the blue line), split by signal distribution. Gray circles correspond to one signal of precision $\rho$ (as in \Cref{fig:main_effects}); purple squares correspond to two signals of precision $\rho$ in one direction and one signal of precision $\rho$ in the opposing direction; green diamonds correspond to three signals of strength $\s/3$, where $\s=\logit\rho$ for precision $\rho$, in the same the direction. This figure shows that participants put less weight on three signals as compared to the weight they put on one signal. Error bars indicate 95\% confidence intervals.
\vspace{-2mm}
\end{scriptsize}
\end{tablenotes}
\end{threeparttable}
\label{fig:multiple}
\end{figure}

Figures~\ref{fig:benjamin_comp}--\ref{fig:multiple} provide additional results discussed in the text.

\newpage

\newpage

\clearpage

\subsubsection*{Further Results: Demand for Information}

\begin{figure}[htb!]
\caption{Number of Signals Purchased}
\begin{center}
\vspace{-4mm}
    \includegraphics[height=.40\textheight]{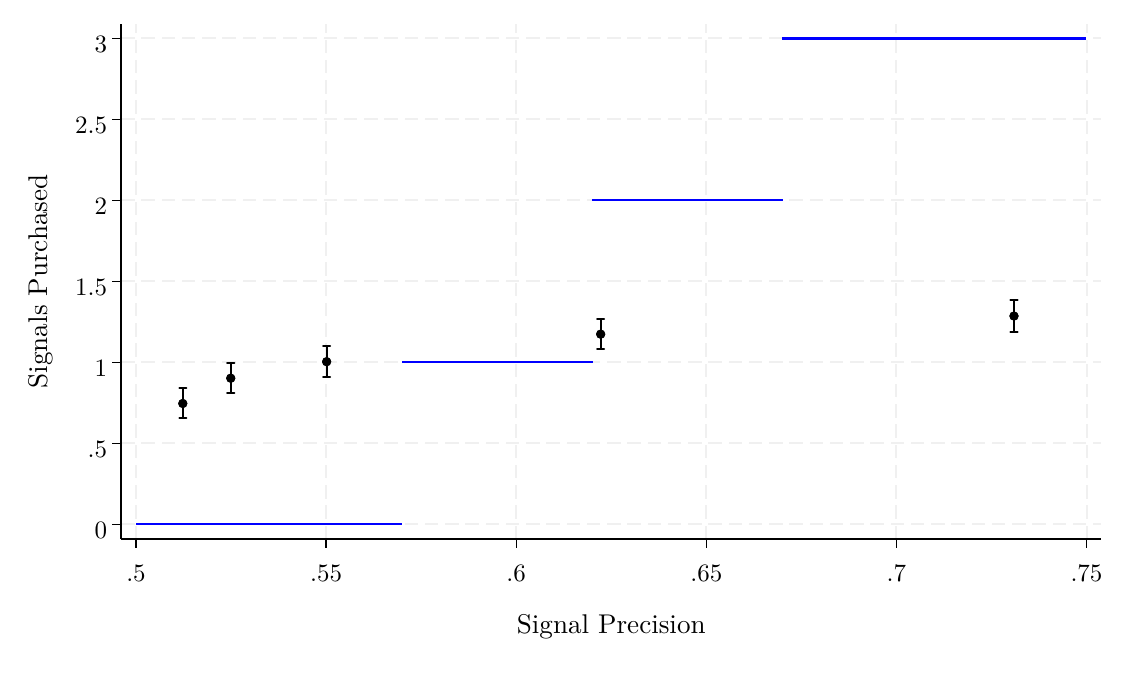}
\end{center}
\begin{threeparttable}
\begin{tablenotes}
\begin{scriptsize}
\vspace{-7mm}
\item \textbf{Notes:} This figure plots the number of signals purchased as a function of signal precision. The blue lines correspond to the payoff-maximizing number of signals purchased. This figure shows that participants over-purchase weak signals and under-purchase strong signals. Error bars indicate 95\% confidence intervals. 
\vspace{-2mm}
\end{scriptsize}
\end{tablenotes}
\end{threeparttable}
\label{fig:wtp}
\end{figure}

Patterns of overinference and underinference can also lead to demand for information that is too high or too low relative to the optimum. \Cref{fig:wtp} plots the average number of signals purchased as a function of each signal strength, comparing participant behavior to the optimal choice if participants were Bayesian and only valued signals for their instrumental value. 

As can be seen in the figure, participants systematically over-purchase weak signals and under-purchase strong signals. The cost of a signal that leads a Bayesian to form a posterior of less than 0.57 outweighs its benefit; however, the majority of participants purchase at least one signal when $p=0.55$ and $p=0.525$. Additionally, 81 percent of participants purchase fewer than the optimal level of three signals when $p=0.73$. Over- and underinference patterns therefore matter not just for stated beliefs; they also lead people to overvalue low-quality information and undervalue high-quality information, as reflected in their purchase decisions.

\clearpage

\subsection*{Study 1b}
\subsubsection*{Grether Regression Approach}

\begin{table}[htb!]
\centering
\begin{threeparttable}%[tb!]
\centering
\onehalfspacing
\begin{small}
\caption{Effect of Logit Prior and Signal Strength on Logit Posterior}
\begin{tabular}{l*{4}{c}}
\toprule
                              &\multicolumn{1}{c}{(1)}&\multicolumn{1}{c}{(2)}&\multicolumn{1}{c}{(3)}&\multicolumn{1}{c}{(4)}\\
                              &\multicolumn{1}{c}{50\% Prior}&\multicolumn{1}{c}{All Priors}&\multicolumn{1}{c}{All Priors}&\multicolumn{1}{c}{All Priors}\\
\midrule
Signal Strength: 0.05         &    2.371&    2.844&    2.800&    2.795\\
                              &  (0.224)&  (0.286)&  (0.179)&  (0.178)\\
Signal Strength: 0.20         &    1.359&    1.554&    1.504&    1.500\\
                              &  (0.072)&  (0.083)&  (0.059)&  (0.059)\\
Signal Strength: 0.50         &    1.176&    1.208&    1.190&    1.191\\
                              &  (0.054)&  (0.041)&  (0.035)&  (0.035)\\
Signal Strength: 1.25         &    0.840&    0.852&    0.857&    0.857\\
                              &  (0.028)&  (0.021)&  (0.020)&  (0.020)\\
Signal Strength: 1.75         &    0.824&    0.768&    0.762&    0.762\\
                              &  (0.021)&  (0.017)&  (0.015)&  (0.015)\\
Logit Prior                   &         &       1 &    0.984&         \\
                              &         &     (.) &  (0.016)&         \\
Logit Prior: Signal=0.05      &         &         &         &    1.028\\
                              &         &         &         &  (0.017)\\
Logit Prior: Signal=0.20      &         &         &         &    1.036\\
                              &         &         &         &  (0.021)\\
Logit Prior: Signal=0.50      &         &         &         &    0.996\\
                              &         &         &         &  (0.023)\\
Logit Prior: Signal=1.25      &         &         &         &    0.944\\
                              &         &         &         &  (0.029)\\
Logit Prior: Signal=1.75      &         &         &         &    0.915\\
                              &         &         &         &  (0.029)\\
Participant FE                    &      Yes&      Yes&      Yes&      Yes\\
Round FE                      &      Yes&      Yes&      Yes&      Yes\\
\midrule
Observations                  &     2500&     7500&     7500&     7500\\
\(R^{2}\)                     &     0.80&     0.60&     0.76&     0.76\\
\bottomrule
%\multicolumn{5}{l}{\footnotesize Standard errors in parentheses}\\
\end{tabular}

\label{tab:expt_results_grether}
\end{small}
\begin{tablenotes}[para,flushleft]
\begin{scriptsize}
\textbf{Notes:} OLS, with standard errors in parentheses clustered at participant level. We regress logit posterior on each signal strength separately. Column (1) restricts to observations where the prior is symmetric (as in Study 1a); other columns use the full dataset. Column (2) assumes that people put weight 1 on their prior. Column (3) allows for misweighting priors overall. Column (4) allows for weights on priors to vary for each signal strength. See main text for discussion. \par
\end{scriptsize}
\end{tablenotes}
\end{threeparttable}
\end{table}

\clearpage

\subsection*{Study 2}
\subsubsection*{Grether Regression Approach}
\begin{table}[htb!]
\centering
\begin{threeparttable}%[tb!]
\centering
\onehalfspacing
\begin{small}
\caption{Weight on Signal and Prior by Quarter of Basketball Game}\label{tab:expt_results_nba}
\begin{tabular}{l*{3}{c}}
\toprule
                              &\multicolumn{1}{c}{(1)}&\multicolumn{1}{c}{(2)}&\multicolumn{1}{c}{(3)}\\
                              &\multicolumn{1}{c}{All Quarters}&\multicolumn{1}{c}{All Quarters}&\multicolumn{1}{c}{All Quarters}\\
\midrule
Quarter 1 x Signal Strength   &    1.344&    1.405&    1.406\\
                              &  (0.108)&  (0.066)&  (0.066)\\
Quarter 2 x Signal Strength   &    1.398&    1.360&    1.359\\
                              &  (0.110)&  (0.059)&  (0.059)\\
Quarter 3 x Signal Strength   &    0.968&    0.929&    0.928\\
                              &  (0.070)&  (0.041)&  (0.040)\\
Quarter 4 x Signal Strength   &    0.735&    0.585&    0.587\\
                              &  (0.037)&  (0.020)&  (0.020)\\
Logit Prior                   &       1 &    0.906&         \\
                              &      (.)&  (0.013)&         \\
Quarter 1 x Logit Prior       &         &         &    1.001\\
                              &         &         &  (0.051)\\
Quarter 2 x Logit Prior       &         &         &    0.948\\
                              &         &         &  (0.031)\\
Quarter 3 x Logit Prior       &         &         &    0.917\\
                              &         &         &  (0.025)\\
Quarter 4 x Logit Prior       &         &         &    0.889\\
                              &         &         &  (0.014)\\
Participant FE                    &      Yes&      Yes&      Yes\\
Round FE                      &      Yes&      Yes&      Yes\\
Quarter FE                    &      Yes&      Yes&      Yes\\
\midrule
Observations                  &     8000&     8000&     8000\\
\(R^{2}\)                     &     0.48&     0.86&     0.86\\
\bottomrule
%\multicolumn{4}{l}{\footnotesize Standard errors in parentheses}\\
\end{tabular}

\end{small}
\begin{tablenotes}[para,flushleft]
\begin{scriptsize}
\textbf{Notes:} OLS, with standard errors in parentheses clustered at participant level. We regress logit posterior on signals in each quarter separately. Column (1) assumes that people put weight 1 on their prior. Column (2) allows for misweighting priors overall. Column (3) allows for weights on priors to vary for each quarter. See main text for discussion. \par
\end{scriptsize}
\vspace{2mm}
\end{tablenotes}
\end{threeparttable}
\end{table}

\clearpage

\subsection{Empirical Evidence}
\label{app:details_full} \label{app:rnd} \label{app:details}

\subsection*{Measurement Details for Risk-Neutral Beliefs}

This subsection describes our use of option-price data, as introduced in Section~\ref{sec:findata}, in greater detail. (Much of this detail is directly from AL \citeyear{AL-WP}.) First, we describe how we clean the option data and then translate the option prices to risk-neutral beliefs. We then detail how we translate from risk-neutral to physical beliefs under different parameterizations for risk aversion.

\subsubsection*{Option Data Cleaning and the Risk-Neutral Distribution}

We start from the OptionMetrics data described in the text, obtaining the end-of-day bid and ask prices for all European call and put options on the S\&P 500 index, for all available strike prices and option expiration dates for trading dates from January 1996 through December 2018. We then average the bid and ask price to obtain the mid price. We also, as in AL (\citeyear{AL-WP}), obtain S\&P 500 index prices  to use when determining the realized index-return state. We first get end-of-day index prices (which we take as well from OptionMetrics, and then augment these with hand-collected settlement values for any options whose settlement value depends on the opening (rather than closing) index price, from the CBOE website.\footnote{The results for the binarized noise-corrected data in \Cref{fig:financeBinary} below also use separate data directly from AL (\citeyear{AL-WP}), so we refer to that paper --- in particular, Section~6 and Online Appendix~C.5--C.6 --- for details on the data and methodology used for the noise estimation (which use intraday option data obtained directly from the CBOE), as well as the conversion of the histogram of risk-neutral beliefs to binarized beliefs.}

To measure the risk-free rate $R_{t,T}^f$ in order to define our excess return space,  we follow \textcite{vanBinsbergen2021} and estimate the risk-free rate from the cross-section of option prices by applying put-call parity. We use their ``Estimator 2,'' which estimates $R_{t,T}^f$ from Theil--Sen (robust median) estimation of the put-call parity relationship. This provides a risk-free rate consistent with observed option prices.

For the OptionMetrics data, we then use the same steps as described in Online Appendix~C.5 of AL (\citeyear{AL-WP}) to clean the data and convert to a risk-neutral distribution. For cleaning, we drop any options with bid or ask price of zero (or less than zero), with uncomputable Black--Scholes implied volatility or with implied volatility of greater than 100 percent, with more than one year to maturity, or (for call options) with mid prices greater than the price of the underlying; we drop any option cross-section (i.e., the full set of prices for the pair $(t,T)$) with no trading volume on date $t$, with fewer than three listed prices across different strikes, or for which there are fewer than three strikes for which both call and put prices are available (as is necessary to calculate the forward price and risk-free rate). 

We then measure the risk-neutral distribution following \textcite{M14}, again as described in Online Appendix~C.5 of AL (\citeyear{AL-WP}): 
\begin{enumerate}
    \item We translate the option mid prices into equivalent Black--Scholes implied volatilities.
    \item We discard the resulting observations for in-the-money calls and puts, so that the remaining steps use data from only out-of-the-money put and call prices. To determine the at-the-money point, we use the strike $K$ at which call and put prices are equal (or closest to each other).
    \item For each trading date--expiration date pair, we fit a clamped cubic spline to the resulting implied volatility curve (i.e., the curve of implied volatility vs.\ strike price).
    \item Evaluate this spline at 1,901 strike prices, for S\&P index values ranging from 200 to 4,000 (so that the evaluation strike prices are $K=200,202,\ldots,4000$), to obtain a set of fitted implied-volatility values across this fine grid of possible strike prices for each $(t,T)$ pair.
    \item Invert the resulting smoothed implied volatility schedule back into call prices $\hat{q}_{t,T,K}$.
    \item Using a discrete-state version of the classic \textcite{BL78} formula, calculate the risk-neutral {CDF} for the date-$T$ index value at strike price $K$ as follows: $\p_t^*(V_{T}<K)=1+R_{t,T}^f (\hat{q}_{t,T,K}-\hat{q}_{t,T,K-2})/{2}$.
    \item Defining $V_{T,j,\max}$ and $V_{T,j,\min}$ to be the date-$T$ index values corresponding to the upper and lower bounds, respectively, of the bin defining return state $\theta_j$ (i.e., the upper and lower end of the five-percentage-point excess-return range defining a given return outcome), calculate the risk-neutral belief that state $\theta_j$ will be realized at date $T$ as $\pi_{t,j}^* = \p_t^*(V_{T}<V_{T,j,\max})-\p_t^*(V_{T}<V_{T,j,\min}).$ (The beliefs for states $\theta_1$ and $\theta_{10}$ then collect the tail probabilities for below -20\% and above 20\% returns, respectively.)
\end{enumerate}
We do this for states $\theta_1,\ldots,\theta_{10},$ where the return states are as defined and described in the text --- i.e., 5-log-point ranges of log excess returns from the first observable option trading date (within a year of expiration) to the expiration date --- for all trading dates under consideration. We then use the resulting histogram of risk-neutral beliefs for our tests.

We note that unlike AL (\citeyear{AL-WP}), we include beliefs over the tail return states $\theta_1$ and $\theta_{10}$, whereas AL discard them before calculating binarized beliefs $\pi_{t,j}^*/(\pi_{t,j}^*+\pi_{t,j+1}^*)$. AL discard them due to concerns over complications from potential changes in risk aversion over tail outcomes; given the binarization, small changes in risk aversion would have large effects on the measured binarized RN beliefs. But this is not the case for our analysis: since we just use the (non-binarized) histogram of beliefs, the tail states have very low probabilities and thus do not meaningfully affect the results. (Results are very similar when only including movement and uncertainty reduction for states 2 through 9.) This is another way in which just using the RN histogram, rather than continuing from above and calculating the binarized beliefs, helps minimize the potential effect of noise on our results.

\subsubsection*{Translating from Risk-Neutral to Physical Beliefs}

Given the RN beliefs as measured from above, we now describe the translation from RN to physical beliefs in greater detail. Assume there exists a representative investor (``the market'') with time-separable utility over the market index value.\footnote{These illustrative assumptions aid in the interpretation of our risk-aversion assumptions, but they are stronger than needed in general; see AL (\citeyear{AL-WP}) for a discussion.} Assume, as above, that the state space (the set of possible terminal index values $V_T$) is discrete, with states indexed by $j$ ($V_T = \theta_j$ for $j=1,2,\ldots,J$), and denote terminal utility by $U(V_T)$. The physical belief regarding the likelihood of state $j$ is $\pi_{t,j}$, and the risk-neutral belief is $\pi^*_{t,j}$. The two are related as follows:%\footnote{This is a multi-state generalization of equation~(5) of AL (\citeyear{AL-WP}), or see equation~(7) of \textcite{BP04}.}
\begin{equation}
\pi^*_{t,j} = \frac{U'(\theta_j)\pi_{t,j}}{\sum_{k} U'(\theta_k)\pi_{t,k}}.
\label{eq:transl}
\end{equation}
(This is a multi-state generalization of equation~(5) of AL \citeyear{AL-WP}, or see equation~(7) of \cite{BP04}.)
Our main translation assumes that $U'(V_T) = V_T^{-\gamma},$ corresponding to the assumption of power utility over the terminal index return, with constant relative risk aversion coefficient of $\gamma$. We then follow \textcite{BP04} in estimating $\gamma$ as the value under which the physical beliefs over the S\&P 500 value at the one-month horizon are well calibrated (i.e., unbiased); see \textcite{BP04} for details on the maximum likelihood estimation procedure.\footnote{Like \cite{BP04}, we obtain reasonable estimates of risk aversion of (with estimated $\hat\gamma<10$) given this calibration procedure.}

We then consider dozens of generalizations of this basic framework. First, we reparameterize \eqref{eq:transl} in terms of the \emph{ratio} of marginal utilities (or SDF realizations) across adjacent index states $\phi_j$, by substituting $U'(\theta_j) = \phi_j U'(\theta_{j-1})$. We then make a range of assumptions on the function $\phi_j$. We assume that $\phi_j$ varies by state $j$, either linearly or quadratically in $V_T$, and we estimate $\phi_j$ by maximum likelihood for each state; we assume that $\phi_j$ varies over time (either linearly or quadratically) or by horizon to expiration (as in \cite{L-WP2}); and then we consider interactions in which $\phi_j$ varies both by bin $j$ and over time. In all cases (as can be seen in Figure~\ref{fig:financealt}, the right panel of which contains one line for each parameterization), the movement and uncertainty reduction statistics are close to unchanged. (This is in contrast to the \emph{physical probabilities}, which do change depending on the parameterization; it is their evolution over time that is unchanged.)

\clearpage

\subsection*{Additional Empirical Results}

\begin{figure}[htb!]
\caption{Movement and Uncertainty Reduction for Sports Betting Data: 12 Time Chunks}
\label{fig:sports12}
\begin{center}
\vspace{-6.5mm}
    \centerline{\includegraphics[width=.925\textwidth]{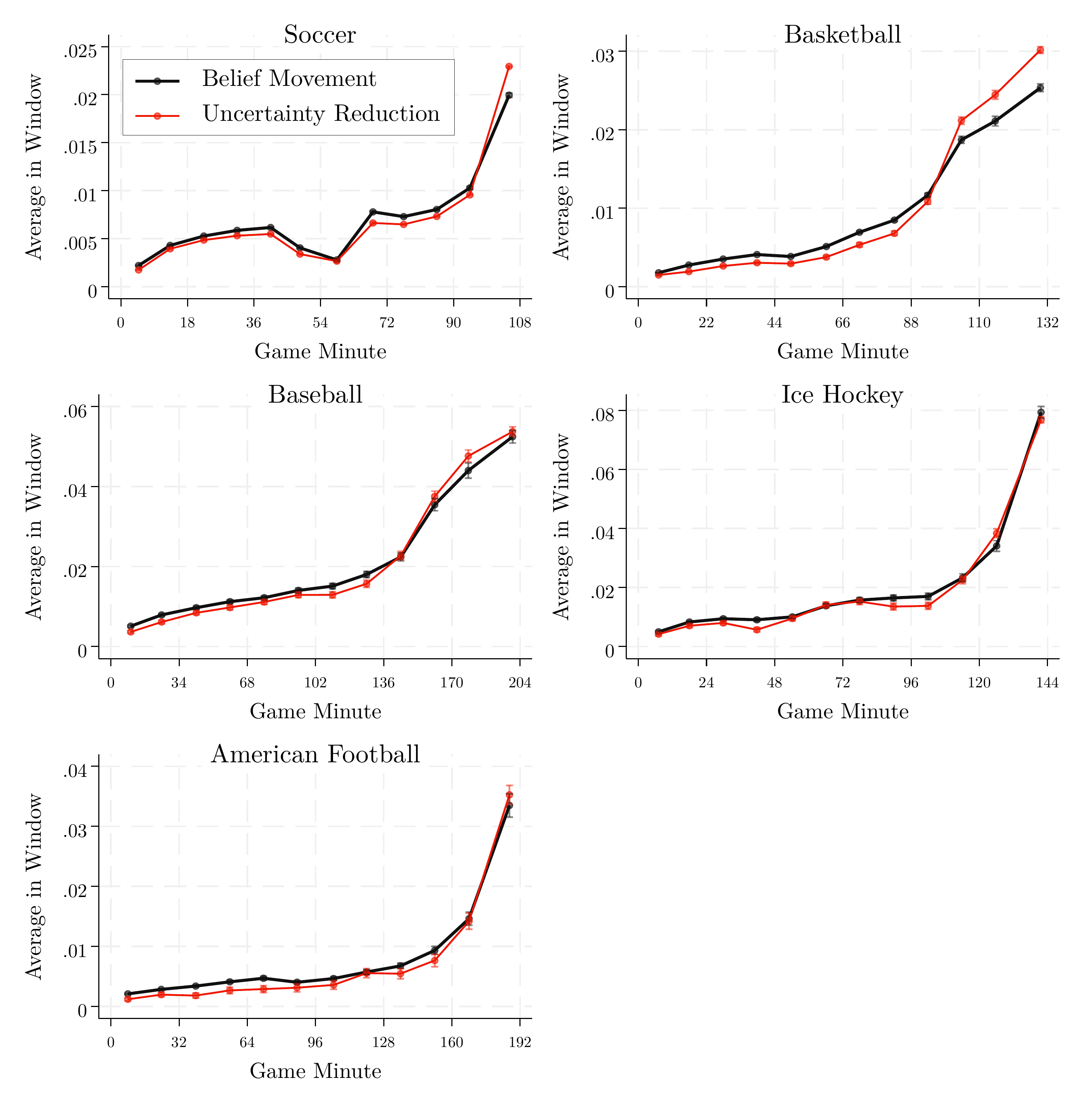}}
\end{center}
\begin{threeparttable}
\begin{tablenotes}
\begin{scriptsize}
\vspace{-15.5mm}
\item \textbf{Notes:} This figure replicates \Cref{fig:sports}, but with 12 equal-length time windows, rather than 24. See that figure's notes for details on construction.
\end{scriptsize}
\end{tablenotes}
\end{threeparttable}
\end{figure}

\begin{figure}[tb!]
\caption{Movement and Uncertainty Reduction for Finance Data: 12 Time Chunks}
\label{fig:finance12}
\begin{center}
\vspace{-6.5mm}
    \centerline{\includegraphics[width=.995\textwidth]{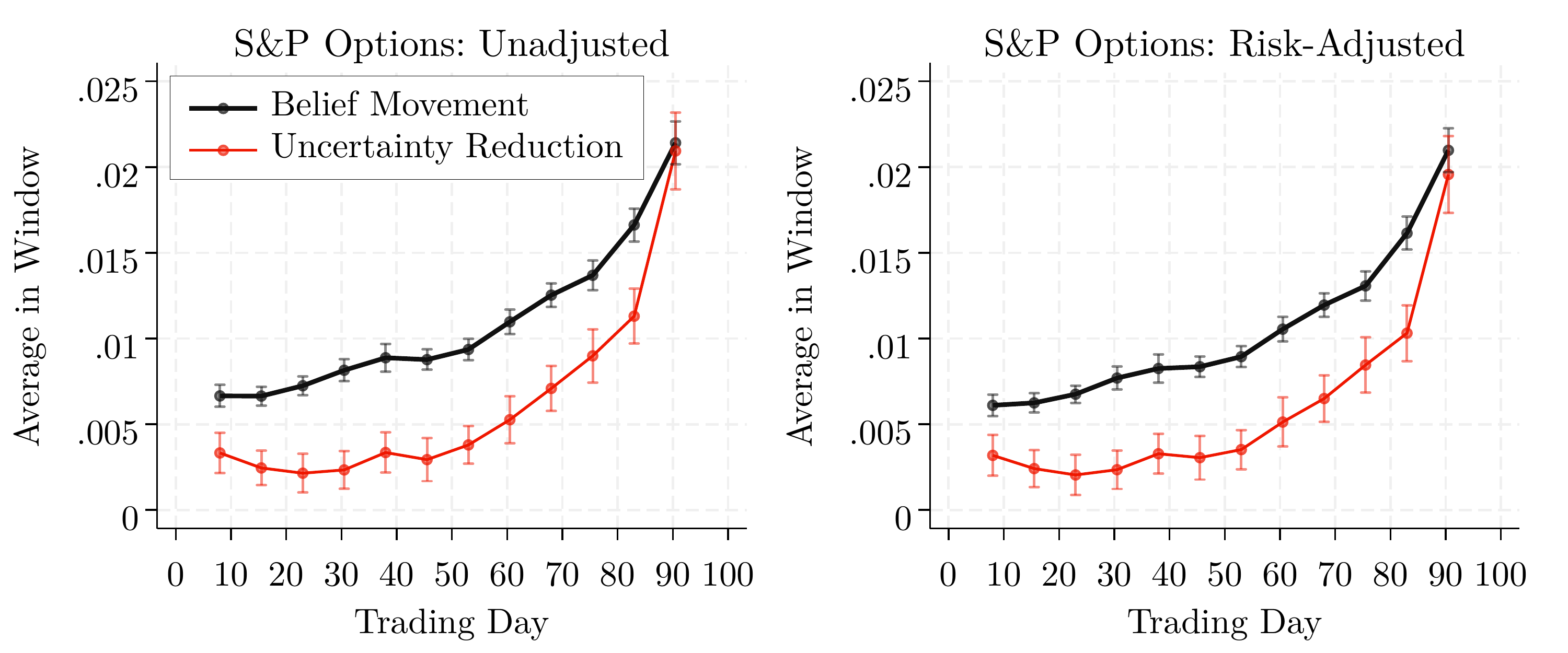}}
    %{combo_mrFinanceWithAll.png}}
\end{center}
\begin{threeparttable}
\begin{tablenotes}
\begin{scriptsize}
\vspace{-12.5mm}
\item \textbf{Notes:} This figure replicates \Cref{fig:finance}, but with 12 equal-length time windows, rather than 24. See that figure's notes for details on construction.
\end{scriptsize}
\end{tablenotes}
\end{threeparttable}
\end{figure}

\begin{figure}[htb!]
\caption{Movement and Uncertainty Reduction for Sports Betting Data: 36 Time Chunks}
\label{fig:sports36}
\begin{center}
\vspace{-6.5mm}
    \centerline{\includegraphics[width=.995\textwidth]{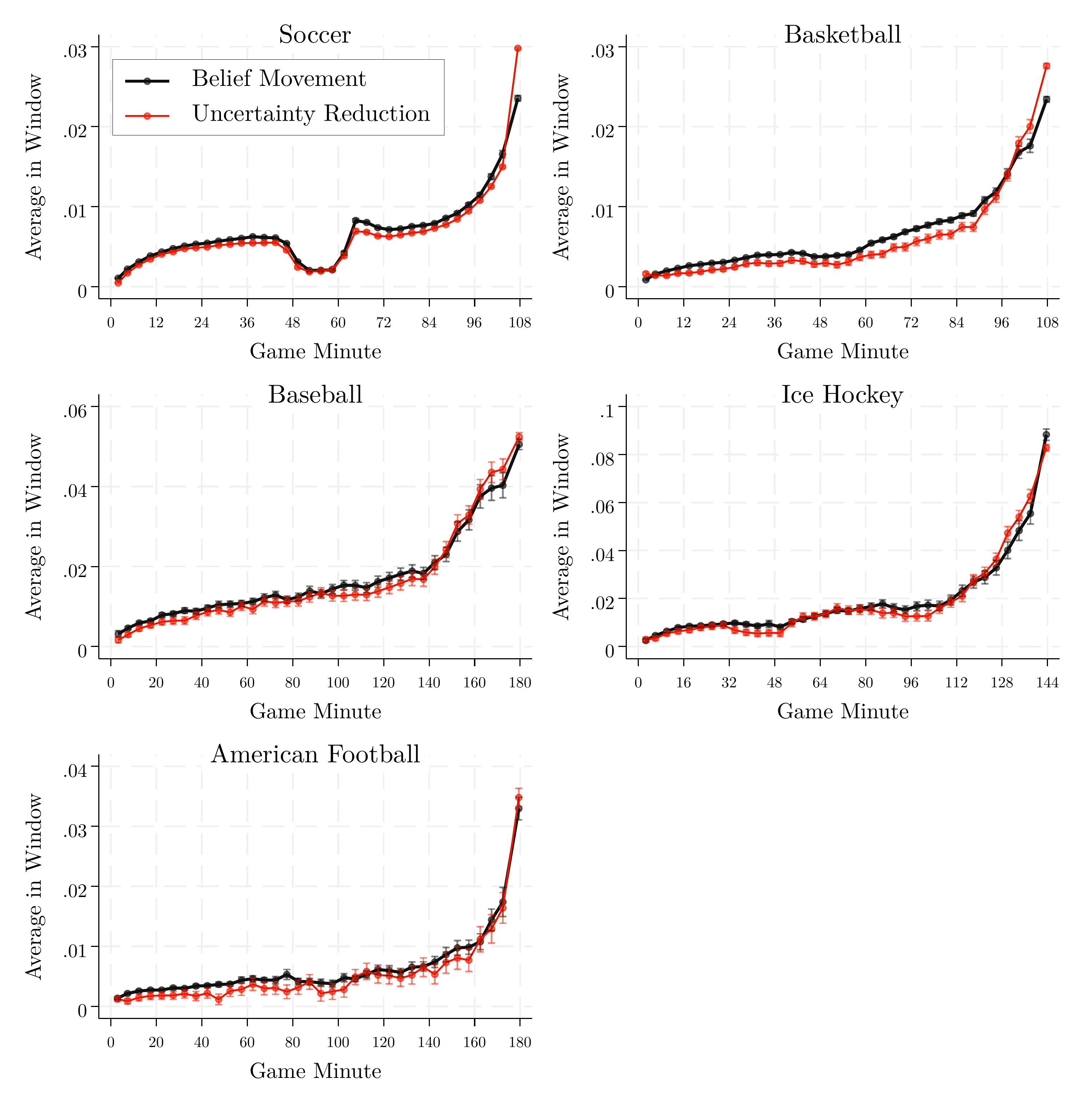}}
\end{center}
\begin{threeparttable}
\begin{tablenotes}
\begin{scriptsize}
\vspace{-15.5mm}
\item \textbf{Notes:} This figure replicates \Cref{fig:sports}, but with 36 equal-length time windows, rather than 24. See that figure's notes for details on construction.
\end{scriptsize}
\end{tablenotes}
\end{threeparttable}
\end{figure}

\begin{figure}[tb!]
\caption{Movement and Uncertainty Reduction for Finance Data: 36 Time Chunks}
\label{fig:finance36}
\begin{center}
\vspace{-6.5mm}
    \centerline{\includegraphics[width=.995\textwidth]{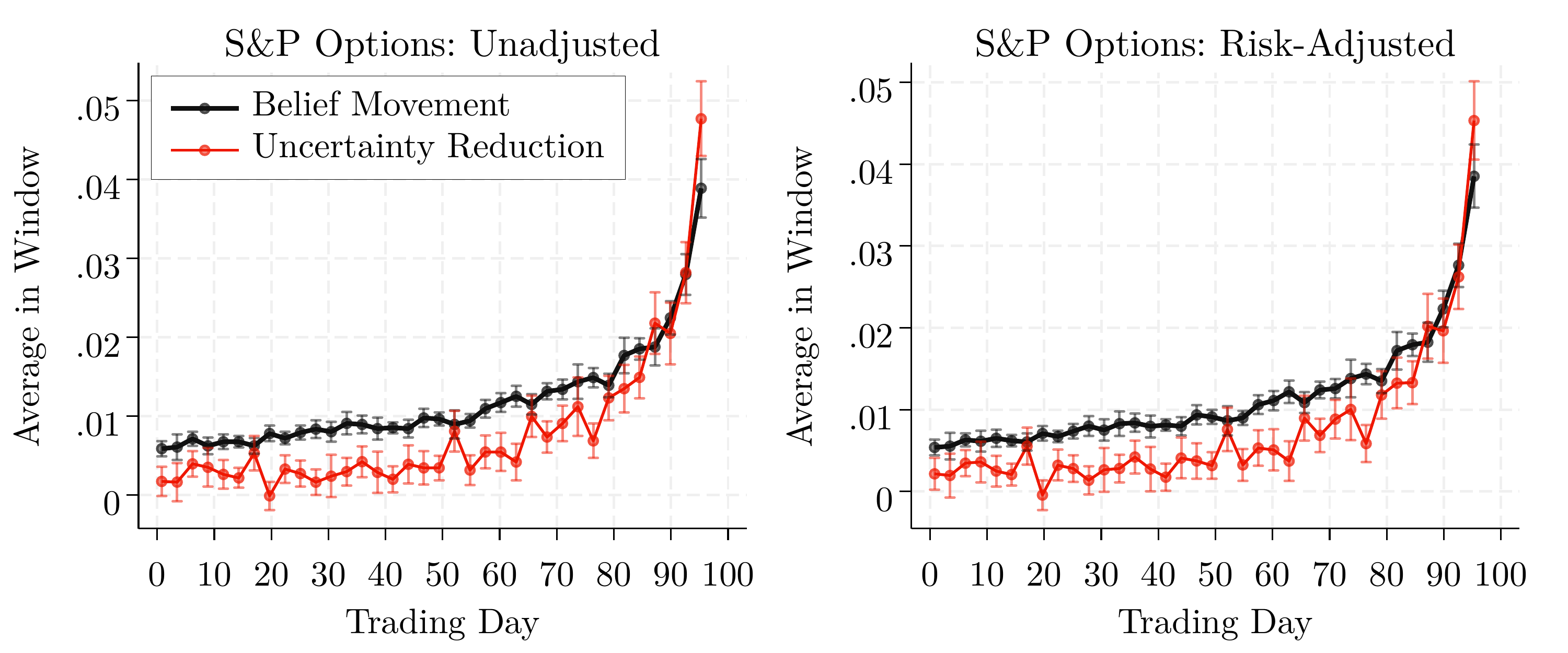}}
    %{combo_mrFinanceWithAll.png}}
\end{center}
\begin{threeparttable}
\begin{tablenotes}
\begin{scriptsize}
\vspace{-12.5mm}
\item \textbf{Notes:} This figure replicates \Cref{fig:finance}, but with 36 equal-length time windows, rather than 24. See that figure's notes for details on construction.
\end{scriptsize}
\end{tablenotes}
\end{threeparttable}
\end{figure}

\begin{figure}[htb!]
\caption{Movement and Uncertainty Reduction for Options: Alternative Adjustments}
\label{fig:financealt}
\begin{center}
\vspace{-4mm}
    \centerline{\includegraphics[width=.95\textwidth]{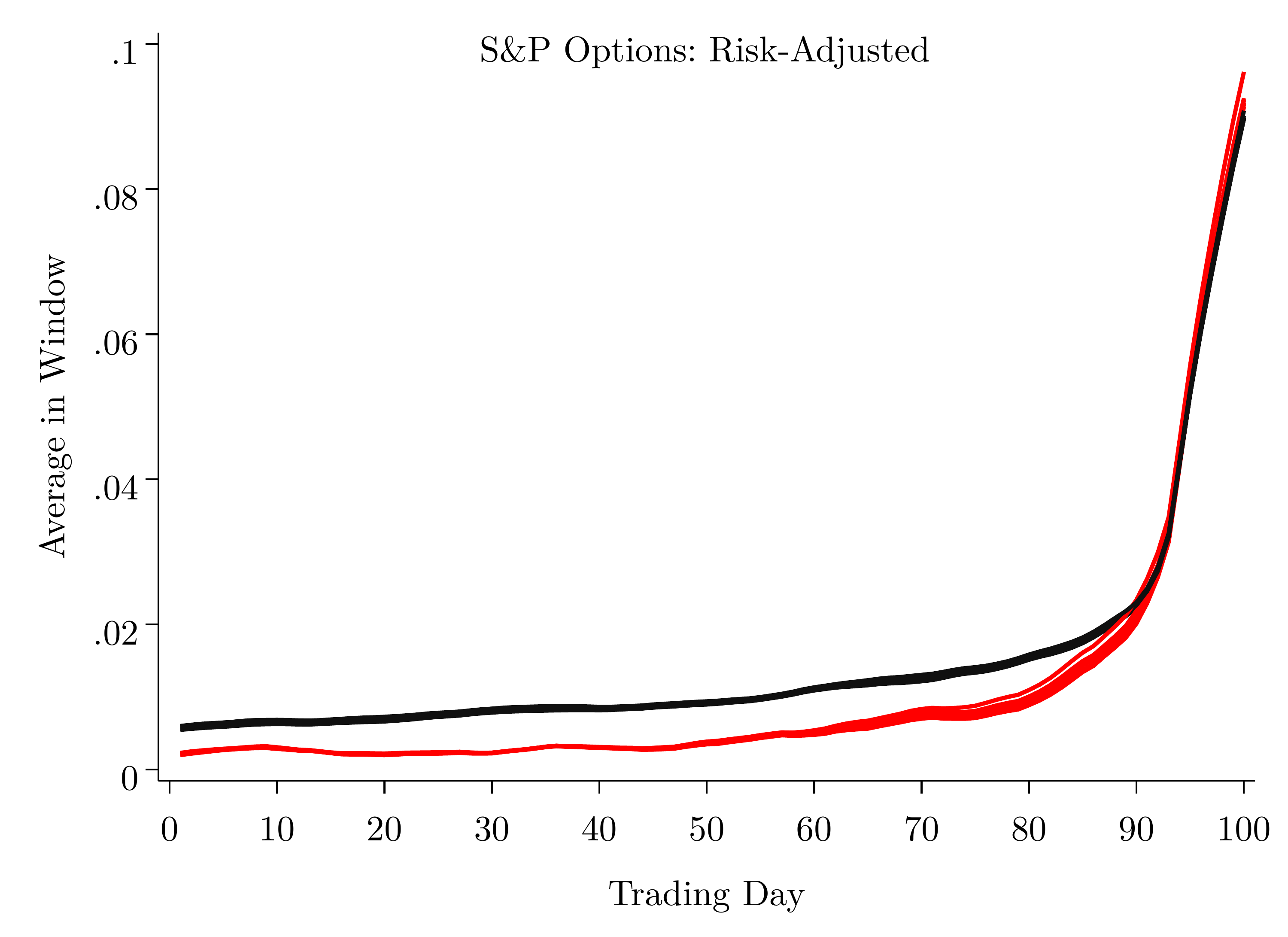}}
\end{center}
\begin{threeparttable}
\begin{tablenotes}
\begin{scriptsize}
\vspace{-10mm}
\item \textbf{Notes:} This figure replicates the right panel of \Cref{fig:finance} to show the smoothed average movement (black lines) and uncertainty reduction (red lines) statistics over time for the beliefs implied by option data, but with alternative risk adjustments.  Each line represents a different method to calculate risk-adjusted beliefs from the raw, unadjusted risk-neutral beliefs, as described in Appendix~\ref{app:details}. Some aspects of the figure (including confidence intervals) are omitted to enable a clear view of the range of plotted lines across risk adjustments. While the different risk-adjustment methods do lead to different inferred beliefs, the broad pattern of movement and uncertainty curves is very similar across the methods, as the curves are close to overlapping in most cases.
\end{scriptsize}
\end{tablenotes}
\end{threeparttable}
\end{figure}

\begin{figure}[tb!]
\caption{Movement and Uncert.\ Red.\ for Finance: Binarized, Noise-Corrected Beliefs}
\label{fig:financeBinary}
\begin{center}
\vspace{-4mm}
    \centerline{\includegraphics[width=.95\textwidth]{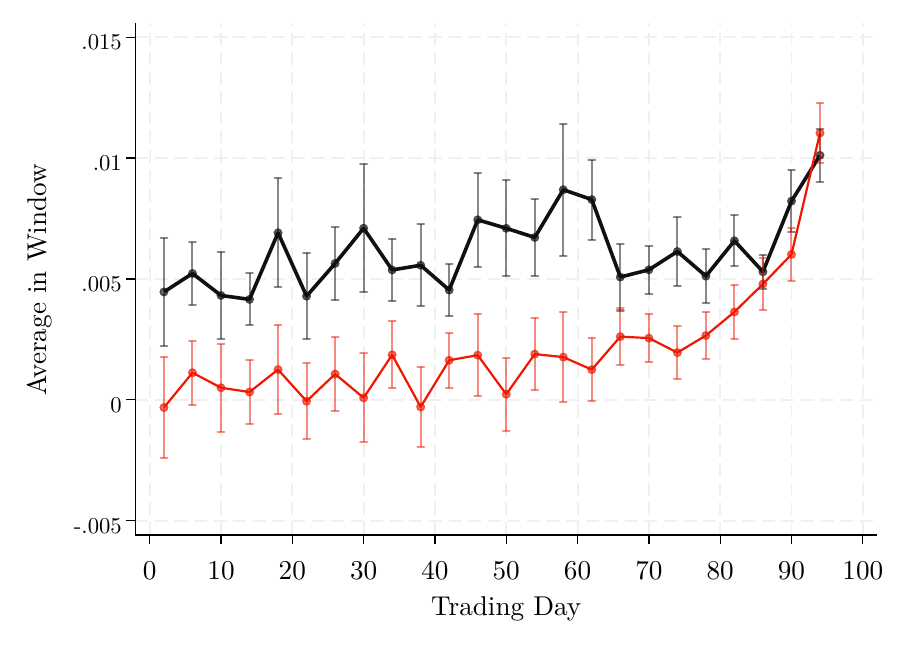}}
    %{combo_mrFinanceWithAll.png}}
\end{center}
\begin{threeparttable}
\begin{tablenotes}
\begin{scriptsize}
\vspace{-14.5mm}
\item \textbf{Notes:} This figure replicates the left panel of \Cref{fig:finance}, but using the binarized and noise-corrected risk-neutral beliefs data from \textcite{AL-WP}. Belief movement is plotted in black, and uncertainty reduction in red. Data are not adjusted for risk aversion. See \Cref{fig:finance} for details on the plot, and see Section~6 and Online Appendix~C.6 of AL (\citeyear{AL-WP}) for details on the noise correction and binarization.
%\par
%The left-hand panel uses the implied unadjusted (risk-neutral) beliefs. The right-hand panel uses a risk adjustment described in the text. Movement is greater than uncertainty at the start of the contract. At the later stages, movement is lower than uncertainty reduction. \par
\end{scriptsize}
\end{tablenotes}
\end{threeparttable}
\end{figure}

\begin{figure}[tb!]
\caption{Movement and Uncertainty Reduction for Finance Data: Post-2000}
\label{fig:financeAfter2000}
\begin{center}
\vspace{-6.5mm}
    \centerline{\includegraphics[width=.995\textwidth]{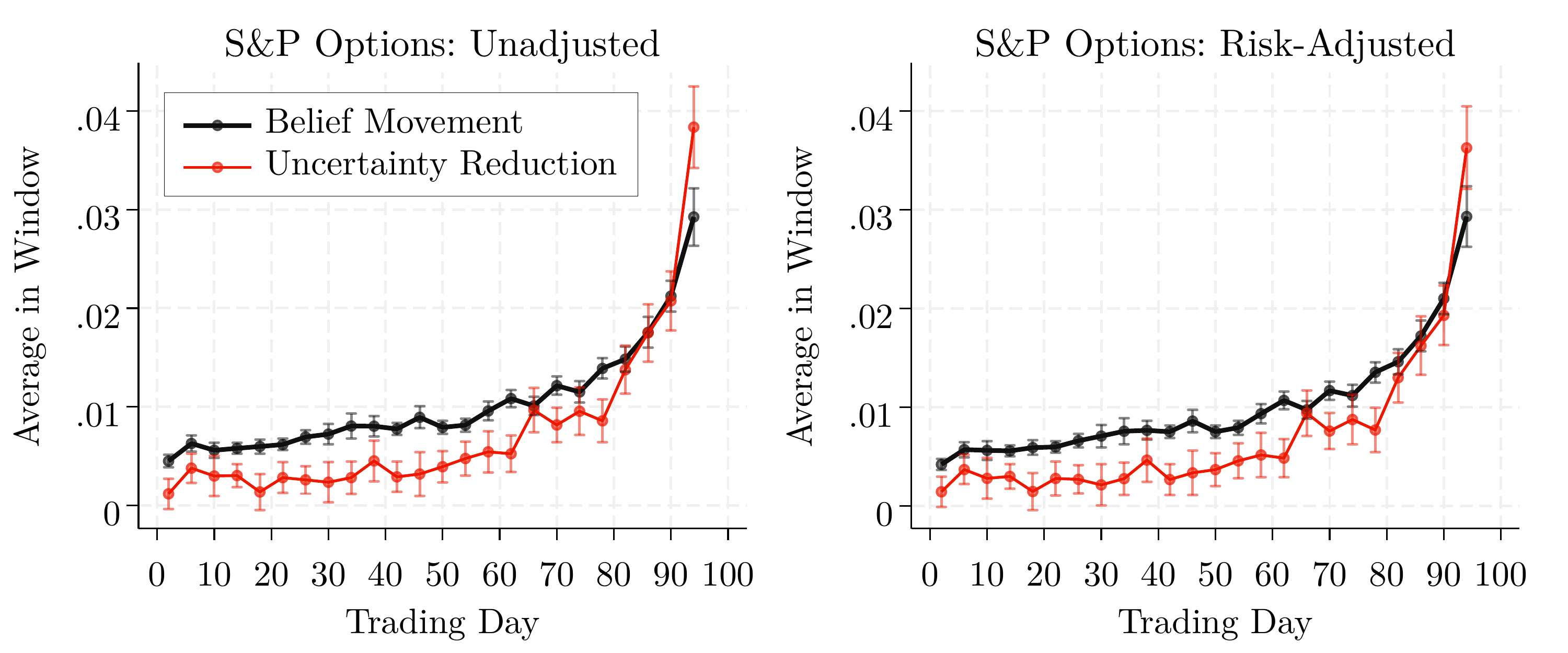}}
    %{combo_mrFinanceWithAll.png}}
\end{center}
\begin{threeparttable}
\begin{tablenotes}
\begin{scriptsize}
\vspace{-12.5mm}
\item \textbf{Notes:} This figure replicates \Cref{fig:finance}, using only data after the year 2000. See that figure's notes for details on construction. The figure demonstrates that our option results are robust to not including the early part of the sample, which contains somewhat noisier option data (see AL \citeyear{AL-WP}).
\end{scriptsize}
\end{tablenotes}
\end{threeparttable}
\end{figure}

\begin{figure}[tb!]
\caption{Movement and Uncertainty Reduction for Finance Data: Post-2010}
\label{fig:financeAfter2010}
\begin{center}
\vspace{-6.5mm}
    \centerline{\includegraphics[width=.995\textwidth]{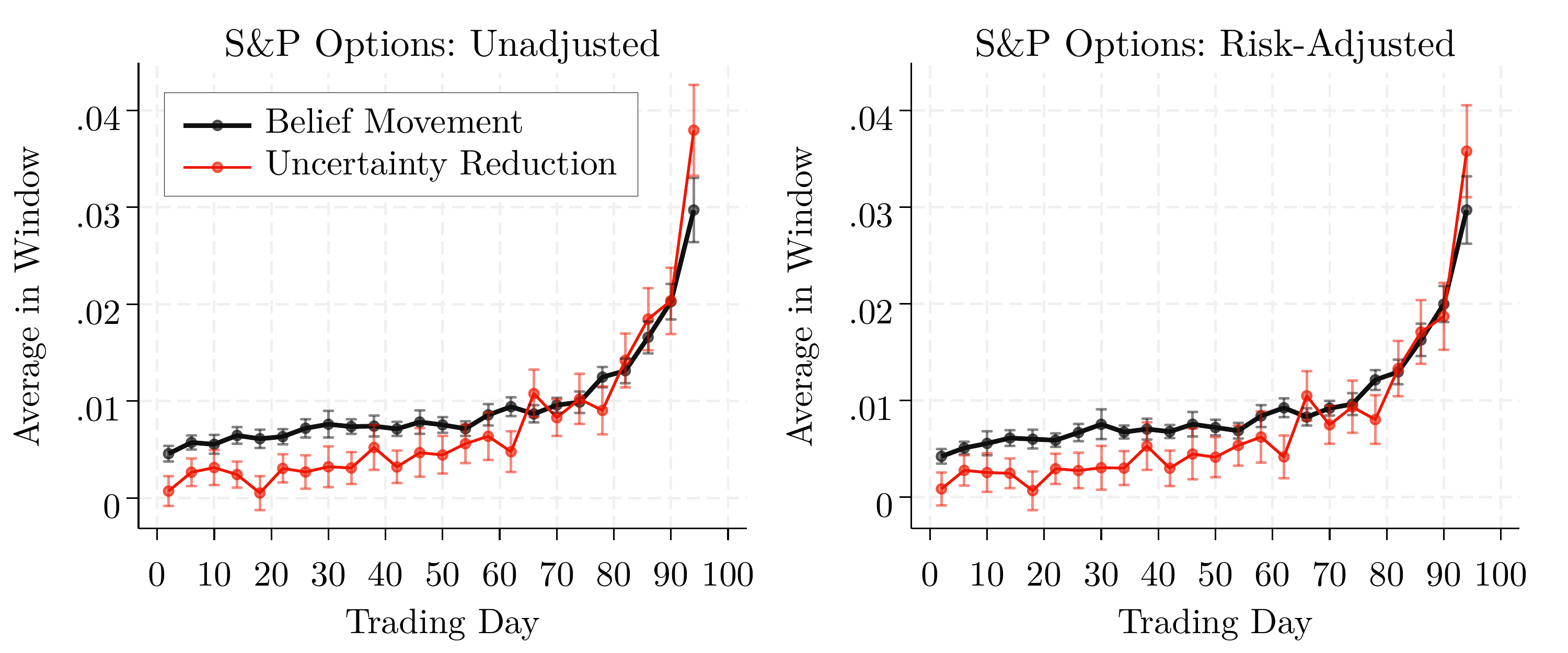}}
    %{combo_mrFinanceWithAll.png}}
\end{center}
\begin{threeparttable}
\begin{tablenotes}
\begin{scriptsize}
\vspace{-12.5mm}
\item \textbf{Notes:} This figure replicates \Cref{fig:finance}, using only data after 2010. See that figure's notes for details on construction. Along with \Cref{fig:financeAfter2000}, this figure demonstrates that our option results are robust across subsamples.
\end{scriptsize}
\end{tablenotes}
\end{threeparttable}
\end{figure}

We now provide a set of robustness results  (Figures \ref{fig:sports12}--\ref{fig:financeAfter2010} and Tables~\ref{tab:mrRegResults12}--\ref{tab:mrRegResults36}) for the betting and finance data. As described in the text, we present figures and regression tables for when the data is split into either 12 or 36 time chunks. For the options data, we also show results of different risk adjustments, use of binarized noise-corrected data, and in subsamples.

\subsubsection*{Different Time Windows}

\clearpage

%\begin{center}
\begin{table*}[t!]
\centering
\begin{threeparttable}%[t!]
\onehalfspacing
\begin{small}
\caption{Regressions of Movement on Uncertainty Reduction: 12 Time Chunks}
\begin{tabular}{l*{8}{c}}
\toprule
  \hspace{-2pt}\raisebox{-7pt}[\height][\depth]{Dep. Var.:}   &\multicolumn{5}{c}{Sports}  &\multicolumn{2}{c}{Finance} \\ \cmidrule(lr){2-6} \cmidrule(lr){7-8}
{Movement} &\multicolumn{1}{c}{Soccer}&\multicolumn{1}{c}{Basketball}&\multicolumn{1}{c}{Baseball}&\multicolumn{1}{c}{Hockey}&\multicolumn{1}{c}{Football} &\multicolumn{1}{c}{Raw} &\multicolumn{1}{c}{Risk-Adj.}\\
\midrule
Constant              &          0.0014&      0.0018&       0.0024&       0.0013&       0.0015&       0.0060&       0.0054\\
                      &        (0.0003)&    (0.0003)&     (0.0004)&     (0.0009)&     (0.0002)&     (0.0005)&     (0.0005)\\
Uncert.\ Red.\       &           0.839&       0.797&        0.903&        0.987&        0.912&        0.796&        0.861\\
                      &         (0.006)&     (0.007)&      (0.012)&      (0.012)&      (0.027)&      (0.054)&      (0.063)\\
\midrule
%$\sqrt{\text{Constant}}$&          0.037&        0.043&        0.049&       0.036&        0.039&        0.067&       0.068\\
$R^2$                   &          0.984&        0.991&        0.996&       0.990&        0.997&        0.945&       0.941\\
Time Chunks      &             12&           12&           12&          12&           12&           12&          12\\
Events                  &          175,026&        48,430&        16,536&       19,445&        3,212&          955&         955\\
Belief Obs.\            &      4,589,289&      867,567&      166,346&     109,751&       86,193&       58,864&      58,864\\
$p$-val: $\!\text{Const} =0$     &         <0.001&       <0.001&       <0.001&      <0.001&        <0.001&       <0.001&      <0.001\\
$p$-val: $\!\text{Slope} =1$     &          <0.001&      <0.001&       <0.001&      0.274&        0.002&       0.004&      0.025\\
\bottomrule
\end{tabular}

\label{tab:mrRegResults12}
\end{small}
\begin{tablenotes}[para,flushleft]
\begin{scriptsize}
 \textbf{Notes:} This table replicates \Cref{tab:mrRegResults}, but with 12 equal-length time windows, rather than 24. See that table's notes for details on estimation and interpretation.
\end{scriptsize}
\vspace{5mm}
\end{tablenotes}
\end{threeparttable}
\end{table*}
%\end{center}

\clearpage

\clearpage

%\begin{center}
\begin{table*}[t!]
\centering
\begin{threeparttable}%[t!]
\onehalfspacing
\begin{small}
\caption{Regressions of Movement on Uncertainty Reduction: 36 Time Chunks}
\begin{tabular}{l*{8}{c}}
\toprule
  \hspace{-2pt}\raisebox{-7pt}[\height][\depth]{Dep. Var.:}   &\multicolumn{5}{c}{Sports}  &\multicolumn{2}{c}{Finance} \\ \cmidrule(lr){2-6} \cmidrule(lr){7-8}
{Movement} &\multicolumn{1}{c}{Soccer}&\multicolumn{1}{c}{Basketball}&\multicolumn{1}{c}{Baseball}&\multicolumn{1}{c}{Hockey}&\multicolumn{1}{c}{Football} &\multicolumn{1}{c}{Raw} &\multicolumn{1}{c}{Risk-Adj.}\\
\midrule
Constant              &          0.0014&      0.0016&       0.0027&       0.0020&       0.0015&       0.0063&       0.0058\\
                      &        (0.0001)&    (0.0001)&     (0.0002)&     (0.0002)&     (0.0001)&     (0.0003)&     (0.0003)\\
Uncert.\ Red.\       &           0.847&       0.849&        0.883&        0.925&        0.920&        0.705&        0.751\\
                      &         (0.003)&     (0.008)&      (0.015)&      (0.013)&      (0.026)&      (0.035)&      (0.040)\\

\midrule
%$\sqrt{\text{Constant}}$&          0.037&        0.043&        0.049&       0.036&        0.039&        0.067&       0.068\\
$R^2$                   &          0.955&        0.974&        0.993&       0.975&        0.982&        0.932&       0.928\\
Time Chunks      &             36&           36&           36&          36&           36&           36&          36\\
Events                  &          175,026&        48,430&        16,536&       19,445&        3,212&          955&         955\\
Belief Obs.\            &      4,589,289&      867,567&      166,346&     109,751&       86,193&       58,864&      58,864\\
$p$-val: $\!\text{Const} =0$    &         <0.001&       <0.001&       <0.001&      <0.001&        0.051&       <0.001&      <0.001\\
$p$-val: $\!\text{Slope} =1$       &          <0.001&      <0.001&       <0.001&      <0.001&        0.054&       <0.001&      <0.001\\
\bottomrule
\end{tabular}

\label{tab:mrRegResults36}
\end{small}
\begin{tablenotes}[para,flushleft]
\begin{scriptsize}
 \textbf{Notes:} This table replicates \Cref{tab:mrRegResults}, but with 36 equal-length time windows, rather than 24. See that table's notes for details on estimation and interpretation.
\end{scriptsize}
\vspace{5mm}
\end{tablenotes}
\end{threeparttable}
\end{table*}

\clearpage

\subsubsection*{Further Robustness Results for Options Data}

%\clearpage

\clearpage

\clearpage

\clearpage

\clearpage

\section*{\LARGE Supplementary Appendix}

\section{Experiment Study Materials}
\label{app:screenshots}

\subsection{Study 1a}
\subsubsection*{Overview and Instructions}
\begin{center}
    \includegraphics[width=\textwidth]{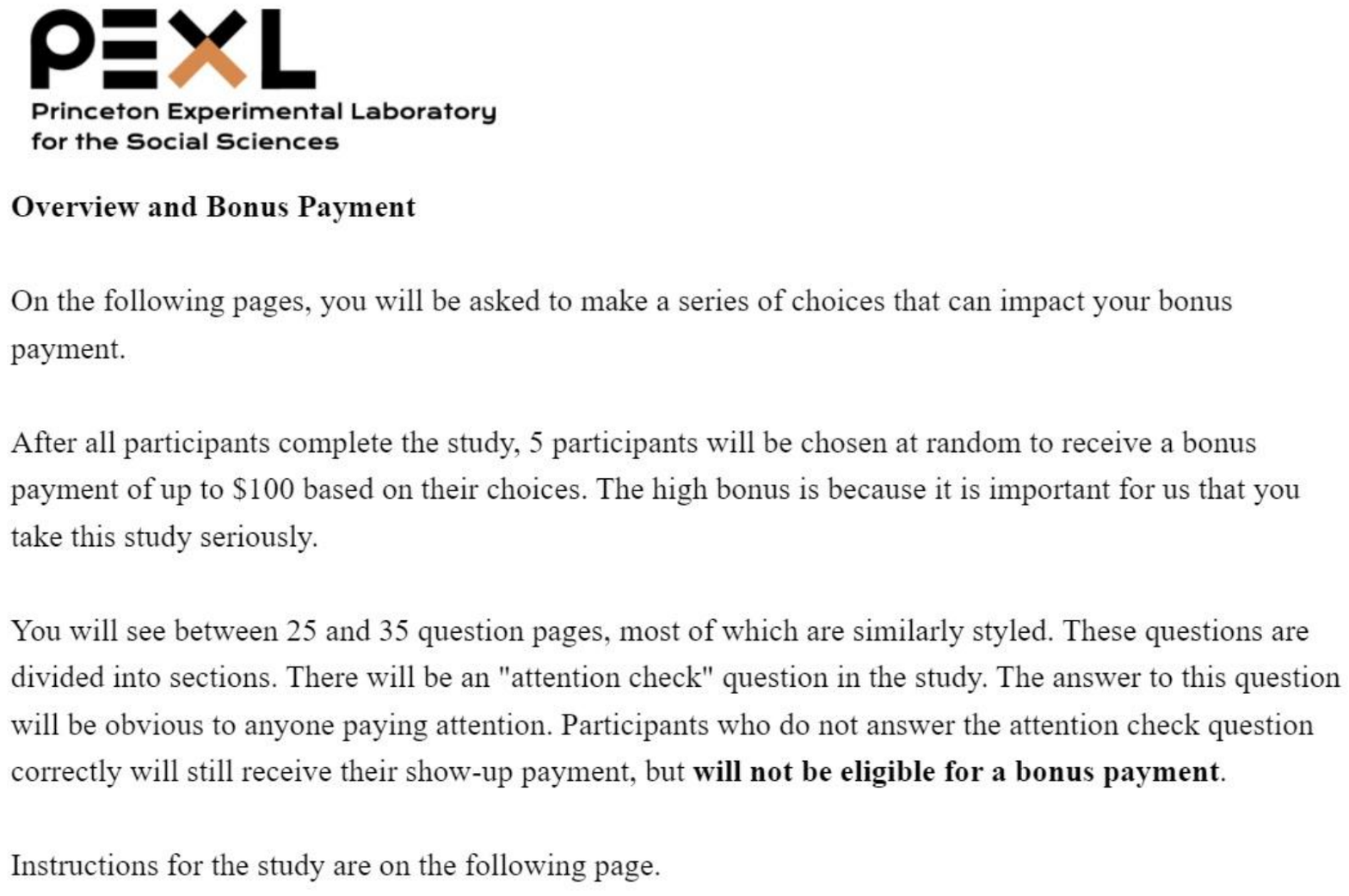}
    \newpage
    \includegraphics[height=0.97\textheight]{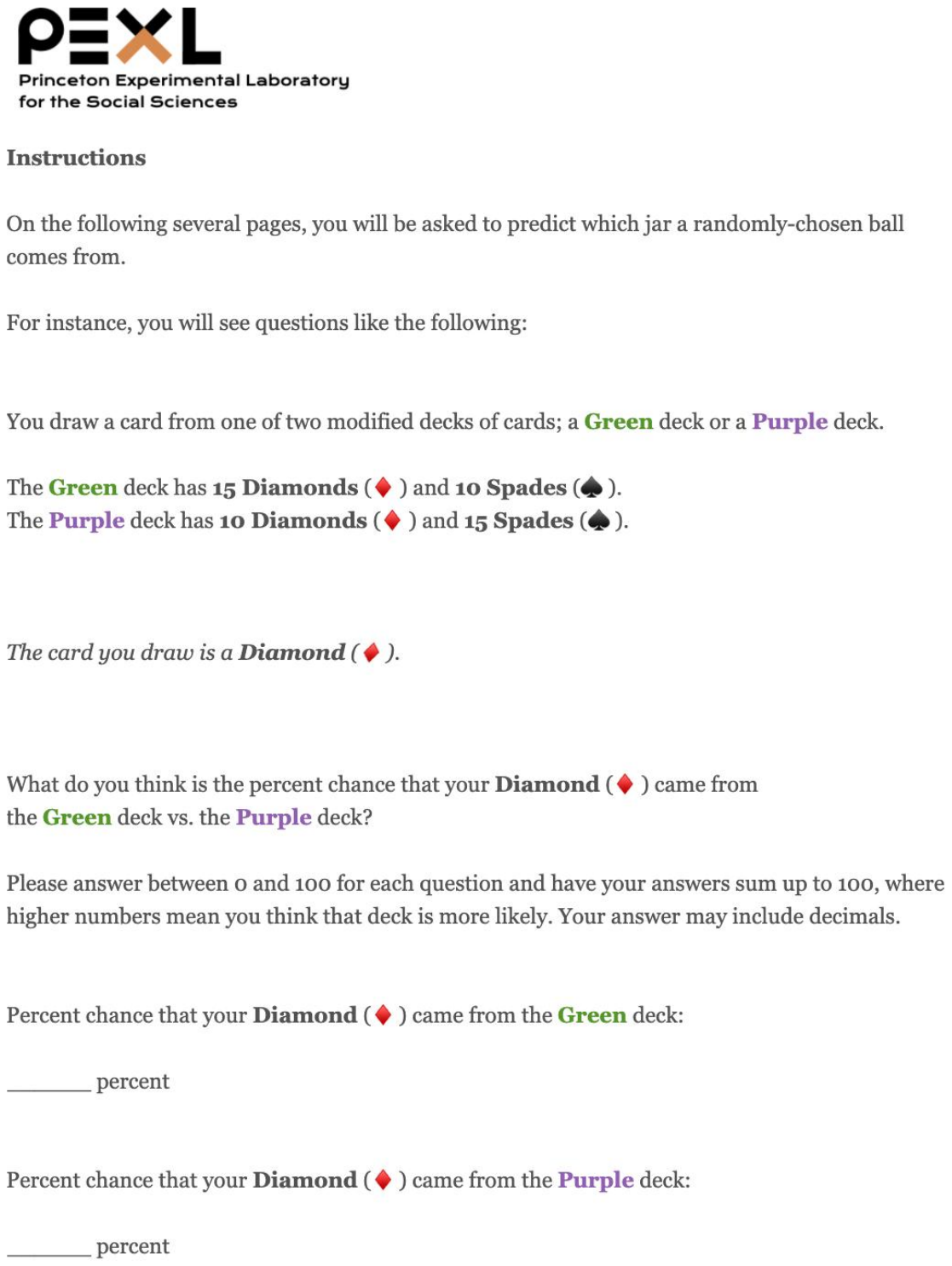}
    \newpage
    \includegraphics[width=\textwidth]{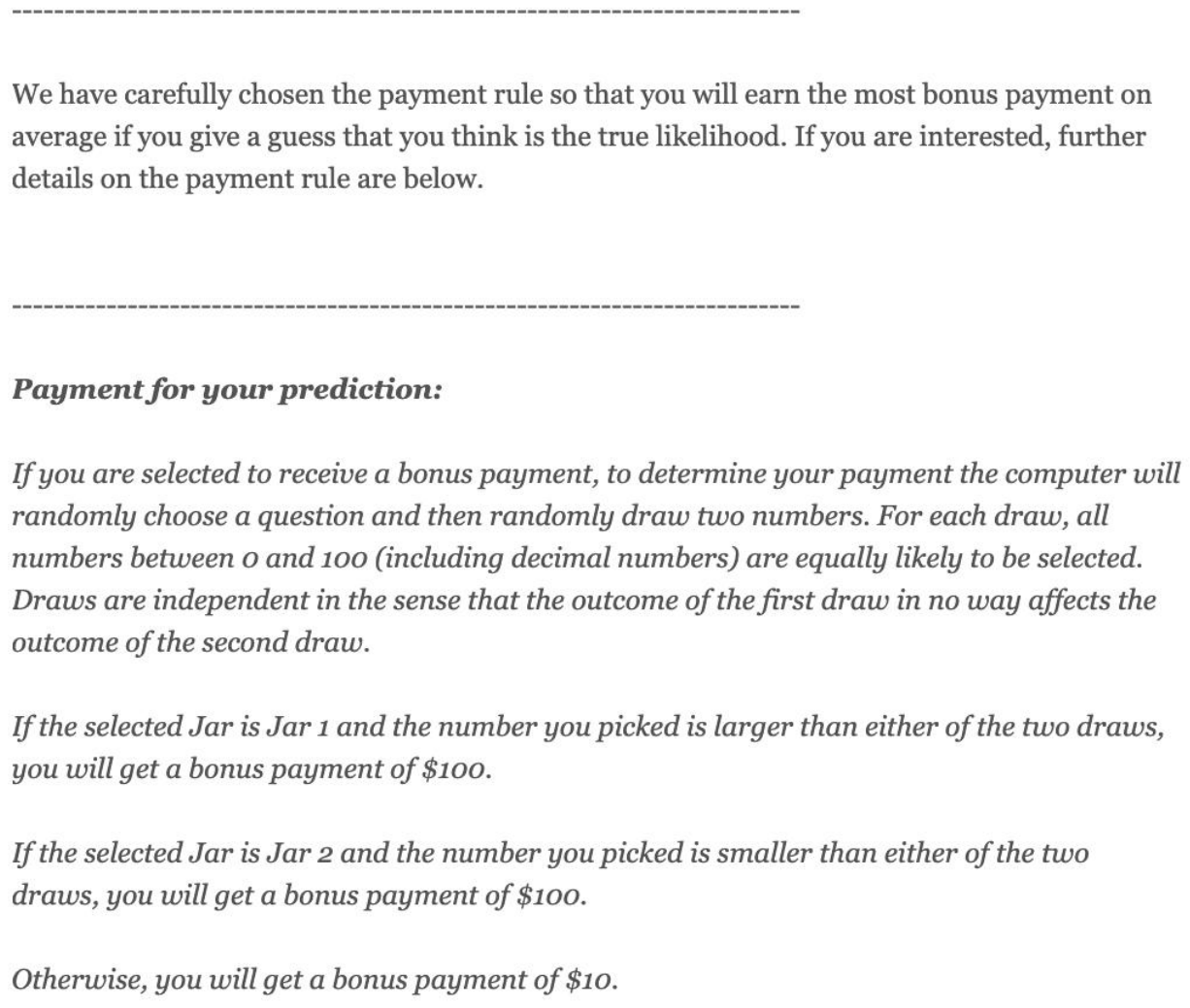}
\end{center}
    \newpage
\subsubsection*{Main Decision Screen}
\begin{center}
    \includegraphics[width=\textwidth]{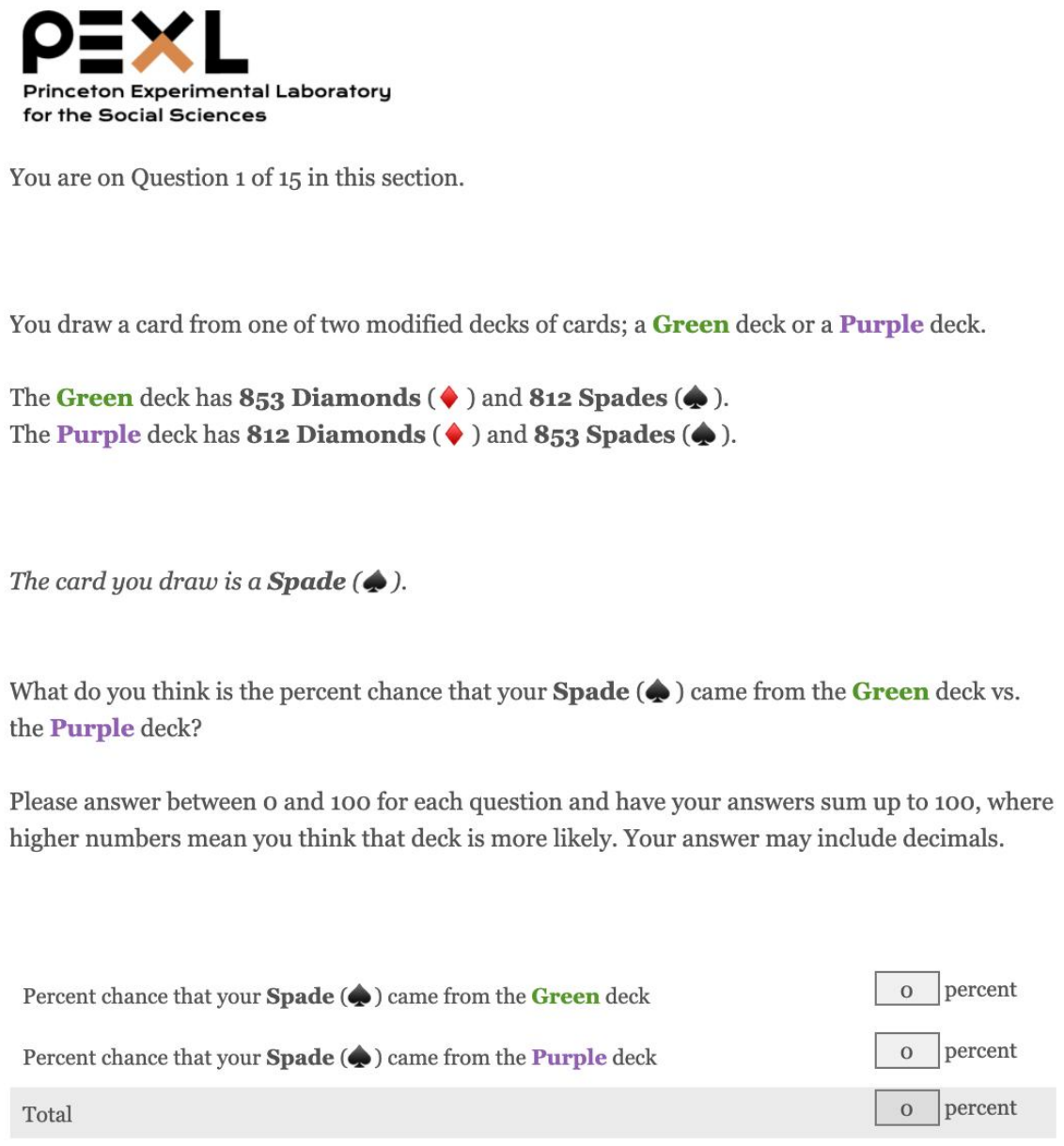}
\end{center}
    \newpage
\subsubsection*{Attention Check}
\begin{center}
    \includegraphics[width=\textwidth]{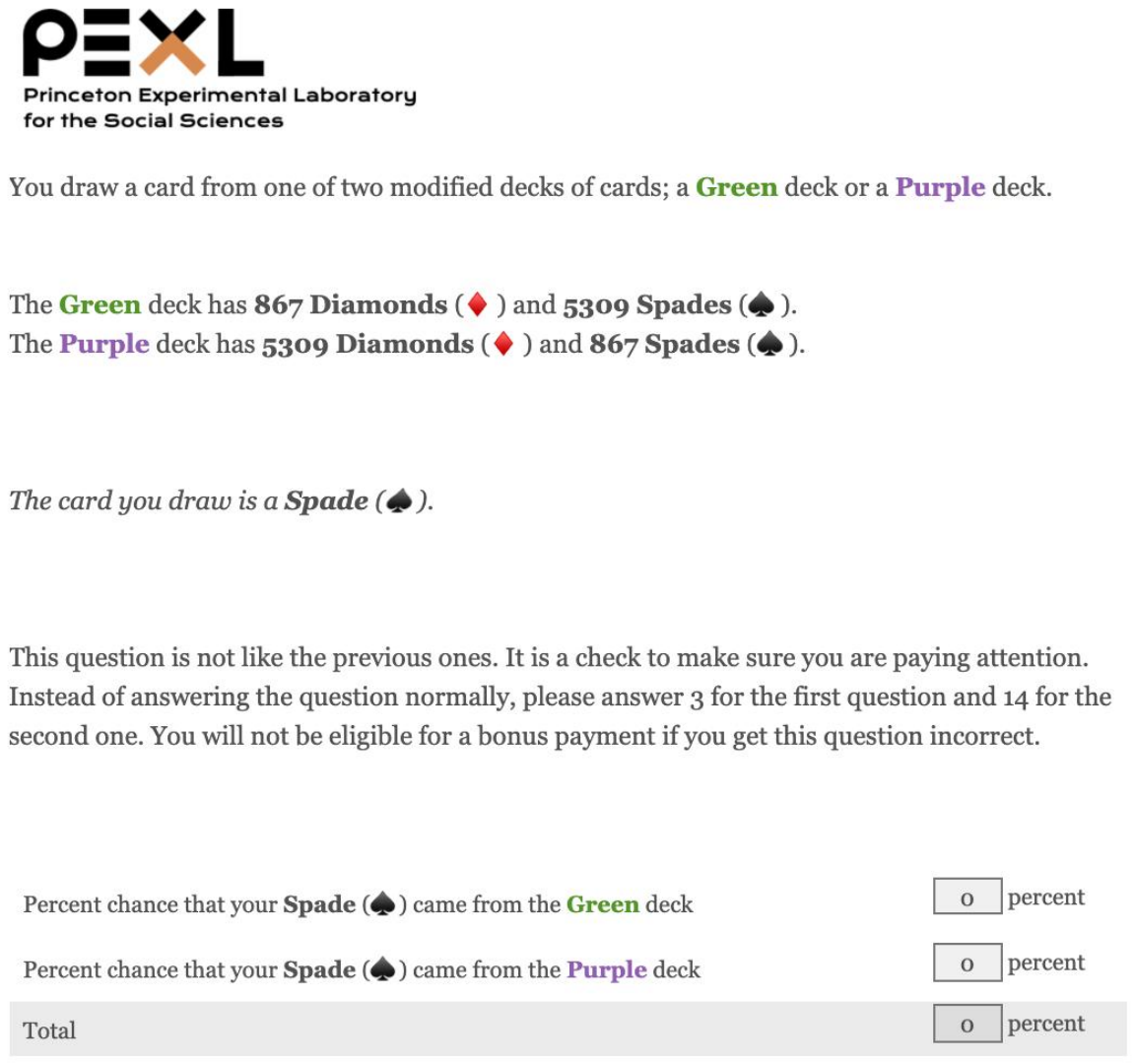}
\end{center}
    \newpage
\subsubsection*{Unknown Signal Strength}
\begin{center}
    \includegraphics[height=\textheight]{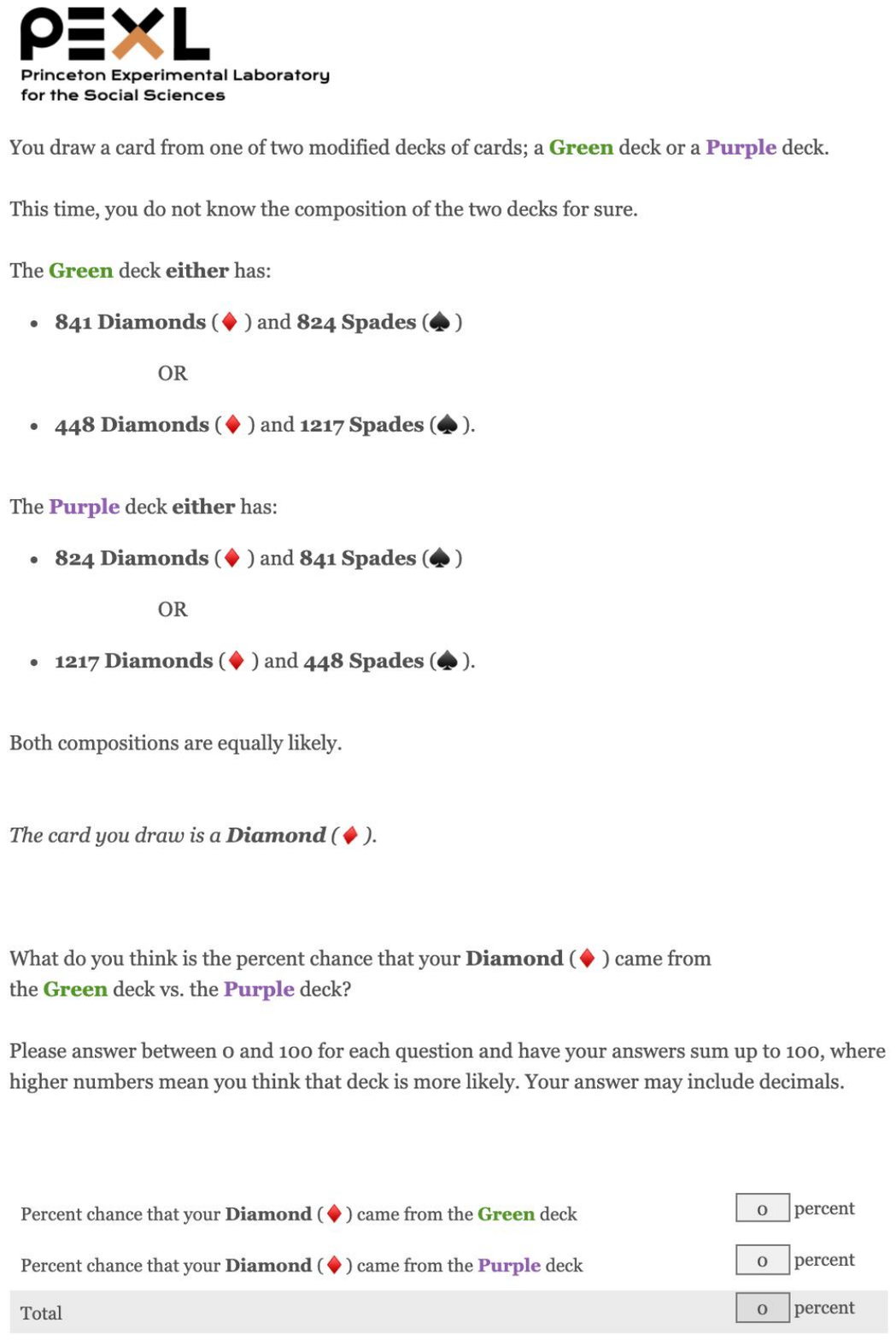}
\end{center}
    \newpage
\subsubsection*{Demand for Information}
\begin{center}
    \includegraphics[height=\textheight]{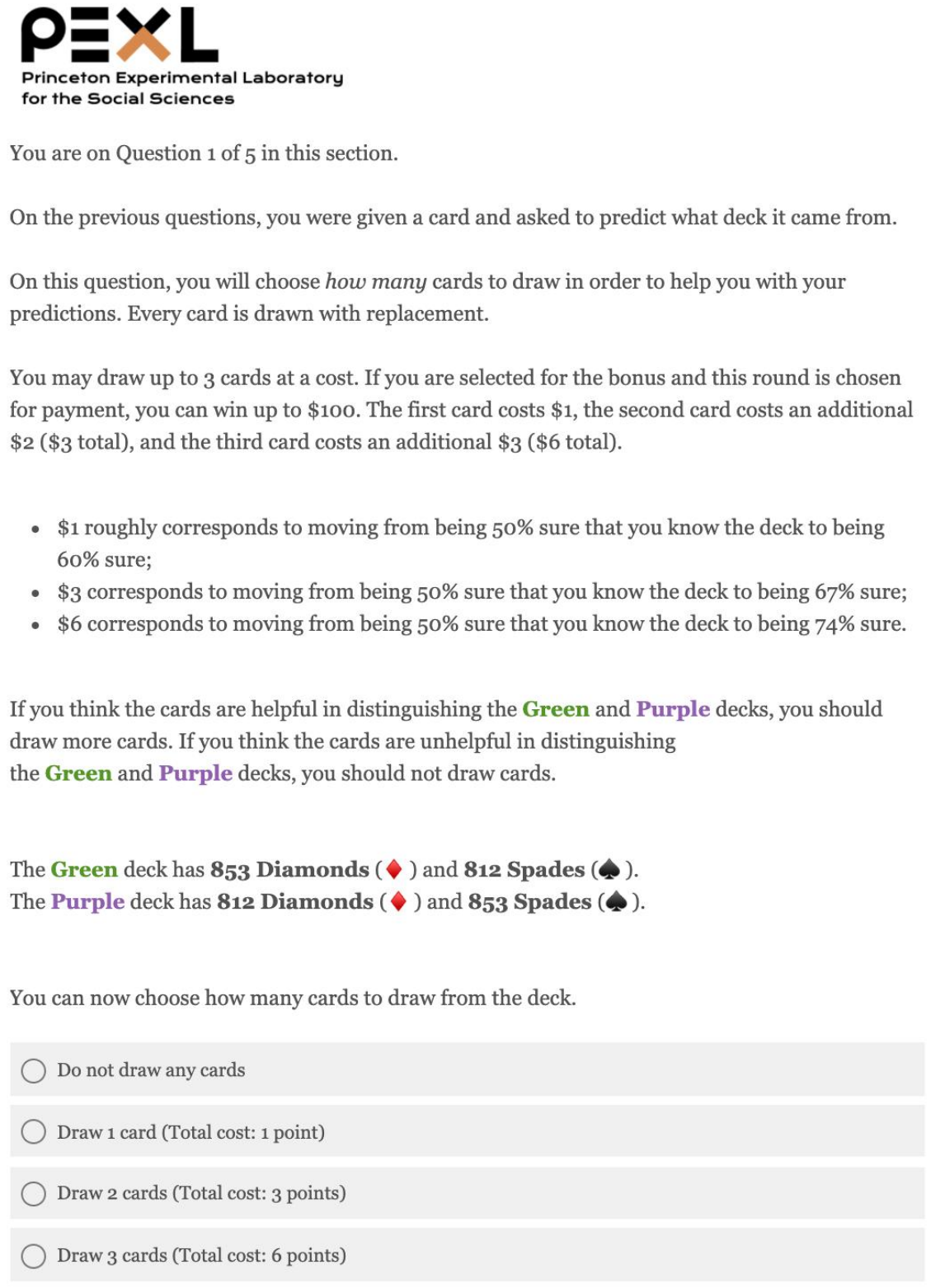}
\end{center}
   \newpage
    \includegraphics[width=\textwidth]{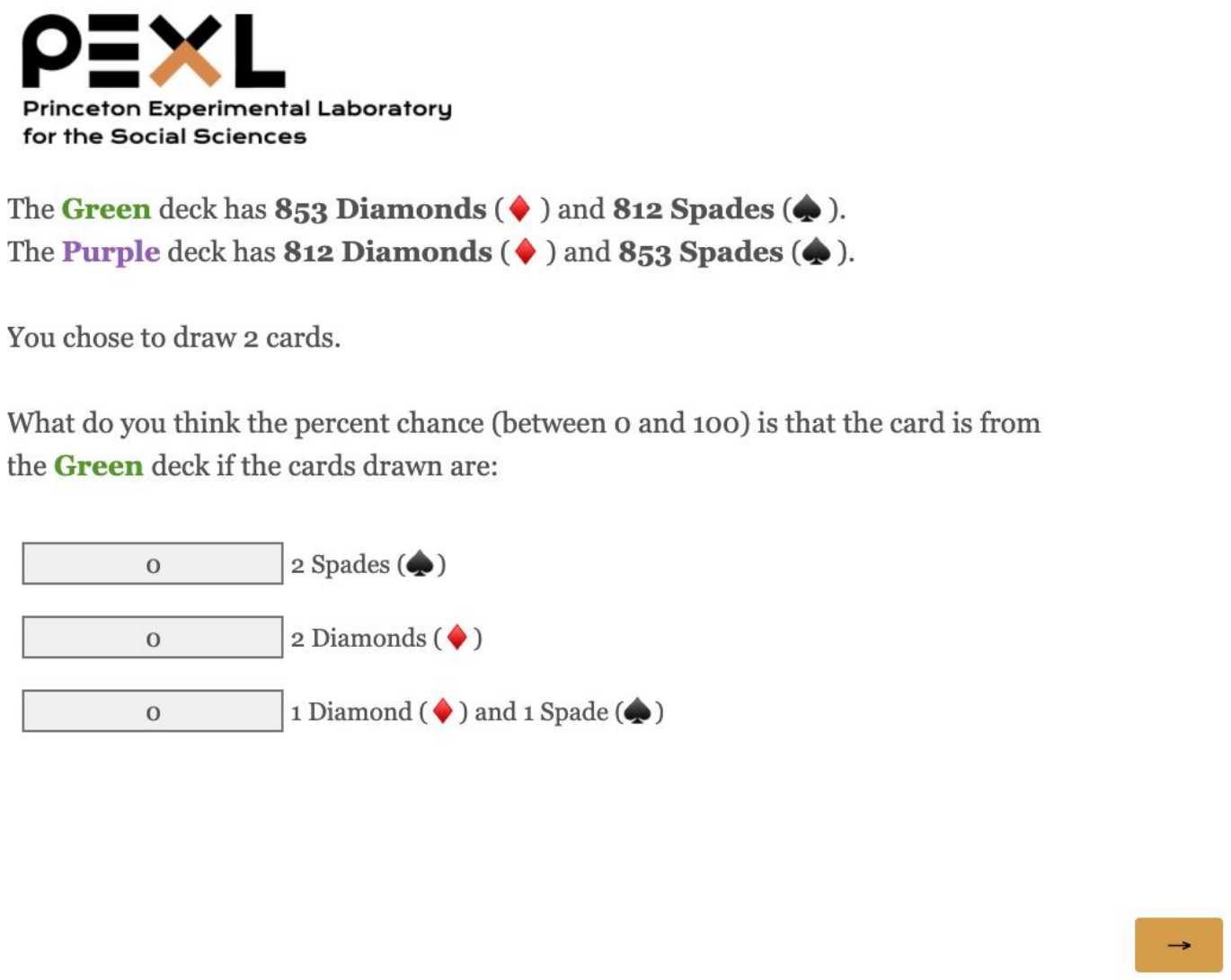}
    \newpage
\subsubsection*{Demographics}
\begin{center}
    \includegraphics[height=1\textheight]{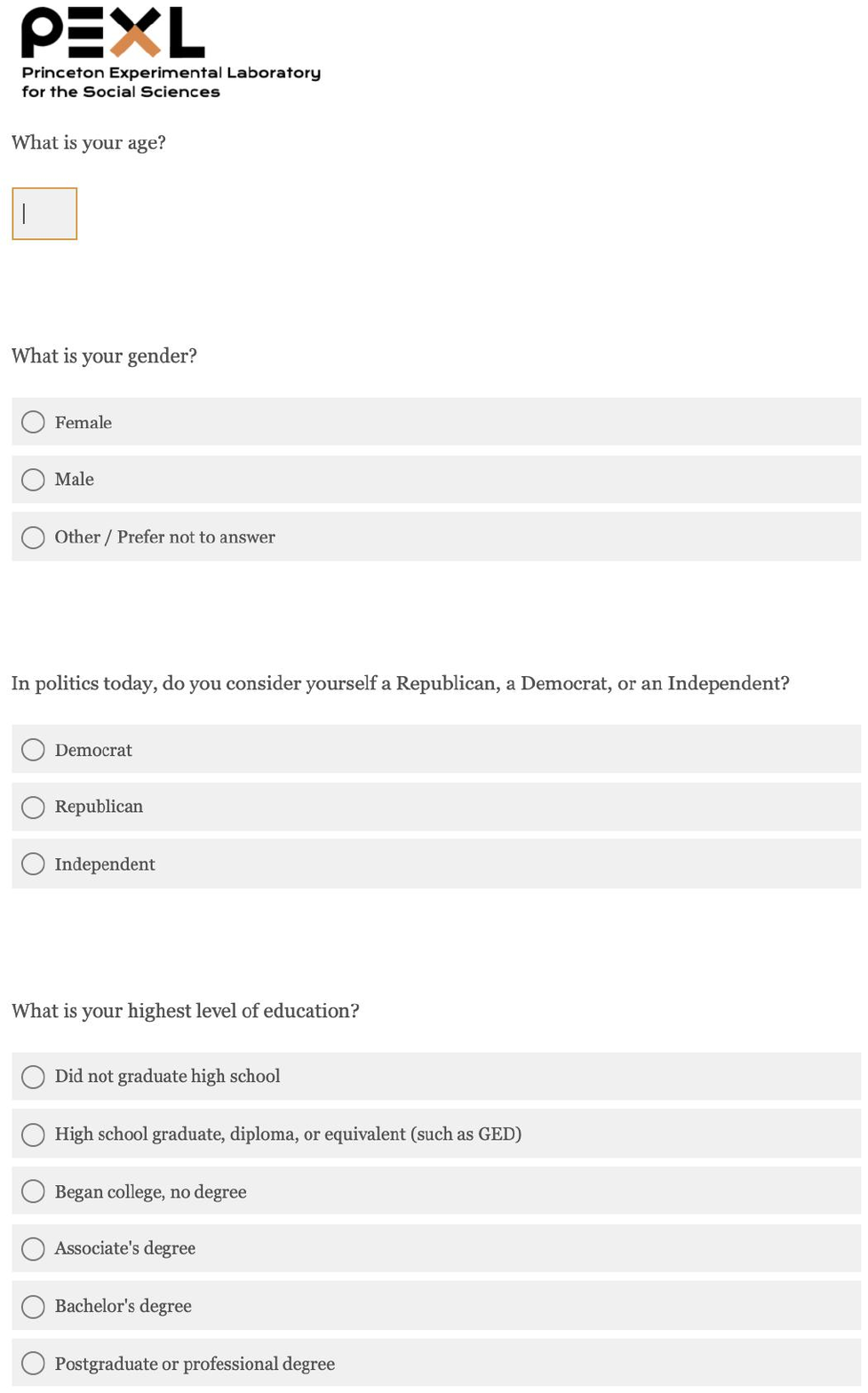}
    \newpage
    \includegraphics[width=.89\textwidth]{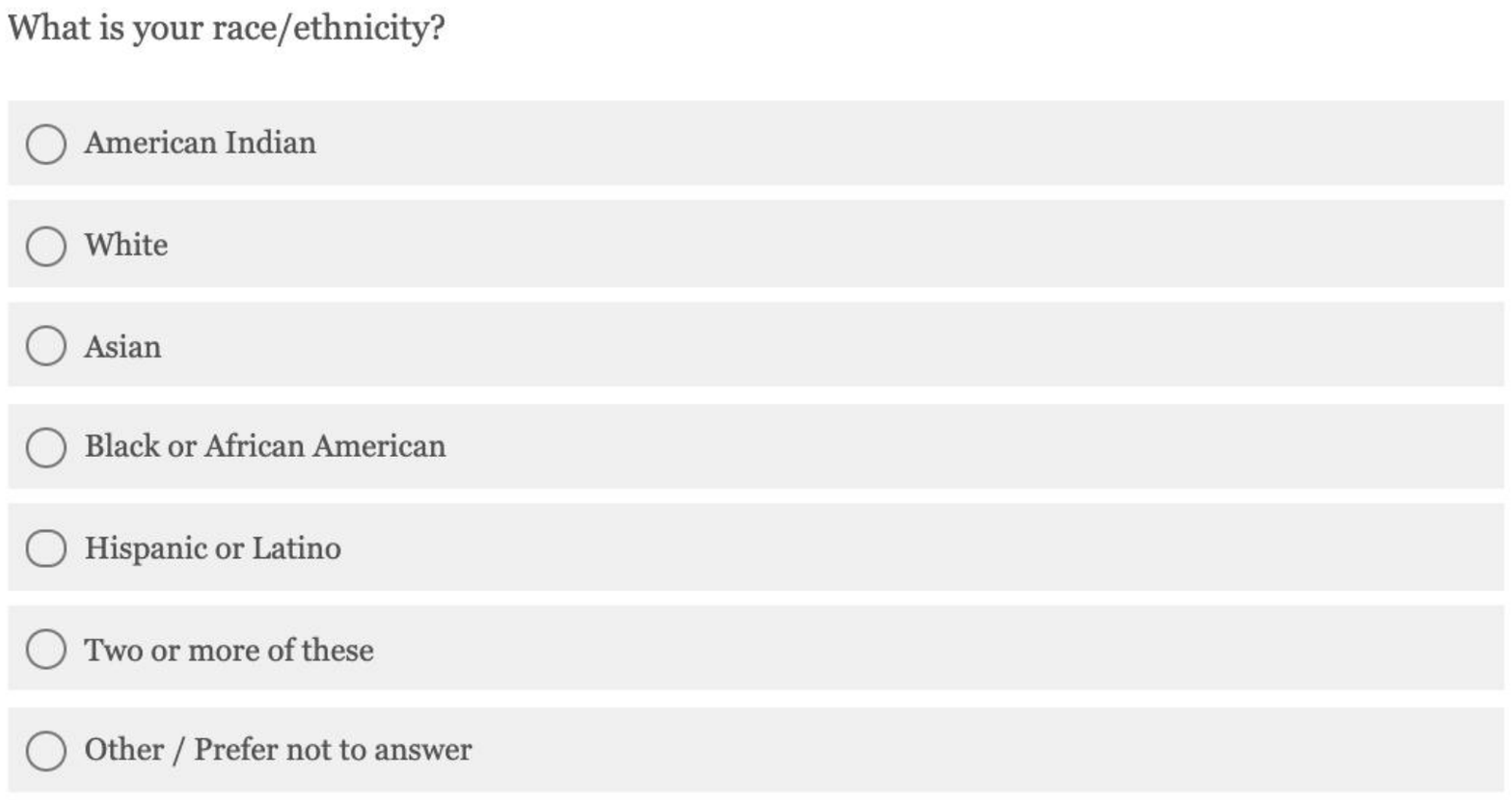}
\end{center}
\newpage
\subsubsection*{Cognitive Reflection Test}
\begin{center}
    \includegraphics[width=\textwidth]{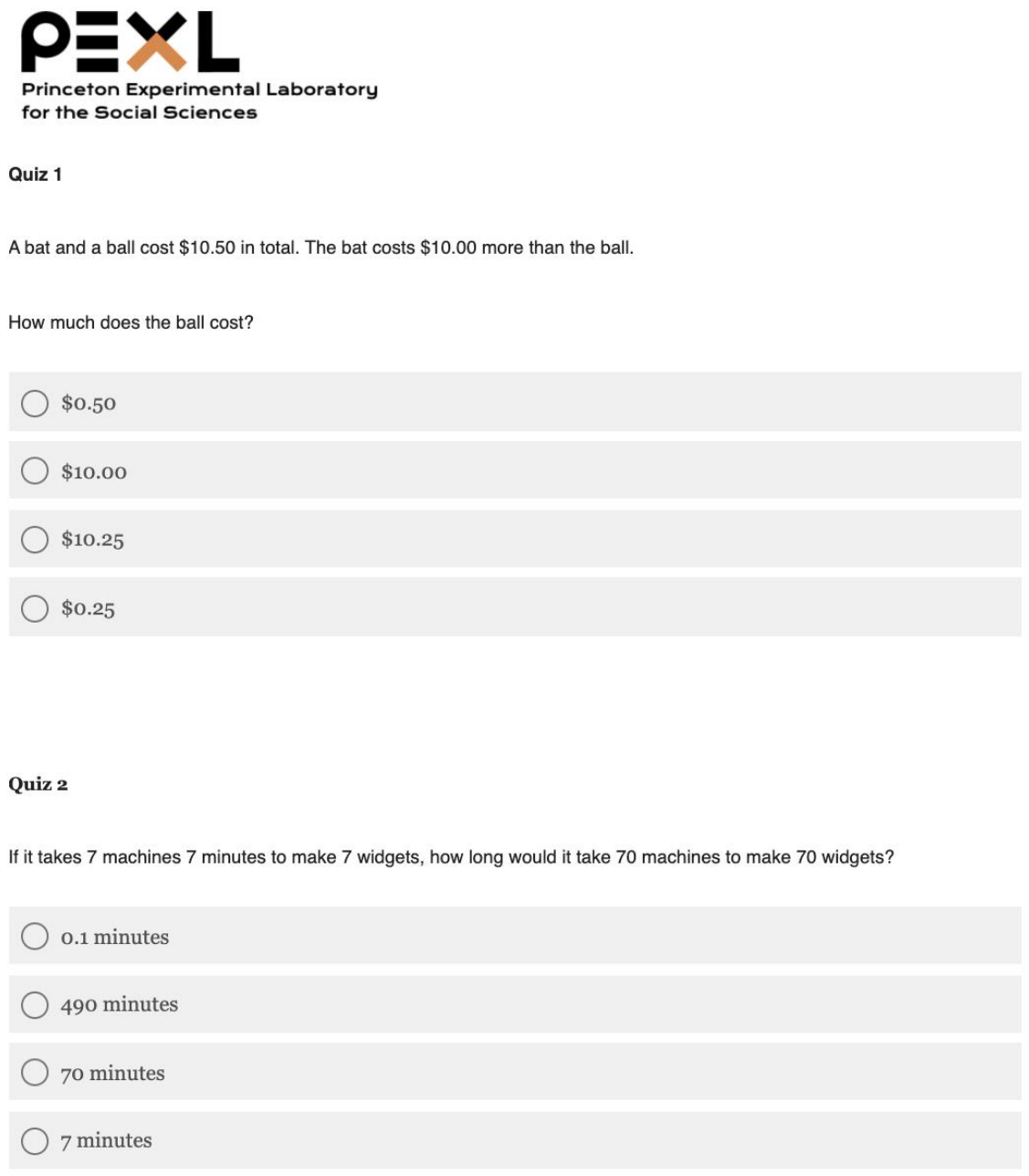}
    \newpage
    \includegraphics[width=\textwidth]{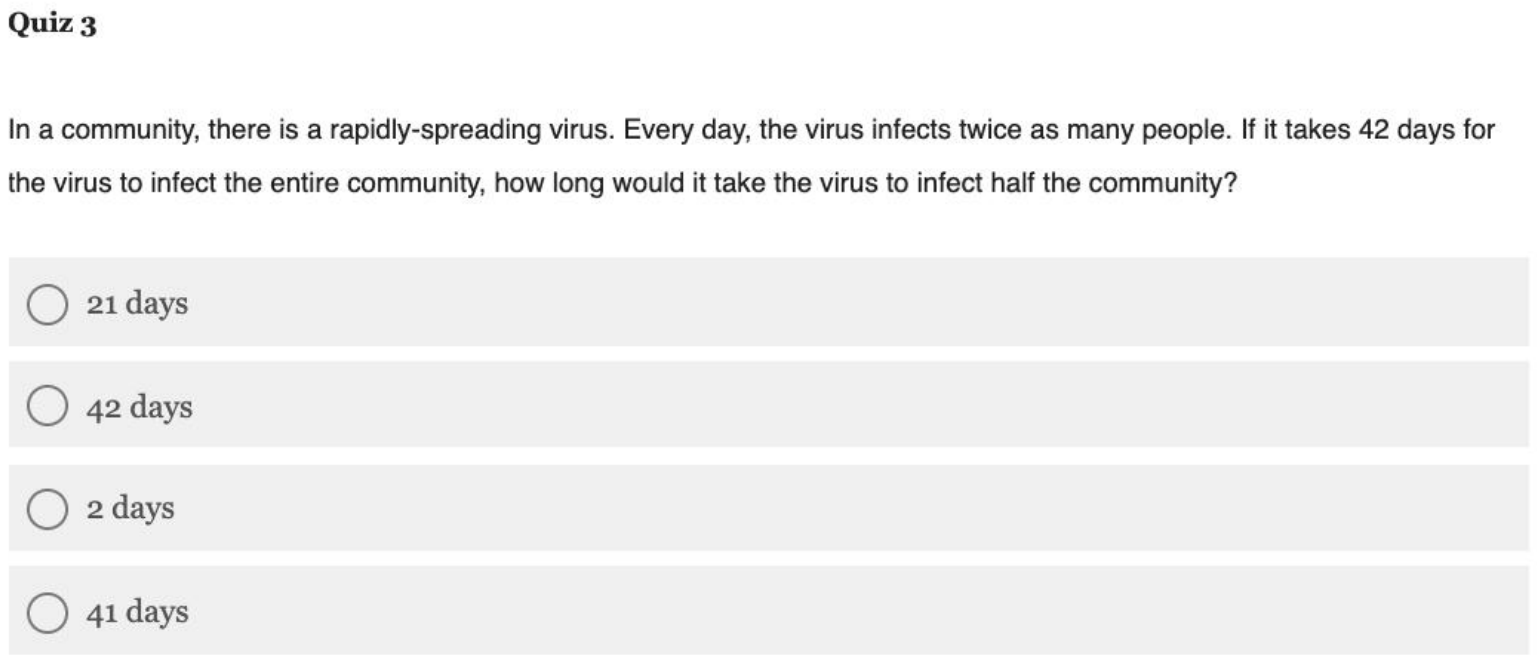}
\end{center}

\clearpage

\subsection{Study 1b}
\subsubsection*{Overview and Instructions}
\begin{center}
    \includegraphics[width=\textwidth]{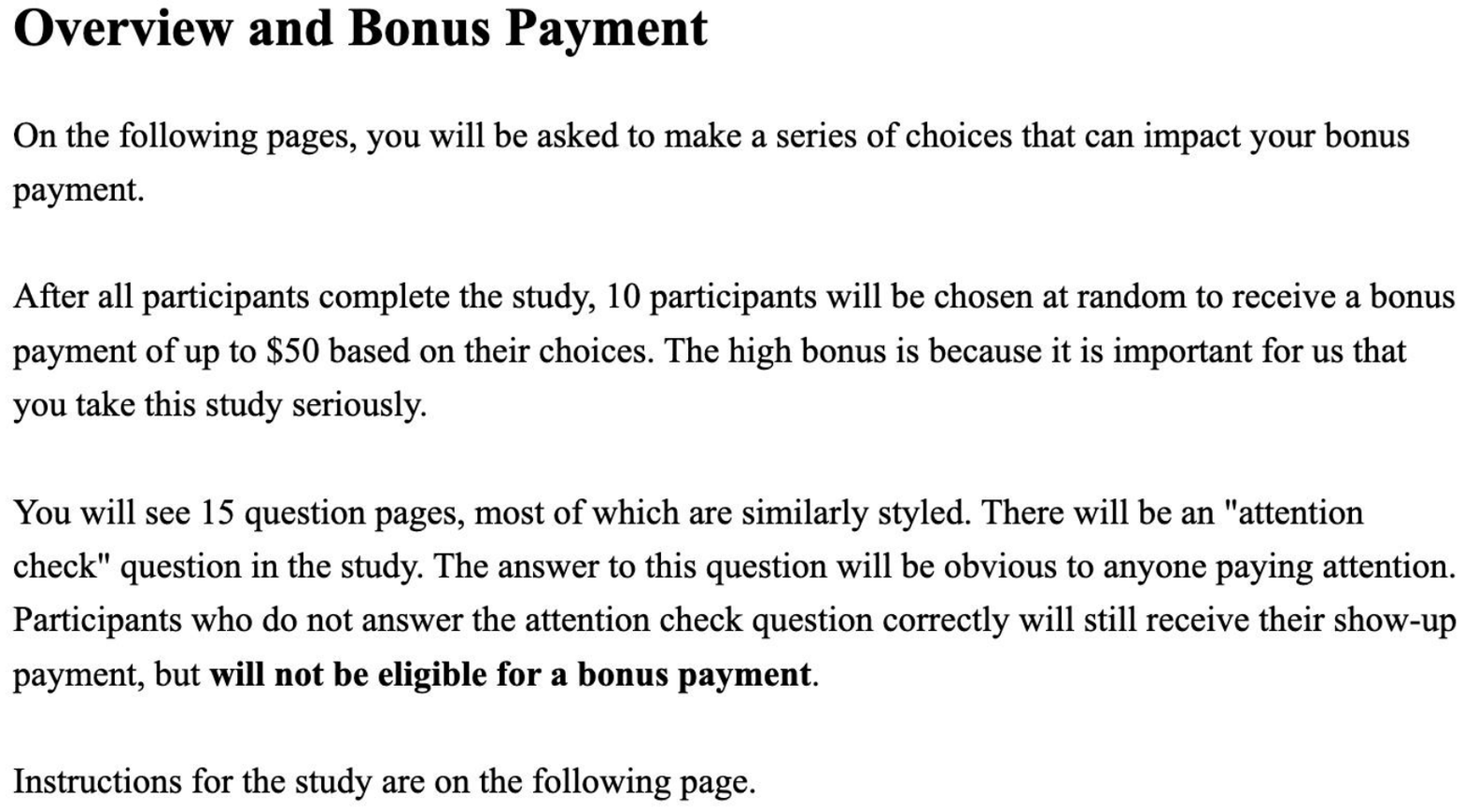}
    \newpage
    \includegraphics[width=\textwidth]{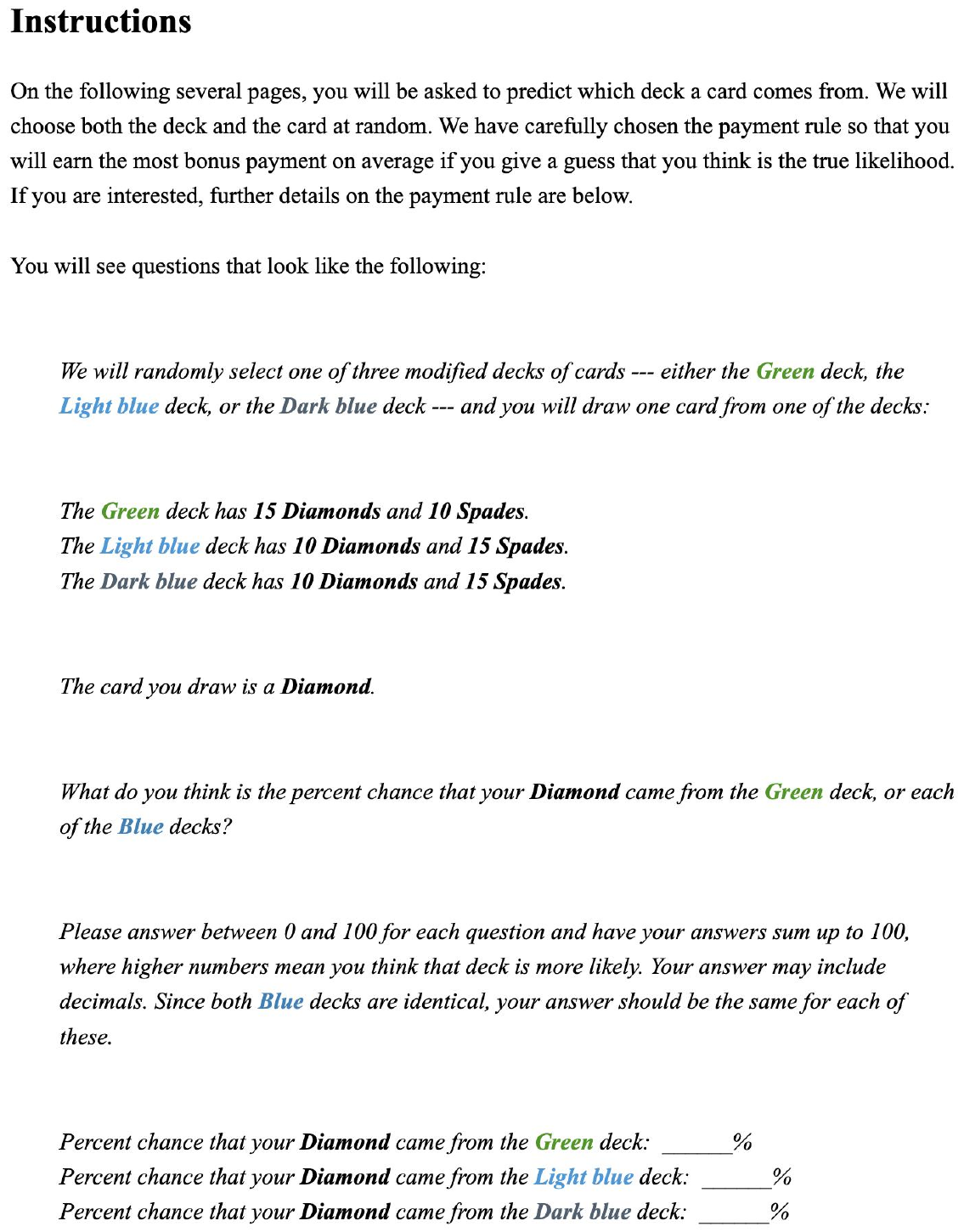}
    \newpage
    \includegraphics[width=\textwidth]{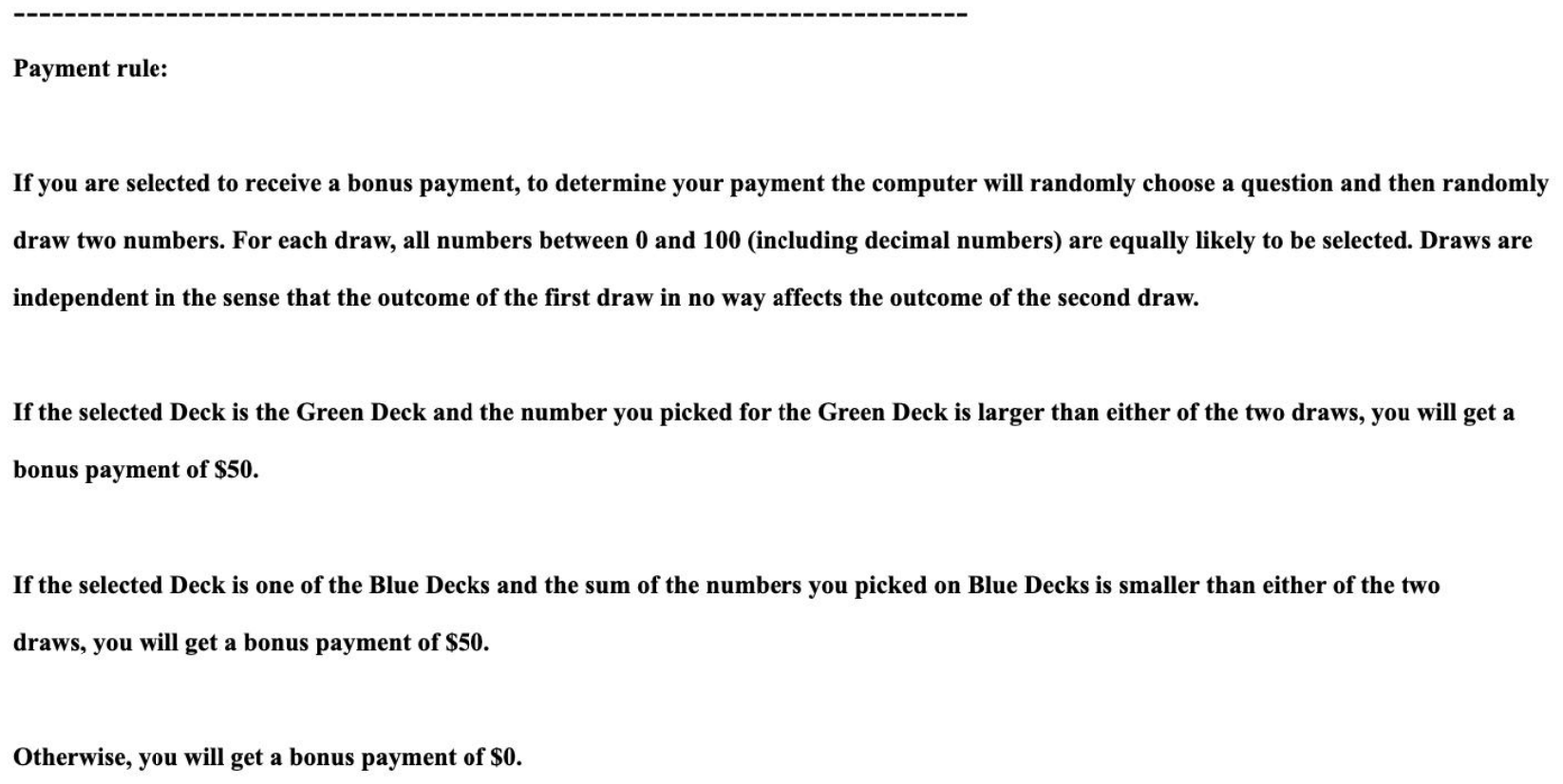}
    \newpage
    \includegraphics[width=\textwidth]{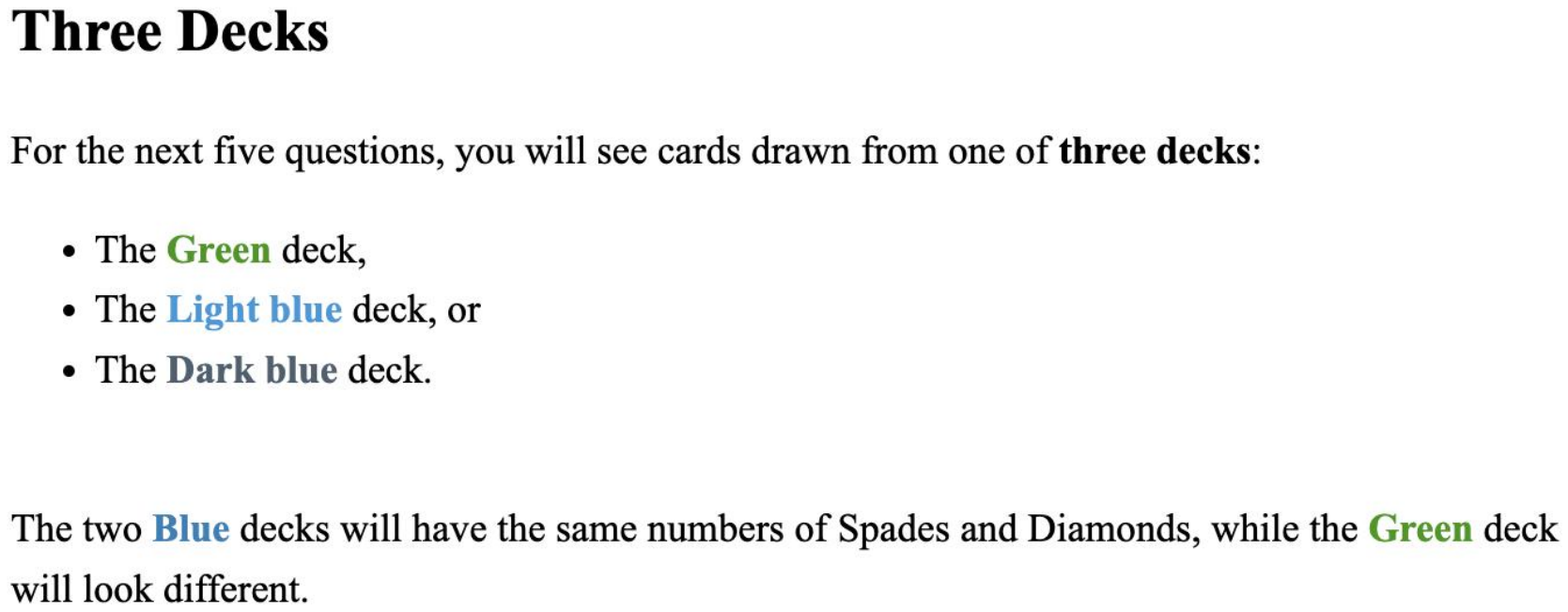}
\end{center}
    \newpage

\subsubsection*{Main Decision Screen}
\begin{center}
    \includegraphics[width=\textwidth]{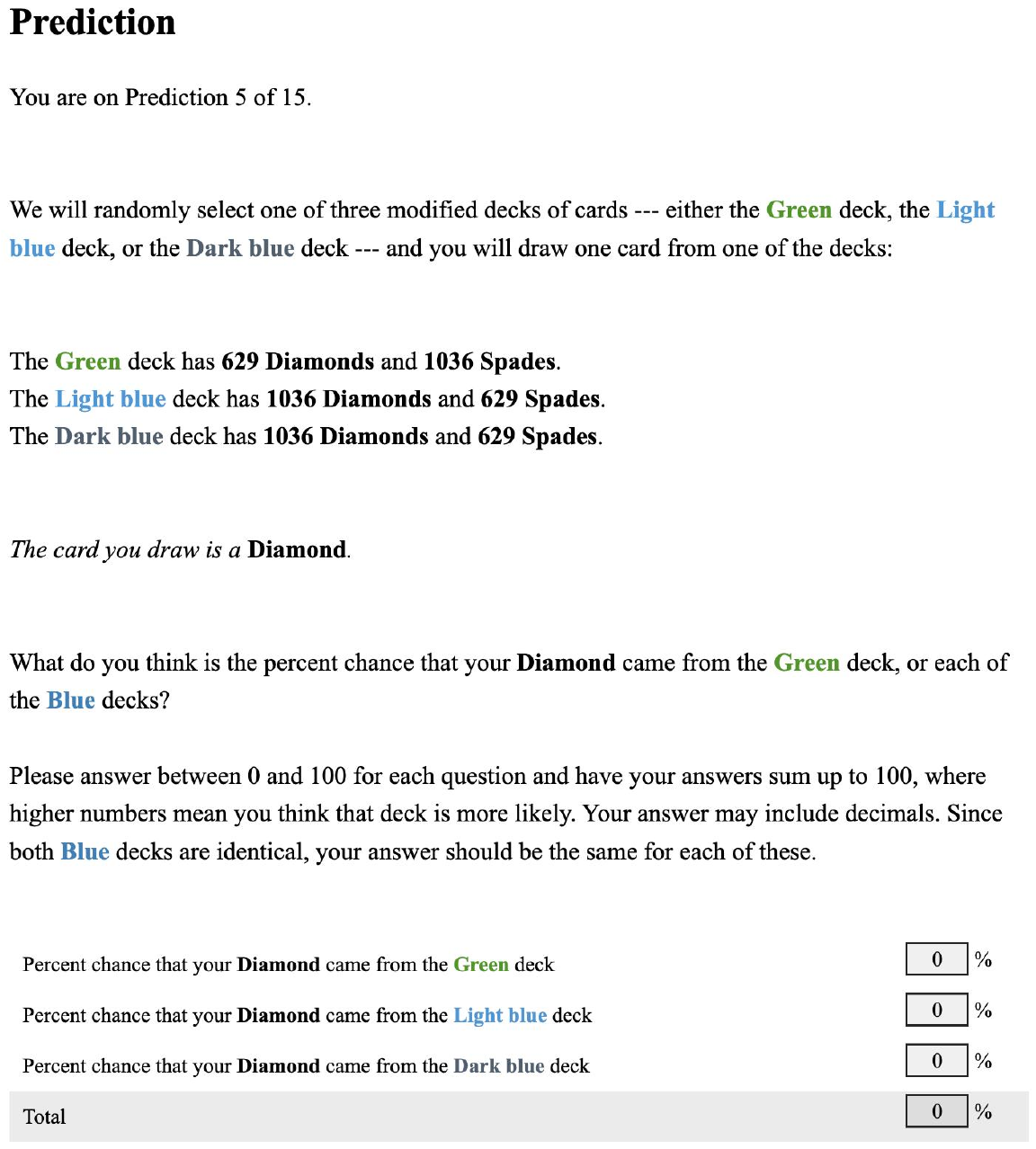}
\end{center}
\newpage

\subsubsection*{Attention Check}
\begin{center}
    \includegraphics[width=\textwidth]{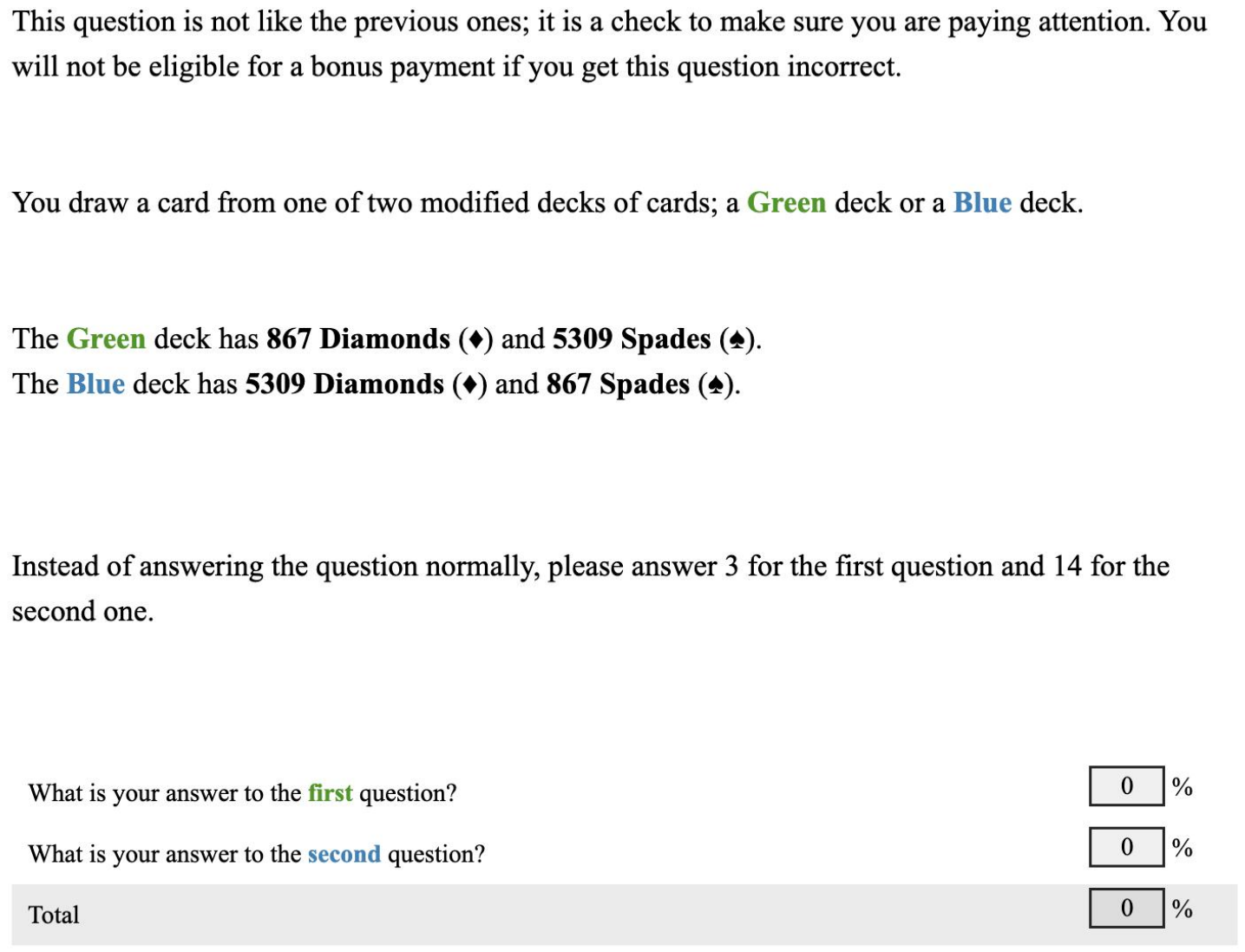}
\end{center}
\newpage

\subsubsection*{Confidence Elicitation}
\begin{center}
    \includegraphics[width=\textwidth]{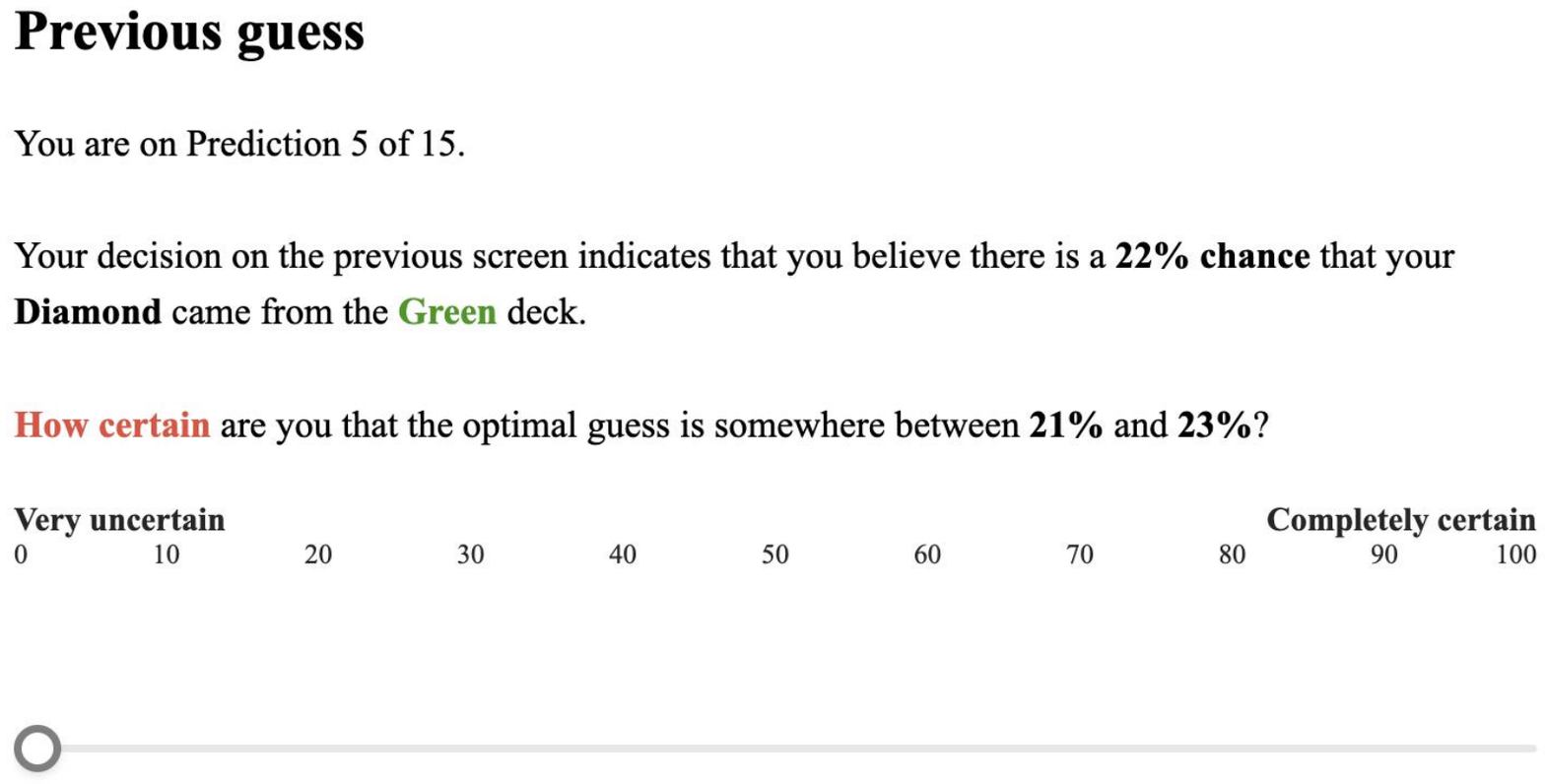}
\end{center}

\clearpage

\subsection{Study 2}
\subsubsection*{Overview and Instructions}
\begin{center}
    \includegraphics[width=\textwidth]{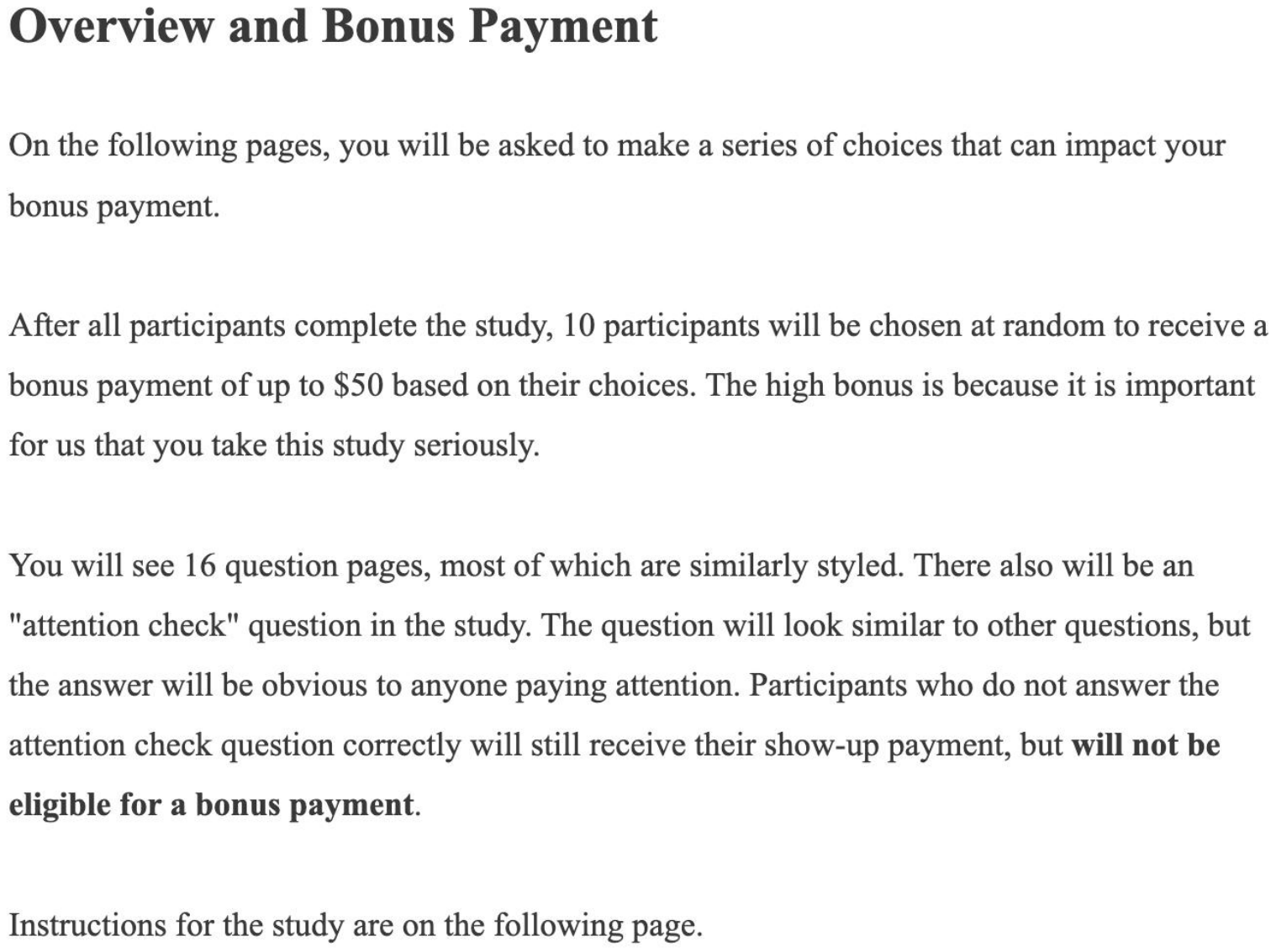}
    \newpage
    \includegraphics[width=\textwidth]{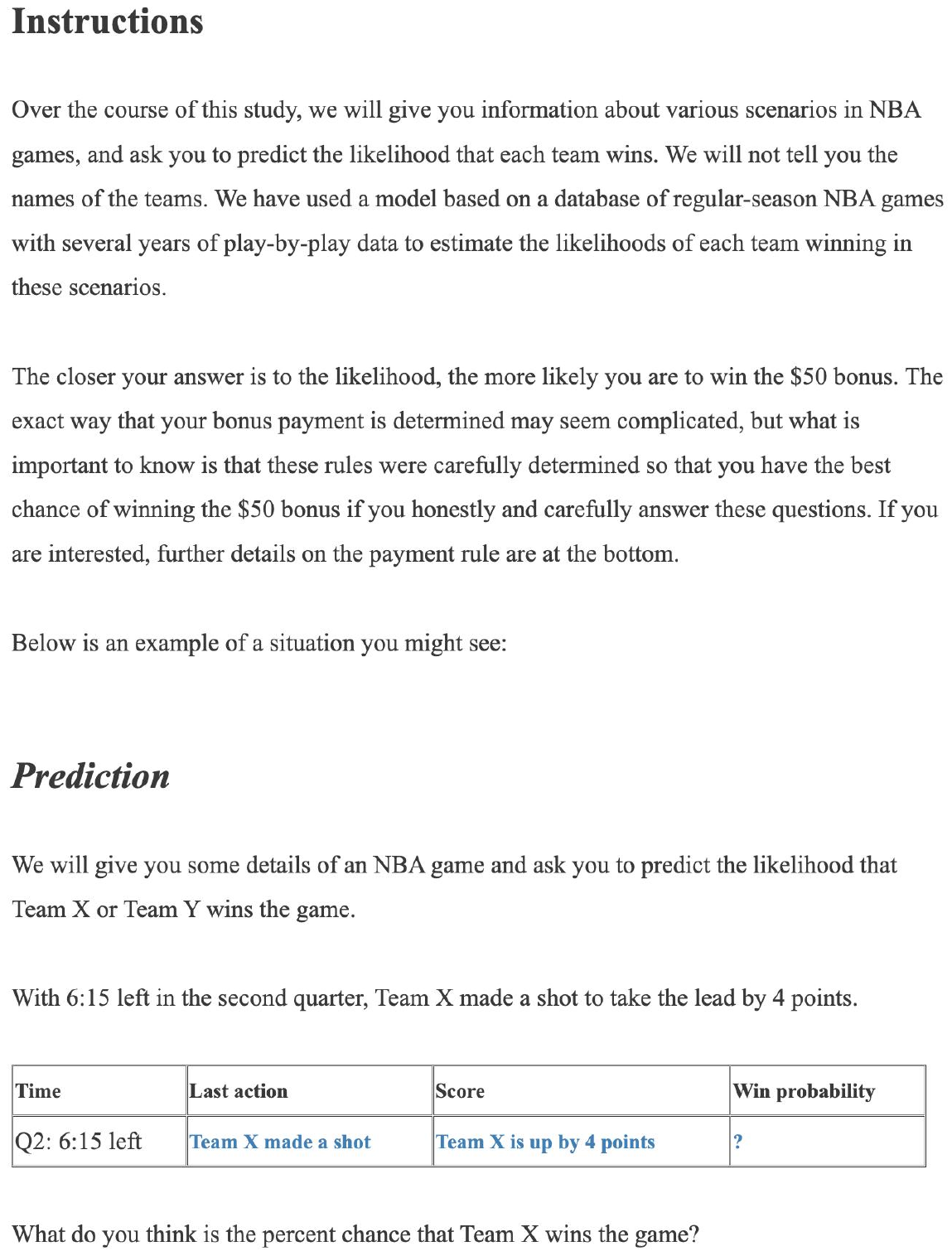}
    \newpage
    \includegraphics[width=\textwidth]{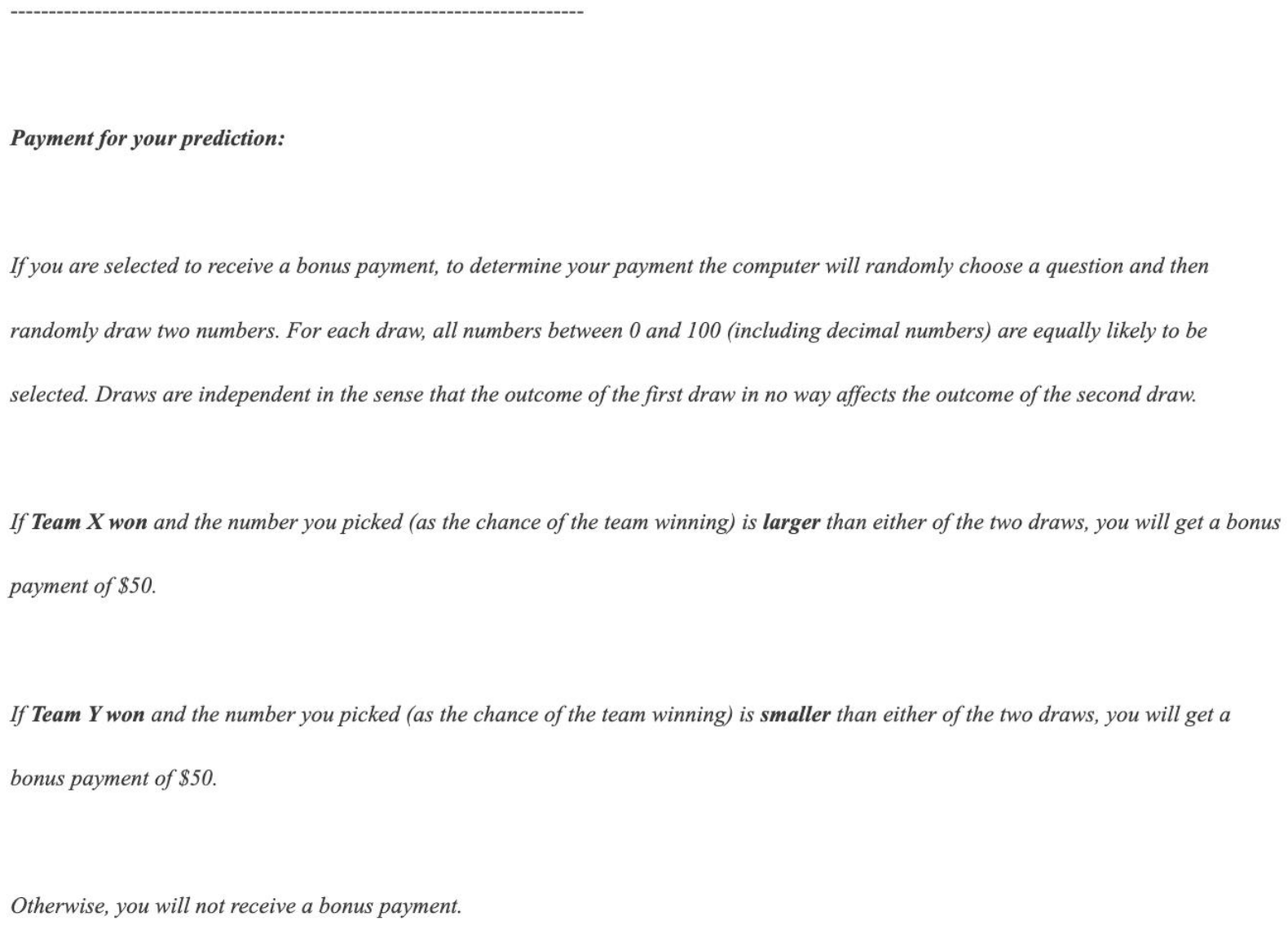}
\end{center}
\newpage

\subsubsection*{Main Decision Screen}
\begin{center}
    \includegraphics[width=.98\textwidth]{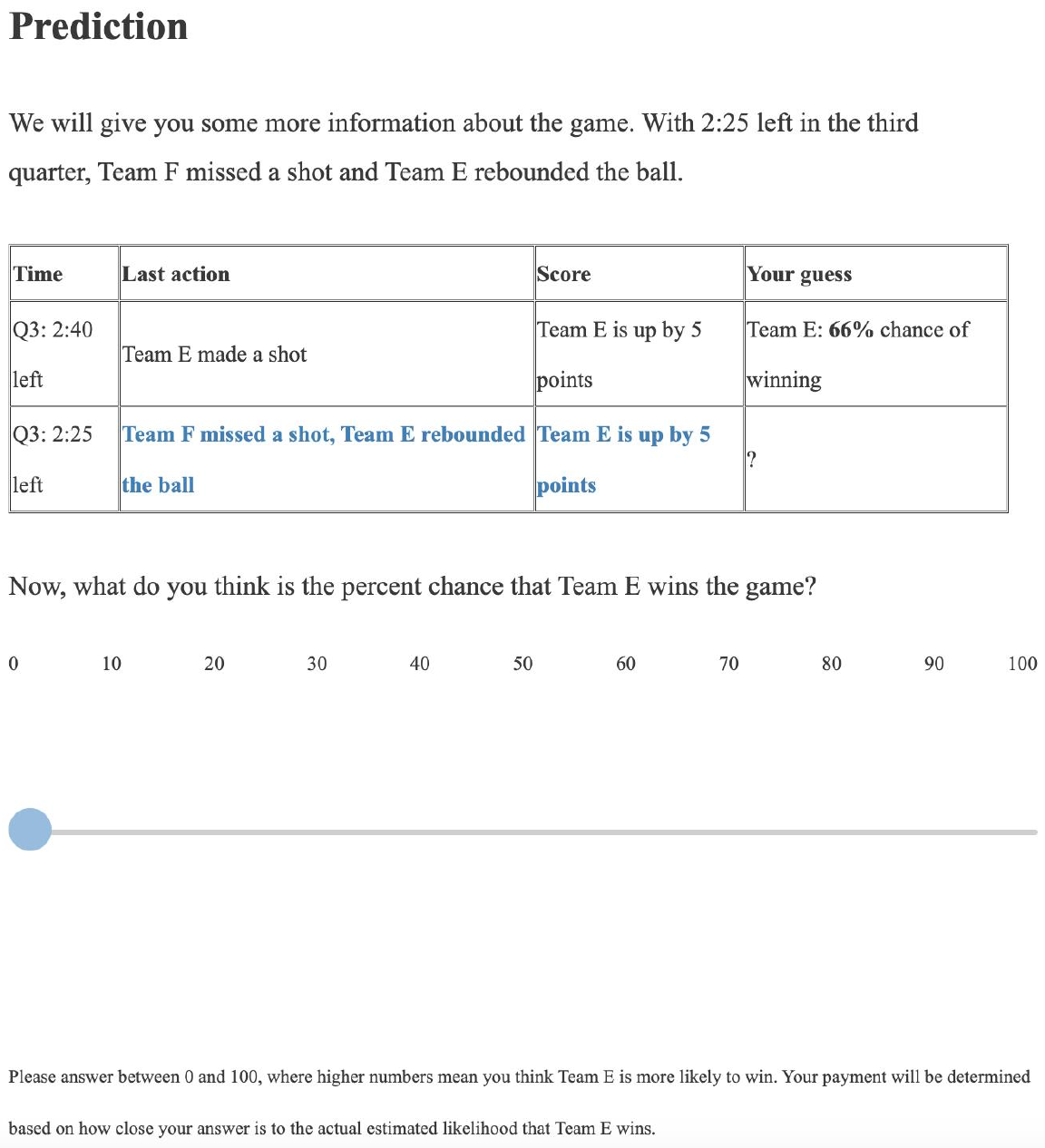}
\end{center}
    \newpage
\subsubsection*{Attention Check}
\begin{center}
    \includegraphics[width=.98\textwidth]{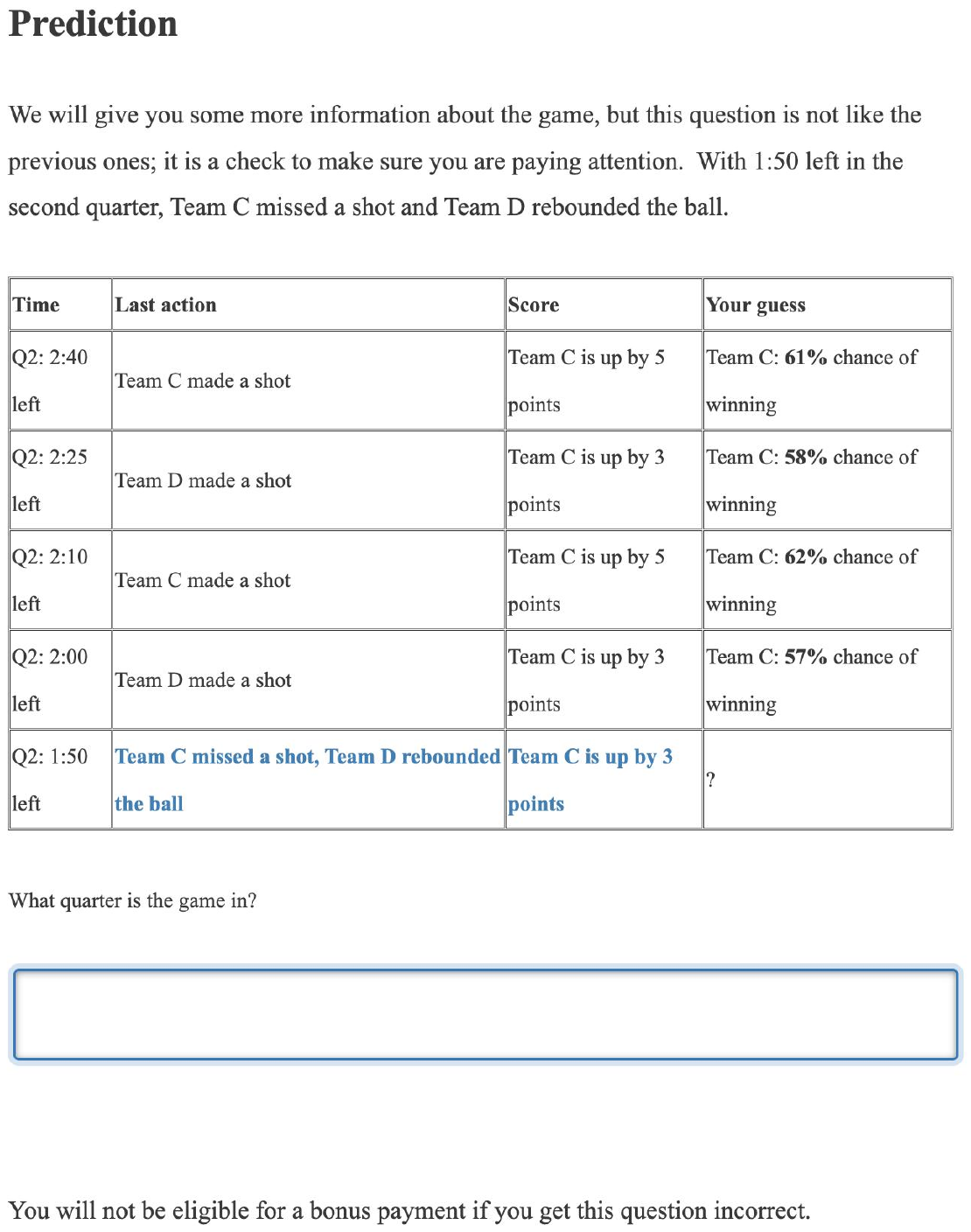}
\end{center}

\end{document}